\documentclass[11pt, oneside]{article}

\usepackage[paper=a4paper, total={16cm,23cm}]{geometry}
\usepackage[T1]{fontenc}
\usepackage{setspace} 
\usepackage[hang]{footmisc} 
\usepackage{titling} 
\usepackage{tocloft} 
\usepackage{parskip} 
\usepackage{comment}
\usepackage{manfnt}
\usepackage{fontawesome}
\newcommand{\mathtree}{\text{\faTree}}
\usepackage{multirow}

\usepackage{amsfonts}
\usepackage{amsmath}
\usepackage{amssymb}
\usepackage{mathrsfs, dsfont, amsthm}
\usepackage{physics}
\usepackage{mathtools}
\usepackage{upgreek}
\numberwithin{equation}{section} 
\usepackage{wasysym}
\usepackage{marvosym}
\usepackage{cancel}

\usepackage{xcolor}
\usepackage{colortbl}
\definecolor{dark-red}{rgb}{0.50,0.12,0.12} 
\definecolor{mblue}{rgb}{0.30, 0.45, 0.70}
\definecolor{mred}{rgb}{0.70, 0.20, 0.20}
\definecolor{mgray}{rgb}{0.63, 0.63, 0.63}

\usepackage{hyperref} 
\hypersetup{colorlinks=true, allcolors=dark-red, linktoc=all}
\usepackage{cite}

\usepackage{graphicx}
\usepackage{tikz}
\usepackage{tikzlings}
\usetikzlibrary{calc, patterns, decorations, decorations.markings, shapes, positioning, intersections, quotes, angles}
\usepackage[subrefformat=parens]{subcaption}
\usepackage{pgfplots}
\pgfplotsset{compat=newest}
\newcommand{\mathdefault}[1][]{}

\def \cA {\mathcal{A}}
\def \cC {\mathcal{C}}
\def \cD {\mathcal{D}}

\def \cF {\mathcal{F}}
\def \cH {\mathcal{H}}
\def \cI {\mathcal{I}}
\def \cJ {\mathcal{J}}
\def \cM {\mathcal{M}}
\def \cN {\mathcal{N}}
\def \cO {\mathcal{O}}
\def \cR {\mathcal{R}}
\def \cS {\mathcal{S}}

\def \rE {\mathrm{E}}
\def \rH {\mathrm{H}}

\def \rK {\mathrm{K}}
\def \rN {\mathrm{N}}

\def \rS {\mathrm{S}}

\def \bbR {\mathbb{R}}

\def \bmu {M}
\def \bnu {N}
\def \brho {K}
\def \bsigma {L}

\def \Osmeared {O}
\def \psib {\psi^{\text{bulk}}}
\def \psibs {\psi^{\text{bulk}*}}
\def \psibb {\psi^{\text{bulk}(*)}}
\def \tpsib {\tilde{\psi}^{\text{bulk}}}
\def \dnPpsi {\qty[\prod \diff P_j\,\psi_j]}
\def \bdypol {\varepsilon}
\def \bulkpol {\xi}

\DeclareMathOperator{\arctanh}{arctanh}

\newcommand{\ep}{\mathrm{e}}
\newcommand{\ic}{\mathrm{i}}
\newcommand{\diff}{\mathrm{d}}
\newcommand{\Diff}{\mathrm{D}}
\newcommand{\fdiff}{\updelta}
\newcommand{\diag}{\mathrm{diag}}
\newcommand{\AdS}{\text{AdS}}

\newcommand{\zh}{z_\mathrm{h}}

\newcommand{\defeq}{\mathrel{\rlap{\raisebox{0.3ex}{$\cdot$}}\raisebox{-0.3ex}{$\cdot$}}=}

\newcommand{\conj}[1]{\mkern 1.3mu\overline{\mkern-1.3mu#1\mkern-1.3mu}\mkern 1.3mu}
\newcommand{\propsiminn}[2]{\mathrel{\vcenter{
  \offinterlineskip\halign{\hfil$##$\cr
    #1\sim\cr\noalign{\kern0pt}#1\propto\cr\noalign{\kern0pt}}}}}

\begin{document}
\begin{titlingpage}
    \vspace*{3em}
    \onehalfspacing
    \begin{center}
    {\LARGE Looking at bulk points in general geometries}
    \end{center}
    \singlespacing
    \vspace*{2em}
      \begin{center}
        \textbf{
        Simon Caron-Huot, Joydeep Chakravarty,
        and Keivan Namjou
        }
    \end{center}
    \vspace*{1em}
    \begin{center}
        \textsl{
        Department of Physics, McGill University \\
        Montr\'eal, QC, Canada \\[\baselineskip]
        }
        \href{mailto:schuot@physics.mcgill.ca}{\small schuot@physics.mcgill.ca},
        \href{mailto:joydeep.chakravarty@mail.mcgill.ca}{\small joydeep.chakravarty@mail.mcgill.ca},
        \href{mailto:keivan.namjou@mail.mcgill.ca}{\small keivan.namjou@mail.mcgill.ca}
    \end{center}
    \vspace*{3em}
    \begin{abstract}    
    The holographic correspondence predicts that certain strongly coupled quantum systems describe an emergent, higher-dimensional bulk spacetime in which excitations enjoy local dynamics. We consider a general holographic state dual to an asymptotically AdS bulk spacetime, and study boundary correlation functions of local fields integrated against wavepackets. We derive a factorization formula showing that when the wavepackets suitably meet at a common bulk point, the boundary correlators develop sharp features controlled by flat-space-like bulk scattering processes. These features extend along boundary hyperboloids whose shape naturally reveals the bulk geometry. We discuss different choices of operator ordering, which lead to inclusive and out-of-time-ordered amplitudes, as well as fields of various spins and masses.
    \end{abstract}
\end{titlingpage}
\tableofcontents
\pagebreak

\section{Introduction}
The striking feature of holograms, in the lay sense of the term, is that no effort is needed to decode them: one simply looks at a holographic printout on a two-dimensional sheet and is fooled into thinking that a three-dimensional object stands in its place. Similarly, a striking feature of holography, in the sense of the AdS/CFT correspondence \cite{Maldacena:1997re, Gubser:1998bc, Witten:1998zw}, is that decoding the hologram requires no particular effort.

To illustrate this, suppose one knew how to realize experimentally a strongly coupled 2+1-dimensional conformal field theory (CFT). We can imagine a sheet of this material hanging on a wall. If the CFT contains operators that couple to our photon and electron, it would be very easy to convince oneself that the CFT is holographic: one would simply throw a charged particle at the sample, shine light at it continuously, and observe the reflected light. The result would be exactly as if the light were reflected off a particle that entered the sample and continued its way into the wall (calculations that demonstrate this and other effects will be discussed in this paper). It would take no effort for our lucky experimentalists to convince themselves that their sample can be equivalently described by a local higher-dimensional theory. In principle, the illusion could be so good that it could be difficult to convince oneself that the sample is really just a sheet, not a portal into a strange new world!

In this experimental sense, one may say that the problem of reconstructing bulk physics from boundary observables is trivial: the system's own dynamics can do it for us for free. Of course, a major difficulty with this kind of experiment is that we do not currently know of a candidate material realizing such a holographic CFT. We will add nothing to this question. Rather, we will consider more thought experiments that explore the self-decoding feature of holographic field theories. Ultimately, we hope these will help us understand the mechanism for the emergence of extra dimensions of space and dynamical gravity from systems that do not have them, with possible lessons for gravity in our world.

Generally, field theory correlators encounter singularities when two points become lightlike. The striking holographic features mentioned above are associated with bulk lightcones and local flat-space processes happening in the bulk \cite{Polchinski:1999ry, Gary:2009ae, Heemskerk:2009pn, Okuda:2010ym, Penedones:2010ue, Fitzpatrick:2011ia, Raju:2012zr, Goncalves:2014ffa, Maldacena:2015iua, Komatsu:2020sag, Caron-Huot:2021enk, Hijano:2019qmi, Duary:2022pyv, Chandorkar:2021viw, Jain:2023fxc, Banerjee:2022oll, vanRees:2023fcf, deGioia:2024yne, Duary:2024kxl}. These features are difficult to explain in terms of boundary kinematics and consequently serve as clear signatures of whether a strongly coupled CFT is holographic or not.

It seems very natural to use these features to map out the geometry of nontrivial bulk spacetimes \cite{Engelhardt:2016wgb, Engelhardt:2016vdk, Engelhardt:2016crc}. The bulk lightcones control the optical properties of holographic theories: why not use optics to decode the hologram? 

The simplest correlators, two-point functions, indeed reveal various interesting signatures notably in black hole backgrounds \cite{Berenstein:2020vlp, Cruz:1994ir, Hashimoto:2018okj, Hashimoto:2019jmw, Kaku:2021xqp, Dodelson:2022eiz, Kinoshita:2023hgc, Dodelson:2023nnr}. However, extracting the full geometry from them amounts to solving an inverse problem that is not entirely straightforward. Using more complex correlators, a method to experimentally measure a bulk (conformal) geometry from $(d+2)$-points correlators has been described in \cite{Engelhardt:2016crc}, which has the advantage of requiring no prior knowledge of bulk equations of motion.

One of the main goals of this paper is to fill this complexity gap: we will derive a factorization formula that predicts sharp features in $n$-point boundary correlators in terms of the bulk geometry and scattering processes, for an arbitrary asymptotically anti-de Sitter spacetime.

Compared with previous studies, we add two ingredients: the use of directional wavepackets and of inclusive (non-time-ordered) observables. These ingredients have appeared in a recent discussion of ``cameras'' \cite{Caron-Huot:2022lff}, which considered certain coincidence limits; here we will discuss general configurations. In the rest of this introduction, we discuss the significance of these two ingredients and summarize our results.

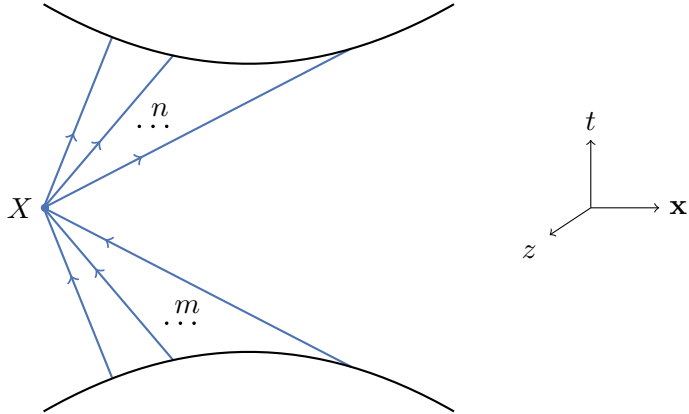
\begin{figure}
    \centering
    \begin{tikzpicture}[scale=.9]
    \draw[color=mblue, thick, decoration={markings, mark=at position 0.7 with \arrow{<}}, postaction=decorate] (1.5,2.347) -- (-3,0);
    \draw[color=mblue, thick, decoration={markings, mark=at position 0.6 with \arrow{<}}, postaction=decorate] (-2.0,2.51) -- (-3,0);
    \draw[color=mblue, thick, decoration={markings, mark=at position 0.6 with \arrow{<}}, postaction=decorate] (-1.1,2.25) -- (-3,0);
    \filldraw (-1.2,1.2) circle (0.5pt);
    \filldraw (-1.4,1.2) circle (0.5pt);
    \filldraw (-1.6,1.2) circle (0.5pt);
    \node [above] at (-1.3,1.2) {$n$};
    \draw[color=mblue, thick, decoration={markings, mark=at position 0.8 with \arrow{>}}, postaction=decorate] (1.5,-2.343) -- (-3,0);
    \draw[color=mblue, thick, decoration={markings, mark=at position 0.6 with \arrow{>}}, postaction=decorate] (-2.0,-2.5) -- (-3,0);
    \draw[color=mblue, thick, decoration={markings, mark=at position 0.6 with \arrow{>}}, postaction=decorate] (-1.1,-2.25) -- (-3,0);
    \filldraw (-0.8,-1.7) circle (0.5pt);
    \filldraw (-1.0,-1.7) circle (0.5pt);
    \filldraw (-1.2,-1.7) circle (0.5pt);
    \node [above] at (-0.9,-1.7) {$m$};
    \filldraw[mblue] (-2.98,0) circle (1.5pt);
    \node [left] at (-3,0) {$X$};
    \draw[thick] (-3,3) to[out=-30,in=-150] (3,3);
    \draw[thick] (-3,-3) to[out=30,in=150] (3,-3);
    \draw[->] (5,0) -- (6,0);
    \draw[->] (5,0) -- (5,1);
    \draw[->] (5,0) -- (4.4,-0.4);
    \node [right] at (6,0) {$\mathbf{x}$}; 
    \node [above] at (5,1) {$t$}; 
    \node [below] at (4.1,-0.4) {$z$}; 
    \end{tikzpicture} \caption{A two-dimensional projection of a $m \to n$ scattering process happening near a bulk point $X$. For massless bulk fields, the boundary endpoints lie along two curves associated with the past and future lightcones of $X$, which we call hyperboloids.}\label{fig:1}
\end{figure}

\paragraph{Using wavepackets to control positions and momenta.}
In strict position space, $(d+2)$ boundary points are needed in order to isolate local physics near a point $X$ in a $(d+1)$-dimensional bulk \cite{Maldacena:2015iua, Engelhardt:2016wgb}. Intuitively, inserting a local operator at a boundary point emits a non-directional pulse that spreads into the bulk at the speed of light (considering massless fields here for definiteness). The intersection of $d+1$ such wavefronts generically singles out a unique bulk point. (This is similar to the number of Global Positioning System satellites needed to uniquely specify one's spacetime coordinate if the receiver does not record the arrival directions of satellite signals.) An additional pulse must then be fine-tuned in order to observe localized features and to force a large amount of energy into all pulses.

In a spacetime with sufficient symmetry, sharp features can also appear in lower-point correlators, however, they then originate from an extended sub-manifold of the bulk \cite{Maldacena:2015iua}.

The restriction to point sources in this context seems artificial since it is relatively easy to control directionality using Fourier transforms. Indeed, wavepackets have been discussed in several early references on bulk-point singularities \cite{Polchinski:1999ry, Giddings:1999jq}. In this paper, we will only consider simple Gaussian wavepackets, which suffice to produce directional beams. A single such beam together with a non-directional pulse can isolate a single bulk point $X$.

The use of wavepackets has a big conceptual payoff: it enables one to map out the bulk geometry in any dimension using low-point correlators. Accounting for bulk momentum conservation, four points will suffice in the general case, while two may suffice if a macroscopic object is already in place in the bulk (one to send light to the object and one to record its reflection). Four is an important psychological threshold that most physicists are not willing to cross. 

\paragraph{Inclusive observables} The other important limitation to overcome can be seen from the need to conserve bulk momentum at the interaction point: $\sum_{i=1}^n P_i=0$. For $m\to n$ ``in-out'' or ``exclusive'' scattering processes of the sort depicted in figure \ref{fig:1}, this precludes access to many interesting regions. We believe this makes it impossible, for example, to access any point inside the photon sphere of a Schwarzschild-AdS black hole using the $(d+2)$-point correlator experiment described in \cite{Engelhardt:2016wgb}. One of the produced particles will necessarily have an inward-directed radial momentum and thus fall into the hole, after which it cannot be recovered using simple measurements at the boundary. Roughly speaking, to conserve momentum in an exclusive experiment around a bulk point $X$, one needs to be able to observe this point from sufficiently many different directions; the set of allowed points seems difficult to describe in general. Certainly, in empty AdS, no such experiment is possible from any compact region of a Poincar\'e patch.

A natural way to evade this constraint is simply to trace out unobserved degrees of freedom. Indeed, the exclusive amplitude in figure \ref{fig:1} is an idealization designed for collider-type situations where one assumes full experimental control over the final state. However, most real experiments are inclusive to some degree and allow unobserved particles to escape. 

As an elementary example, when our eyes record light scattered off an object, what we are really measuring is an inclusive cross-section where all possible final states of the object are summed over. An in-out amplitude which would prescribe the object's final microstate would be far more complex and hardly more useful. Importantly, this tracing out does not need to be taken at the cross-section level and some quantum coherence can be retained. For example, the phase of reflected light could potentially be measured by interfering it with light reflected off a nearby mirror.

Quantum field theory can naturally handle such observables by paying attention to the ordering of operators. As an illustrative example, the amplitude of a coherent light source reflected off a tree can be expressed by the following four-point vacuum correlator:
\begin{equation}\label{schematic correlator intro}
    \expval{\gamma(x_{\smiley}) \, \gamma^\dagger(x_{\sun})}{\mathtree,{\rm in}} = \expval{{\mathtree}(x_1) \, \gamma(x_{\smiley}) \, \gamma^\dagger(x_{\sun}) \, {\mathtree}^\dagger(x_1)}{0},
\end{equation}
where ${\mathtree}^\dagger$ is an operator which creates the tree in some microstate near $x_1$, $\gamma^\dagger(x_{\sun})$ creates a photon in the source, and $\gamma(x_{\smiley})$ absorbs it at the observer's location. Note that \eqref{schematic correlator intro} specifies nothing about the tree's final state, not even whether it survives the experiment (although that is hoped for): the operator ${\mathtree}$ on the left absorbs the \emph{same} tree microstate as was created on the right. One could take linear combinations of \eqref{schematic correlator intro} to turn the tree's initial state into a density matrix, however, the measurement and formula still entail no measurement of its final state.\footnote{Strictly speaking, observables should involve unitaries, for example, $\expval{U^\dagger \gamma(x_{\smiley}) U}{\mathtree,{\rm in}}$ where $U = T \exp(\ic \int \diff x\, \psi^\mu(x)A_\mu(x))$ uses a real test function $\psi^\mu$ to create a photon near the source. To obtain \eqref{schematic correlator intro}, expand in small $\psi$ and discard the ``wrong-order'' term $A_\mu\gamma(x_{\smiley})$, which vanishes for practical purposes.}

The defining feature of the correlator \eqref{schematic correlator intro} is that it is not time ordered: inserting a time ordering symbol would turn it into an uninteresting quantity, where the tree is annihilated before reflecting light. One could of course formally relate it to in-out amplitudes by inserting a complete basis of out (late-time) states:
\begin{equation} \label{schematic correlator 2}
    \mbox{equation~\eqref{schematic correlator intro}} = \sum_{\rm out} \langle \mathtree,{\rm in}|{\rm out}\rangle \, \langle {\rm out}| \gamma(x_{\smiley}) \, \gamma^\dagger(x_{\sun})|\mathtree,{\rm in}\rangle.
\end{equation}
However, this is not necessarily useful. The point is that the matrix element $\langle \mathtree,{\rm in}|\rm out\rangle$ is not simple since the initial microstate is typically not an energy eigenstate (a tree is a living organism that constantly radiates heat, etc.). Thus, instead of formal expressions involving practically incalculable in-out amplitudes, we prefer to directly deal with the in-in correlator in \eqref{schematic correlator intro}.

According to the Schwinger-Keldysh formalism, \eqref{schematic correlator intro} can be calculated using a path integral where it is now be viewed as a \emph{contour-ordered} correlator on a time-folded contour:
\begin{equation}
    \begin{tikzpicture}[scale=0.9]
    \node [right] at (0,-0.15) {$\quad$}; 
    \draw (-1,0.1)--(-1,0) -- (5,0) -- (5,-0.3) -- (-1,-0.3)--(-1,-0.4) ;
    \node [above] at (-0.3,0) {${\mathtree}^\dagger(x_1)$}; 
    \node [above] at (1.1,0) {$\gamma^\dagger(x_{\sun})$}; 
    \node [below] at (-0.3,-0.3) {${\mathtree}(x_1)$};
    \node [above] at (4.2,-0.0) {$\gamma(x_{\smiley})$};
    \node [right] at (5,0.1) {\footnotesize I}; \node [right] at (5,-0.4) {\footnotesize II}; 
    \filldraw (0.7,0) circle (2pt); \filldraw (-0.3,-0.0) circle (2pt); \filldraw (-0.3,-0.3) circle (2pt); \filldraw (4.2,-0.0) circle (2pt);
    \draw (6,0.35) -- (6,-0.15) -- (6.5,-0.15);
    \node [above] at (6.25,-0.15) {$t$}; 
    \end{tikzpicture}\label{moonn}
\end{equation}
While the calculation of these amplitudes may be less familiar from quantum field theory textbooks than that of time-ordered amplitudes (see \cite{Kamenev_2011, Chou:1984es, Calzetta:2008iqa, Haehl:2017qfl} for some reviews), we would like to highlight the conceptual simplicity of the observables being considered. Inclusive amplitudes like \eqref{schematic correlator intro} have been studied scientifically long before any time-ordered amplitude.\footnote{As far as we know, the first scientific measurements of (the square of) a time-ordered amplitude by humanity happened with Rutherford probing the atom. While planets and other macroscopic bodies can be approximated as point particles when their microstates are traced over, their \emph{actual} in-out amplitudes are never measured nor calculated in practice.}

It is easy to see why inclusive correlators circumvent the difficulties of accessing certain regions mentioned above: there is no obstruction to sending a probe toward a black hole and measuring how it reflects light. At the risk of belaboring the point, when we communicate with spacecraft to learn about the environment in the outer Solar System, we get data without recovering the spacecraft. Technically, the correlator in \eqref{schematic correlator intro} still needs to conserve momentum, but this becomes easier to satisfy because the energy-momentum of the leftmost operator counts with a minus sign.

Inclusive amplitudes, and more general out-of-time-ordered amplitudes, have been discussed recently in an S-matrix context \cite{Caron-Huot:2023vxl}, where it was shown how to make sense of strings of creation/annihilation operators acting in the past and future in arbitrary order. This was needed, notably, to understand crossing symmetry at higher multiplicity \cite{Caron-Huot:2023ikn}. In the setup considered in this paper, the ordering of bulk creation-annihilation operators will be determined simply by the ordering of boundary operators.

While the ingredients we will add to the bulk{-}point story---wavepackets and operator orderings---are conceptually straightforward, we find that they make the story more broadly appealing and motivate us to revisit it with increased generality and detail. 

\begin{figure}[t!]
    \centering
    \begin{tikzpicture}[scale=0.75]
    \draw[color=mblue, decoration={markings,mark=at position 0.5 with \arrow{>}},postaction=decorate,thick] (-2,-0.5) -- (-3.5, 1) arc (225:135:.05) coordinate (turning);
   \draw[dashed] (turning)+(-1,-1)-- ++(0.95,0.95); 
    \draw[color=mblue!50, decoration={markings,mark=at position 0.75 with \arrow{>}},postaction=decorate, thick] (turning) arc (135:45:.05) -- ++(1.40,-1.40) ++(0.2,-0.2)--(-.24,-2.13);
    \draw[color=mblue,decoration={markings,mark=at position 0.7 with \arrow{<}},postaction=decorate, thick] (0,2.12) -- (-2,-0.5);
    \draw[color=mblue,decoration={markings,mark=at position 0.25 with \arrow{>}},postaction=decorate, thick] (-2.2,-2.62) -- (-2,-0.5);
    \draw[color=mblue,decoration={markings,mark=at position 0.5 with \arrow{>}},postaction=decorate, thick] (-1.4,-2.32) -- (-2,-0.5);
    \draw (-1.7,-0.5) node[anchor = west]{$X$};
    \draw (0, 2.12) node[anchor = south]{$\gamma_3$};
    \draw (-.24,-2.13) node[anchor = north]{$O_4$}; 
    \draw (-2.2,-2.62) node[anchor = north]{$O_1^\dagger$}; 
    \draw (-1.4,-2.32) node[anchor = north]{$\gamma_2^\dagger$}; 
    \draw[thick] (-3,3) to[out=-30,in=-150] (3,3) node[anchor = north west]{$H^+(X)$};
    \draw[thick] (-3,-3) to[out=30,in=150] (3,-3) node[anchor = south west]{$H^-(X)$};
    \end{tikzpicture}
    \caption{The expectation value $\expval{O_4(x_4) \gamma_3(x_3) \gamma_2^\dagger(x_2) O_1^\dagger(x_1)}{\Psi}$ displays sharp features when $x_1$, $x_2$, and $x_4$ approach the past lightcone and $x_3$ approaches the future lightcone of a common bulk point $X$. The dashed line indicates the (arbitrary) late time at which the ``out'' states are summed over in \eqref{schematic correlator 2}.}\label{fig:incl-exp}
\end{figure}
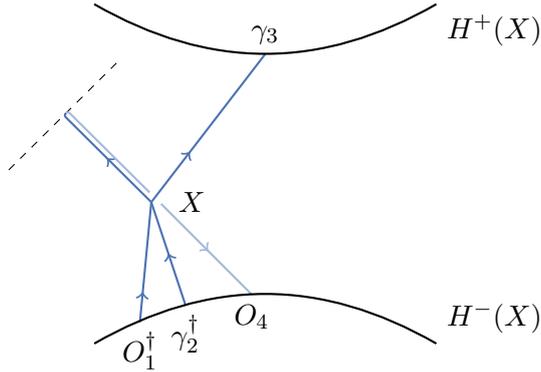

\subsection{Methods and results}
The main goal of this paper is to derive a factorization formula characterizing the ``singularities,'' or more precisely sharp features, of boundary correlation functions (in-in or in-out) in holographic theories in terms of flat-space-like scattering amplitudes in the bulk. Our method is to use the bulk equations of motion (linearized around a given background) to study the propagation of high-frequency waves from the boundary to an interaction point.

Using a semiclassical WKB analysis, we will show how a boundary wavepacket, characterized by a central position $x$, boundary momentum $p$, and width $\sigma$, aimed toward a bulk point $X$, creates near $X$ a superposition of plane waves with approximately definite bulk momentum $P$:
\begin{equation} \label{O vs a intro}
 O_{x,p,\sigma}^\dagger \quad \text{on the boundary} \quad \leftrightarrow \quad
 \int \diff (\fdiff P) \, \psib_{x,p,\sigma;X}(\fdiff P) \, a^\dagger_{X,P + \fdiff P} \quad \text{in the bulk.}
\end{equation}
Here the bulk creation-annihilation operators are defined by applying a naive mode decomposition in a flat neighborhood of the bulk point $X$ (see equation~\eqref{eq:mode-exp} below). The bulk momentum $P$ is related to the boundary momentum $p$ by solving geodesic equations in the underlying smooth geometry. A specific Gaussian wavepacket in the boundary thus leads to a specific Gaussian wavepacket in the bulk, whose parameters account for the spreading of the wave and other propagation effects, as well as for the impossibility of defining bulk momentum to infinite accuracy using a finite neighborhood.

In fact, we will get four variants of \eqref{O vs a intro}: boundary operators can either create or absorb bulk particles (as determined by the sign of the energy), and these can arrive near $X$ either from the past or future (as determined by the insertion time). These will map respectively to bulk creation/annihilation operators $a^\dagger$ or $a$ (that act in the past) or $b^\dagger$ or $b$ (future).

A well-known technical subtlety is that the WKB approximation fails near the AdS boundary, where physical momenta go to zero. Our main assumption (which could potentially be generalized) is that the geometry is asymptotically AdS. This enables us to treat the non-WKB region exactly, and we can then match onto the WKB region. We will obtain versions of \eqref{O vs a intro} for both scalar and spinning particles and for massless and massive bulk fields. For heavy bulk fields (dual to operators of dimension $\Delta\gg 1$), the amplitude will be suppressed by a quantum tunneling factor, reflecting the fact that a massive particle cannot classically reach the boundary.

Our working assumptions will be that the bulk excitations carry large energies compared to the ambient geometry, that their wavefunctions do not overlap except near a certain bulk point $X$, and that all interactions away from that point can be neglected. In this case, \eqref{O vs a intro} makes it straightforward to relate a $n$-point boundary correlator to another $n$-point bulk amplitude, schematically:
\begin{equation} \label{intro boundary vs bulk}
\begin{array}{c}
    \expval{\text{(product of $\Osmeared_{x,p,\sigma}, \Osmeared^\dagger_{x,p,\sigma}$'s)}}{\Psi} \\[.5em]
    \updownarrow \\[.5em]
    \int \prod_{i=1}^n \diff (\fdiff P_i) \, \psib_i(\fdiff P_i)
    \expval{\text{(product of $a$, $a^\dagger$, $b$, $b^\dagger$'s)}}{0}^{\text{bulk}}_{\text{near X}}.
\end{array}
\end{equation}
The bulk amplitude on the right could be equivalent to a standard in-out amplitude (as is always automatically the case for $n=4$ \cite{Caron-Huot:2022lff}) or it could be a generalized non-time-ordered amplitude. Precise examples are given in section \ref{s:dynamics:sing}; some situations where the stated assumptions are not satisfied are discussed in \S\ref{ssec:landau} and \S\ref{ssec:discussion}.

\begin{figure}[t!]
    \centering
    \begin{tikzpicture}[scale=.9]
        \filldraw[mblue!50] (0.38,-0.6) -- (.98,1.05) -- (.9,1.06) -- (.3,-.6);
        \filldraw[mblue!50] (0.38,-0.6) -- (-0.82,.88) -- (-0.9,.9) -- (.3,-.6);
        \filldraw[mblue!50] (0.38,-0.6) -- (.67,-1.72) -- (0.59,-1.73) -- (.3,-.6);
        \filldraw[mblue!50] (0.38,-0.6) -- (-.6,-2.36) -- (-.68,-2.37) -- (.3,-.6);
        \draw (-2,2.25) -- (-2.0,-3.3)  (2,-3.3) -- (2,2.25);
        \draw[dashed] (2,-3.3) arc (0:180:2 and 0.25);
        \draw (2,-3.3) arc (0:-180:2 and 0.25);
        \draw (2, 2.25) arc (0:360:2 and 0.25);
        \draw[thick] (-1.99,1.2) node[anchor=east]{\footnotesize $H^+(X)$} .. controls (-1.6,0.9) and (1.6,0.5) .. (1.99,0.7);
        \draw[thick,dashed] (-1.99,1.2) .. controls (-1.6,1.5) and (1.6,1.1) .. (1.99,0.7);
        \draw[thick,dashed] (-1.99,-2.3) node[anchor=east]{\footnotesize $H^-(X)$} .. controls (-1.6,-2.6) and (1.6,-2.2) .. (1.99,-1.8);
        \draw[thick] (-1.99,-2.3) .. controls (-1.6,-1.9) and (1.6,-1.5) .. (1.99,-1.8);
        \node[right] at (0.34,-0.6) {$X$};
        \filldraw[mblue] (0.34,-0.6) circle (2pt);
    \end{tikzpicture}
    \caption{Boundary curves lightlike to the bulk point $X$. At the center of empty global AdS are concentric circles, but for a general point not at the center, the boundary curves are ellipsoids.}\label{fig:gads}
\end{figure}
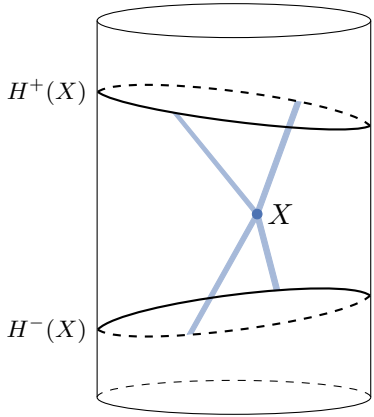

As far as the boundary is concerned, the loci of the resulting signals are controlled by the geometry of null geodesics in the bulk and the induced boundary hyperboloids. Perhaps the simplest example is that of empty AdS in global coordinates, where the hyperboloids associated with a point at the center are simply concentric circles (see figure~\ref{fig:gads}). The singularities that occur as multiple operators approach these circles have been discussed in \cite{Maldacena:2015iua}, and more recently related to measurements along a small ``celestial sphere'' surrounding $X$ in \cite{Alday:2024yyj}. Hyperboloids in more general geometries are described in our companion paper \cite{Caron-Huot:2025she}, which discusses various specific examples as well as their general properties such as parallax and implications of causality.

{\textbf{Notation and convention:}} We will use capital letters to label bulk quantities and lowercase letters for boundary quantities. For instance, we use $X$ and $P$ for the bulk position and the bulk momentum, whereas $x$ and $p$ denote the boundary position and the momentum respectively. When bulk and boundary quantities agree, such as in the presence of translation symmetries, we write both in lowercase. Capital Latin indices $(M, N, \dots)$ will be used for bulk $(d+1)$-dimensional vectors while Greek indices ($\mu, \alpha, \dots$) are used for boundary $d$-vectors or the non-radial part of bulk vectors. Finally, we set the $\AdS$ length scale to $R_\AdS = 1$ unless otherwise specified.

\section{Bulk and boundary observables}\label{s:overview}
Given an arbitrary geometry, our objective is to study the physics inside a small local neighborhood of a given bulk point. We will be interested in essentially flat-space scattering processes happening within this neighborhood. To this end, in this section, we provide an overview of different bulk aspects in a language that will conveniently relate to boundary correlators.

\subsection{Quantum fields near a bulk point} \label{ssec:near X}
We first analyze the behavior of a bulk field in the vicinity of a bulk point $X$ within an asymptotically AdS$_{d+1}$ spacetime. We assume that sufficiently close to $X$ we have an approximate translation invariance that allows us to write a mode expansion in terms of plane waves.

For concreteness, let us discuss here a scalar field $\Phi(X)$ which is minimally coupled to the background geometry,
\begin{equation}\label{eq:scalaraction}
    S[\Phi] = - \frac{1}{2} \int \diff^{d+1}X \, \sqrt{-g} \, \qty(g^{\bmu\bnu} \partial_\bmu \Phi \partial_\bnu \Phi + m^2 \Phi^2)+S_{\rm int}[\Phi],
\end{equation}
where we will neglect the interactions until we consider multiple excitations. Near $X$, we can assume that the metric is constant up to the second order:
\begin{equation}
    g_{MN}(X + \fdiff X) = g_{MN}(X)+\mathcal{O}(\fdiff X^2).
\end{equation}
In geodesic normal coordinates, for example, the quadratic correction can be written in terms of the curvature tensor at $X$. Such curvature corrections will be beyond the accuracy of our approximations. One could furthermore choose local coordinates such that the constant matrix $g_{MN}(X)$ is equal to the $(d+1)$-dimensional Minkowski metric, however for our applications it will be preferable to keep the expressions covariant.

To describe scattering states covariantly we start by imagining a foliation of the spacetime (in a neighborhood of $X$) into time slices $X^\bmu = (T, \mathbf{X})$ where $\mathbf{X}$ denotes the spatial components in the bulk, i.e., both the radial and the transverse coordinates. We use the corresponding ADM decomposition of the metric,
\begin{equation} \label{ADM}
    \diff s^2 = -\rN^2 \diff T^2 + h_{ij} \qty(\diff \mathbf{X}^i + \rN^i \diff T) \qty(\diff \mathbf{X}^j + \rN^j \diff T).
\end{equation}
This decomposition of the metric allows us to write its determinant in the simple form $\sqrt{-g} = \rN \sqrt{h}$. We define a future-directed unit covector $n_M = \rN \delta^0_M$ which satisfies $n \cdot n = -1$.

We want to perform a mode decomposition of the free scalar field in terms of plane waves. We can normalize the creation and annihilation operators $a^\dagger_{X, P}$ and $a_{X, P}$ to satisfy the following commutation relation:
\begin{equation} \label{eq:a comm}
    \big[ a_{X,P}, a^\dagger_{X,P'} \big] = 2 \sqrt{h} \, n_\bmu P^\bmu \, (2\pi)^{d} \delta^{d}\qty(\mathbf{P}_N - \mathbf{P}_N'),
\end{equation}
where each $P$ satisfy the mass shell condition $g^{MN}(X)P_MP_N+m^2=0$. Using $\sqrt{-g} = \rN \sqrt{h}$, it can be checked that this commutation relations is invariant under spatial diffeomorphisms, even time-dependent ones. For the Minkowski metric, this reduces to the usual relativistic commutation relation. Within the considered flat region the field can then be expanded as\footnote{Global mode expansions in AdS have been discussed in many references \cite{El-Showk:2011yvt, Papadodimas:2012aq, Papadodimas:2015jra, Gadde:2022ghy}. Here we are specifically interested in describing the local physics near a bulk point using the simplest possible functions, i.e., $\ep^{\pm \ic P \cdot \fdiff X}$ plane waves, rather than hypergeometric functions.}
\begin{equation}
    \Phi(X + \fdiff X) = \int \frac{\diff^{d} \mathbf{P}_\bullet}{(2\pi)^{d}} \, \frac{1}{2 \sqrt{h} \, n_\bmu P^\bmu} \qty( a_{X, P} \, \ep^{\ic P \cdot \fdiff X} + a^\dagger_{X, P} \, \ep^{- \ic P \cdot \fdiff X}),
\end{equation}
or equivalently,
\begin{equation}\label{eq:mode-exp}
    \Phi(X + \fdiff X) = \int \frac{\diff^{d+1} P_\bullet}{(2\pi)^{d}}\frac{\delta(P^2+m^2)\theta(P^0)}{\sqrt{-g}}  \qty( a_{X, P} \, \ep^{\ic P \cdot \fdiff X} + a^\dagger_{X, P} \, \ep^{- \ic P \cdot \fdiff X}).
\end{equation}
The notation $P_\bullet$ emphasizes that the integral measure is over the variables with lower indices, ensuring the Jacobian of the transformation is properly included via $1/\sqrt{-g}$.

We stress that the modes are defined specifically about $X$. Under a change of coordinates, $P_\bmu$ transforms like a cotangent vector at $X$. The second form of the mode expansion, which is equivalent to the first upon integrating out $P_0$, is manifestly covariant. The definitions can be checked to be compatible with the equal time canonical commutation relation on a Cauchy slice:
\begin{equation}\label{eq:canonical-commutation}
    \qty[\Phi(T, \mathbf{X}),\, - n^\bmu \partial_\bmu \Phi (T, \mathbf{X}')] = \frac{\ic \delta^d (\mathbf{X} - \mathbf{X}')}{\sqrt{h}}.
\end{equation}

As we will show in the next section, the momentum $P$ will be related to boundary data via geodesic evolution. For instance, in empty AdS in Poincar\'e coordinates $X^\bmu = (X^\mu, z)$, $P$ takes the following form:
\begin{equation} \label{eq:mass shell}
    P_M = \qty(- E, \mathbf{p}, \pm \sqrt{E^2 - \mathbf{p}^2-m^2/z^2}),
\end{equation}
with $p_\mu=(-E, \mathbf{p})$ the boundary momentum. In general, for metrics that admit isometries, momentum components that correspond to Noether charges can be matched directly between the bulk and boundary.

In the presence of interactions, it will be important to distinguish whether the creation or annihilation operators are applied (slightly) in the past or future of the interaction point. Following \cite{Caron-Huot:2023vxl}, we denote as $a / a^\dagger$ the operators that appear when the mode expansion \eqref{eq:mode-exp} is used on an early time slice, and we label as $b / b^\dagger$ the operators that appear when the same mode expansion is applied on a late time slice:
\begin{equation}\begin{aligned}
  a_{X,P}^\dagger/a_{X,P}:& \mbox{ create/absorb particles in the past of $X$}, \\
  b_{X,P}^\dagger/b_{X,P}:& \mbox{ create/absorb particles in the future of $X$}.
\end{aligned}\end{equation}
These correspond to the ``in'' and ``out'' operators of old scattering theory. The operators $b^\dagger$ and $b$ satisfy among themselves the same commutation relations as the $a$'s (cf. \eqref{eq:a comm}), whereas commutators between $a$'s and $b$'s are not simple.

In this language, for example, the conventional $m\to n$ scattering amplitude for generic momenta is the matrix element
\begin{equation} \label{eq: ordinary amp}
    \expval{b_{P'_{m+n}}\cdots b_{P'_{m+1}}\, a^\dagger_{P_m}\cdots a^\dagger_{P_1}}{0} = \sqrt{-g}\, (2\pi)^{d+1}\delta^{d+1}\qty(\textstyle{\sum}_{j=1}^{m+n} \alpha_j P_j) \ic \cM\qty(s_{ij}=-(\alpha_iP_i{+}\alpha_j P_j)^2),
\end{equation}
where we use $\alpha_j=+1$ for outgoing momenta (associated with $a$ and $b$ operators) and $\alpha_j=-1$ for incoming momenta ($a^\dagger$ and $b^\dagger$ operators).
Note that, in accordance with the equivalence principle, the amplitude depends on Mandelstam invariants which should be computed using the local metric $g_{\bmu\bnu}(X)$.

As discussed in \cite{Caron-Huot:2023vxl}, it makes sense to also consider more general amplitudes defined by arbitrary strings of $a$, $b$, $a^\dagger$ and $b^\dagger$ operators, not necessarily time-ordered. These include the various inclusive observables sketched in the introduction. Whether a given boundary excitation reaches $X$ from the past or future determines which bulk operator to use, as will be exemplified below.

The justification for using flat-space formulas in the curved bulk spacetime is that we will consider excitations of $\Phi$ that are actually superpositions of the plane waves discussed above. The important assumptions, as stated above \eqref{intro boundary vs bulk}, are that i) these form localized wavepackets that only overlap in a flat neighborhood $\cR$ of $X$, and, ii) interactions happening outside $\cR$ can be neglected. Possible generalizations of ii) are discussed in \S\ref{ssec:discussion}. In our final formulas, \eqref{eq: ordinary amp} will always be integrated against wavepackets that account for the momentum uncertainty caused by the finite size of the region $\cR$.

\subsection{Boundary wavepackets}\label{ssec:wavepackets}
From the perspective of the boundary field theory, we will be considering correlation functions of local operators $O(x)$ integrated against wavepackets that have both a well-defined position and momentum. The idea is to create bulk excitations that follow well-defined paths. For convenience, we will restrict attention to Gaussian wavepackets 
\begin{equation}
    \psi_{p,\sigma}(\fdiff x) = \exp(\ic p_\mu \fdiff x^\mu - \frac{1}{2} \qty(\sigma^{-1})_{\mu\nu} \fdiff x^\mu \fdiff x^\nu), \qquad \det \sigma > 0,
\end{equation}
which smear the operator in the position space. Here $\sigma^{\mu\nu}$ is a positive matrix which defines the wavepacket's time and spatial widths. It could be, for example, diagonal:
\begin{equation}
    \qty(\sigma^{-1})_{\mu\nu} \fdiff x^\mu \fdiff x^\nu = \frac{(\fdiff t)^2}{\sigma_{tt}^2} + \sum_{i=1}^{d-1} \frac{(\fdiff x_i)^2}{\sigma_{ii}^2} \qquad \mbox{(diagonal example).}
\end{equation}
Thus one can choose to adjust the standard deviations of wavepackets in position and time separately using the matrix $\sigma$.

Taking the convention that the boundary momentum always has a positive energy: $p^0>0$, and assuming (with no loss of generality) that the considered operator $O$ is Hermitian, we can then distinguish two types of operators according to whether they absorb or create excitations:
\begin{equation}\label{eq:bdryops}
    \begin{split}
    \Osmeared_{x,p,\sigma} &= \int \diff^d \qty(\fdiff x) \, \psi^*_{p,\sigma}(\fdiff x) \, O(x + \fdiff x), \\
    \Osmeared^\dagger_{x,p,\sigma} &= \int \diff^{d} \qty(\fdiff x) \, \psi_{p,\sigma}(\fdiff x) \, O(x + \fdiff x),
    \end{split}
\end{equation}
such that $O^\dagger_{x,p,\sigma} = (\Osmeared_{x,p,\sigma})^\dagger$. We stress that there is no mass shell condition on boundary momenta, although we normally assume that $p$ is timelike so that the corresponding excitation can become on-shell in the bulk (see equation~\eqref{eq:mass shell}). The above definitions include standard plane waves as a special case with $\sigma=\infty$, which we will often abbreviate by dropping the $\sigma$ and $x$ subscripts: $O_p$ and $O_p^\dagger$.

Alternatively, one could define \eqref{eq:bdryops} in Fourier space and treat each wavepacket as a superposition of plane waves with definite momentum. This perspective, combined with the WKB approximations detailed in section \ref{s:dynamics}, will make it straightforward to relate the boundary wavepackets to the bulk plane waves introduced in \S\ref{ssec:near X}.

A given bulk point $X$ can be reached either by inserting boundary operators in its past or future lightcone. When relevant, we distinguish these two options by a subscript on the coordinate: $x_\pm$. Table \ref{t:modes} displays the general map between boundary and bulk operators. For example, the inclusive expectation value shown in figure \ref{fig:incl-exp} will map to a bulk amplitude as follows:
\begin{equation} \label{incl example}
    \expval{O_{x_4-,p_4,\sigma_4} O_{x_3+,p_3,\sigma_3} O_{x_2-,p_2,\sigma_2}^\dagger O_{x_1-,p_1,\sigma_1}^\dagger}{\Psi} \quad\longrightarrow \quad \expval{a_4 b_3 a^\dagger_2 a^\dagger_1}{0}^{\rm bulk}.
\end{equation}

\begin{table}[t]
    \centering
    \renewcommand{\arraystretch}{1.5}
    \setlength{\tabcolsep}{12pt}
    \arrayrulecolor{gray}
    \begin{tabular}{|c|c|c|c|c|}
    \hline
    Bulk operator & Boundary operator & Process \\ \hline
    $a_{X,P}$ & $\Osmeared_{x_-,p,\sigma}$ & Early absorption \\ \hline
    $a^\dagger_{X,P}$ & $\Osmeared^\dagger_{x_-,p,\sigma}$ & Early emission \\ \hline
    $b_{X,P}$ & $\Osmeared_{x_+,p,
    \sigma}$ & Late absorption \\ \hline
    $b_{X,P}^\dagger$ & $\Osmeared^\dagger_{x_+,p,\sigma}$ & Late emission \\ \hline
\end{tabular}
\caption{Correspondence between in and out bulk creation/annihilation operators in the vicinity of a bulk point $X$ and boundary operators inserted in its past or future.}\label{t:modes}
\end{table}

Finally, while we will not perform calculations in the boundary theory in this paper, let us briefly review the standard procedure to account for operator orderings in field theory \cite{Streater:1989vi} (see \cite{Cornalba:2006xk, Hartman:2015lfa, Maldacena:2015waa, Haehl:2017qfl} for discussions in related contexts). Starting with the Euclidean correlator (or a Lorentzian correlator with spacelike separated insertions) denoted by
$$\expval{\, O_n(x_n) \dots O_1(x_1) \, }{\Psi},$$
one simply continues the time $x_i$ keeping a small imaginary shift in time:
\begin{equation}
    t_i \to t_i - \ic \epsilon_i, \qquad \epsilon_{i} < \epsilon_{i+1}.
\end{equation}
Thus the operator ordering is specified by the ordering of the imaginary part of times---the largest $\epsilon$ placed to the extreme left. The prescription is valid when $| \Psi \rangle$ is an arbitrary state, including our cases of interest such as the vacuum state or the thermal state. A simple way to visualize it is to imagine that the time contour (see \eqref{moonn}) is always slightly tilted downward; this ensures convergence of the path integral.

This prescription works for an arbitrary number of time-folds. An example of an observable whose path-integral calculation requires two time-folds is the out-of-time-order correlator:
\begin{equation} \label{otoc example}
    \expval{O_{x_4-,p_4,\sigma_4} O_{x_3+,p_3,\sigma_3} O_{x_2-,p_2,\sigma_2}^\dagger O_{x_1+,p_1,\sigma_1}^\dagger}{\Psi} \quad\longrightarrow \quad \expval{a_4 b_3 a^\dagger_2 b^\dagger_1}{0}^{\rm bulk}.
\end{equation}
The double-coincidence limit of this correlator ($x_3\to x_1$, $x_4\to x_2$) was shown in \cite{Caron-Huot:2022lff} to be controlled by two bulk geodesics which interact by exchanging a bulk field, such as the graviton. Singular features when geodesics intersect can provide ``camera'' images of moving bulk particles.

In fact, the bulk amplitudes in \eqref{incl example} and \eqref{otoc example} are both equal to the same ordinary $2\to 2$ amplitude $\expval{b_4b_3a_2^\dagger a_1^\dagger}{0}$. This is due to the trivial evolution (stability) of one-particle states, as was explained in \cite{Caron-Huot:2023vxl}. For four-point conformal correlators in the vacuum state, the equivalence can also be seen directly on the CFT side by tracking the complex branch of cross-ratios \cite{Caron-Huot:2022lff}. In excited states, equations \ref{incl example} and \ref{otoc example} represent genuinely distinct CFT correlators that will be controlled by the same bulk amplitudes.

As a final technical comment, when considering correlators with different operator orderings, we take identical copies of the geometries to extend the Schwinger-Keldysh time-fold into the bulk, as prescribed in \cite{Herzog:2002pc, Son:2009vu, Skenderis:2008dg}. For a thermal state, there is an additional thermal circle in the Euclidean direction, which will play no role in this paper; since our bulk point will always lie outside the horizon, we expect this procedure to give equivalent answers as for the holographic Schwinger-Keldysh prescriptions of \cite{Glorioso:2018mmw, Chakrabarty:2019aeu, Loganayagam:2022zmq}.

While the constructions described in this paper are rather general, we will often illustrate them on a generic translation-invariant and spatially isotropic geometry:
\begin{equation}\label{eq:pbhmetric}
    \diff s^2 = \frac{1}{z^2} \qty(-A(z) \, \diff T^2 + \delta_{ab} \, \diff X^a \diff X^b + \frac{\diff z^2}{B(z)}).
\end{equation}
The boundary of the spacetime is at $z = 0$, where $A=B=1$. This geometry includes empty AdS in Poincar\'e coordinates, where $A=B=1$ everywhere, and planar black holes, where $A(z) = B(z)=1 - \frac{z^d}{\zh^d}$. We will generally keep $A$ and $B$ unspecified in order to illustrate and test the generality of the WKB formalism.

\section{Dynamics toward and near a bulk point}\label{s:dynamics}
The key step of our calculation will be to understand the bulk excitation $\Phi(X)$ created by a boundary wavepacket \eqref{eq:bdryops}:
\begin{equation}\label{eq:bdry-to-bulk-cfn}
    \Phi_{x,p,\sigma}(X) = \expval{\Phi(X) \, \Osmeared_{x, p,\sigma}^\dagger }{\Psi}.
\end{equation}
We assume that the state $|\Psi\rangle$ is dual to a classical asymptotically AdS bulk geometry and use the linearized equations of motion in that background to propagate $\Phi$. We will focus on situations where the wavepacket is rapidly oscillating compared with other bulk scales, justifying a WKB approximation.

For a massless scalar field, for example, if a null geodesic $X(\lambda)$  passes through a bulk point $X$, then near this point we will find that the phase of the wavefunction \eqref{eq:bdry-to-bulk-cfn} behaves as:
\begin{equation}
    \Phi_{x,p,\sigma}(X + \fdiff X) \propto \exp(\ic P_\bmu \fdiff X^\bmu),
\end{equation}
where the bulk momentum $P_\bmu$ is obtained from the boundary data as
\begin{equation}\label{eq:bulkmomentumdef}
    P_\bmu = g_{\bmu\bnu} \frac{\diff X^\bnu}{\diff \lambda}, \qquad \text{with} \qquad \lim_{z \to 0} P_\mu = p_\mu, \quad \text{and} \quad \lim_{z \to 0} P_z = \sqrt{-\eta^{\mu\nu} p_\mu p_\nu}.
\end{equation}
Here $g_{\bmu\bnu}$ is the local bulk metric at the point $X$, $p$ is the momentum on the boundary at $z = 0$, and $\lambda$ is an affine coordinate along the geodesic. This framework enables a systematic investigation of bulk scattering dynamics near the point $X$ in terms of approximate plane wave states and creation/annihilation operators defined in \eqref{eq:mode-exp}.

The quantity $P_\bmu$ can be understood as the momentum of the particle at any point along its trajectory. This interpretation remains valid in any background, independent of the existence of Killing symmetries. While Killing vector fields and conserved charges allow for the global definition of momenta, these are not essential to our framework. Near the boundary of asymptotically AdS spacetime, this definition coincides with the momenta derived from the Killing vectors of empty AdS (and its boundary).

There is, of course, a long history of using WKB approximations and geodesic motion in the context of AdS/CFT. Often the approximation is justified by considering bulk fields that have a large mass, which has been used for example to explore the black hole singularity  \cite{Grinberg:2020fdj, Festuccia:2005pi, Amado:2008hw, Horowitz:2023ury, Singhi:2024sdr}. Here instead, as in geometric optics, the approximation will be justified by the rapid oscillations of the considered wavepackets (compared with the distance between the inserted operators and other scales in the geometry).

\begin{figure}
    \centering
    \begin{tikzpicture}[scale = 1.3]
    \draw[fill=mblue!10,opacity=0.3] (0.5,-0.5) -- (3,-1) -- (3,4) -- (0.5,4.5) -- cycle;
    \draw[thick,color=mred] (2,0.25) -- (1.25,1);
    \draw[color=mblue,decorate,decoration={snake,amplitude=0.5mm,segment length=1mm}] 
    (1.25,1) .. controls (.75,1.5) and (-.25,1.75) .. (-0.75,3);
    \draw[dashed,thick,color=mblue] (-0.75,3) -- (-1,4);
    \filldraw[color=mred] (2,0.25) circle (1pt) node[anchor=north west,font=\footnotesize,color=black]{$x,p$} (1.25,1) circle (1pt) node[anchor=south west, font=\footnotesize,color=black]{$z \sim 1/\abs{p}$};
    \filldraw[color=mblue] (-0.75,3) circle (1pt) node[anchor=south west,font=\footnotesize,color=black]{$X,P$};
    \end{tikzpicture}
    \caption{Evolution of a wavepacket into the bulk: the wavy blue line represents the WKB regime, where the field undergoes rapid oscillations, while the red line marks the region near the boundary where the WKB approximation breaks down. The WKB approximation is used in the blue region to compute the propagator in non-trivial geometries.}
    \label{fig:wavepacket}
\end{figure}
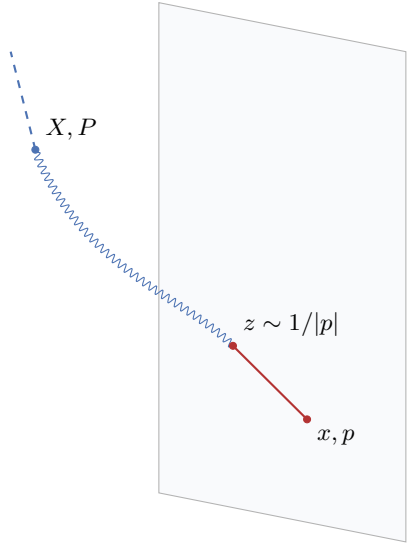

It is well known that the WKB approximation fails near the boundary of AdS. There, physical momenta become small in local units. (This is an essential ingredient in the original AdS/CFT decoupling limit \cite{Maldacena:1997re}!) We deal with this by approximating the spacetime near the boundary as empty AdS and matching the exact solution there with the WKB solution that is valid deeper in the bulk. This transition is illustrated in figure \ref{fig:wavepacket}, where the WKB region is represented by the wavy blue line and the near-boundary region by the dashed red line. This is the only place where we will use that the spacetime is asymptotically AdS. The same method will work for parametrically heavy fields, although we postpone their discussion to \S\ref{s:ms:mass}.

In this section, we begin by motivating our approach through an examination of the worldline action for relativistic particles in \S\ref{s:hj-wkb} and corresponding WKB expansion. We then fix normalizations in \S\ref{s:dynamics:bdrytobulk} and confirm the universality of the formula by considering a non-trivial bulk geometry. Finally, in \S\ref{s:dynamics:wavepackets} we obtain a general map between Gaussian wavepackets in the boundary and Gaussian wavepackets in the bulk. The final result, \eqref{eq:dictionary wavepacket}, is largely a covariant extension of the WKB approximation of \cite{Polchinski:1999ry}.\footnote{The wavepackets we consider have time spread much smaller than $\pi$ in global coordinates. For this reason, we believe they avoid the issues regarding normalizable versus non-normalizable frequencies in \cite{Giddings:1999jq}.}

\subsection{Hamilton-Jacobi formalism and the WKB approximation}\label{s:hj-wkb}
We consider a single particle following a geodesic trajectory. The on-shell value of a worldline action provides the leading WKB approximation to the particle's wavefunction. We briefly review the relevant construction based on the Hamilton-Jacobi formalism (referring to \cite{Goldstein} for more details) and then discuss the first subleading corrections, which will determine the overall normalization of correlators.

We start with the well-known worldline action for a relativistic particle of mass $m$:
\begin{equation}\label{eq:ein-action}
    S[X,\eta] = \frac{1}{2} \int \diff \sigma \, \qty( \frac{g_{\bmu\bnu}}{\eta} \frac{\diff X^\bmu}{\diff \sigma} \frac{\diff X^\bnu}{\diff \sigma} - \eta m^2 ),
\end{equation}
which is invariant under reparametrizations $\eta \, \diff \sigma = \eta' \, \diff \sigma'$. Eliminating the einbein $\eta$ using its equation of motion would reduce this to the more familiar worldline action, namely (minus) the mass times the proper time along the trajectory. The form \eqref{eq:ein-action} will be advantageous to discuss the massless limit.

Let us focus on the value of the action \eqref{eq:ein-action} evaluated on a classical solution. A general variation yields the (vanishing) equations of motion plus boundary terms:
\begin{equation}
    \fdiff_X S[X,\eta] = \frac{1}{\eta} \qty[g_{\bmu\bnu} \frac{\diff X^\bmu}{\diff \sigma} \fdiff X^\bnu]_{X_i}^{X_f} \qquad \mbox{(around a classical solution)}.
\end{equation}
In the Hamilton-Jacobi formalism, classical trajectories are parametrized in terms of their two endpoints, thus defining a function $S(X_i;X_f)$. Its variations are, therefore:
\begin{equation}\label{eq:wl-action-on-var}
    \fdiff S(X_i; X_f) = P_f \cdot \fdiff X_f - P_i \cdot \fdiff X_i,
\end{equation}
where the momentum is defined as:
\begin{equation}\label{eq:momentum-def}
    P_\bmu = g_{\bmu\bnu} \frac{\diff X^\bnu}{\diff \lambda} \quad\mbox{where}\qquad \diff \lambda =\eta \, \diff \sigma.
\end{equation}
The equations of motion of \eqref{eq:ein-action} ensure that $\lambda$ is an affine parameter along the geodesic. In the massless case, it is only defined up to an overall proportionality factor, which we can choose such that the boundary limit of $P_\bmu$ agrees with the boundary momentum \eqref{eq:bulkmomentumdef}.
Generally, the einbein equation of motion gives the bulk mass-shell condition:
\begin{equation}\label{eq:mass-shell}
    g^{\bmu\bnu} P_\bmu P_\bnu + m^2 = 0.
\end{equation}

The position-space Hamilton-Jacobi function $S(X_i;X_f)$ is not very useful for massless particles because it is only defined on a codimension-one subspace in which $X_i$ and $X_f$ are lightlike separated; when defined, it is identically zero. However, as observed in \S\ref{ssec:wavepackets}, wavepacket scattering will be more easily described by fixing boundary momenta rather than position. The relevant Hamilton-Jacobi-function is then a canonical transformation of the above:\footnote{If the boundary is a nontrivial manifold, $x^\mu$ should be viewed as local coordinates in some neighborhood of the considered path.}
\begin{equation} \label{SXp}
    S(p; X) \equiv \qty[S(x; X) + p_\mu x^\mu]_{x^\mu = x^\mu(p;X)}
\end{equation}
which is computed by finding an initial boundary point $x^\mu(p;X)$ such that the geodesic with initial boundary momentum $p_\mu$ passes through a given bulk point $X$. (For massive fields, the relevant geodesics will be complex near the boundary, see \S\ref{s:ms:mass}.)

As detailed in \cite{Caron-Huot:2025she}, the function $S(p;X)$ encodes all relevant properties of bulk lightcones as seen by the boundary. Its derivatives are
\begin{equation}
    \frac{\partial}{\partial p_\nu}S(p; X) = x^\mu(p;X),\qquad
    \frac{\partial}{\partial X^\bmu}S(p; X) = P_\bmu(p;X).
\end{equation}

\subsubsection{First subleading correction}
The precise connection between the on-shell action and the correlator \eqref{eq:bdry-to-bulk-cfn} comes from the path integral. For the next few steps, we restrict for notational simplicity to a momentum eigenstate along the boundary---localized wavepackets will be easily recovered in \S\ref{s:dynamics:wavepackets} by superposing such eigenstates. Schematically, the expected relation is:
\begin{equation} \label{path integral}
    \Phi_p(X)\equiv \expval{\Phi(X) \, \Osmeared_{p}^\dagger }{\Psi} \propto \int \diff^d x \, \ep^{\ic p \cdot x/\hbar} \int_{X_i=(x,z=0)}^{X_f=X} \frac{\Diff X \, \Diff \eta}{V_{\mathrm{diff}}} \, \ep^{\ic S[X,\eta] / \hbar}.
\end{equation}
We have reinstated powers of $\hbar$ to organize the approximations, which in practice will be related to the high-frequency expansion, $\hbar\leftrightarrow 1/\abs{p}$.

The inner path integral in \eqref{path integral} describes the propagation of a particle between specific positions. We see that the Fourier transform over the boundary point turns it into a single path integral that involves the Legendre-transformed action \eqref{SXp} and a now unconstrained endpoint $x_i^\mu$.
At leading semiclassical order, the amplitude \eqref{path integral} is thus controlled by the classical action: $\Phi_p(X)\propto \ep^{\ic S(p; X) / \hbar}$. We would like to determine the proportionality factor.

We will focus here on the $X$ dependence of the result, and use the path integral as a heuristic guide to an answer whose overall normalization will be determined by other means. In this way, details on how diffeomorphism redundancies are fixed will not be important.

In general, the normalization can be calculated as a one-loop determinant of small fluctuations. Fortunately, in quantum mechanics there exists a beautiful general result: the path integral with fixed endpoints is proportional to (the square root of) the van Vleck-Morette determinant \cite{VanVleck:1928zz, Morette:1951zz, Visser:1992pz}:
\begin{equation}
    \cD(X_i;X_f) = \frac{1}{\sqrt{(-g(X_i))(-g(X_f))}} \det\qty(\frac{\partial^2S}{\partial X_{i}^\bmu\partial X_{f}^\bnu}).
\end{equation}
This formula transforms naturally under canonical transformation, so we can try to guess the analogous determinant in our case:
\begin{equation}
    \cD(p; X) \stackrel{?}{=} \frac{\sqrt{-g(x)}}{\sqrt{-g(X)}} {\det}' \qty(\frac{\partial^2S(p; X)}{\partial p_\mu \partial X^\bmu}).
\end{equation}
Using the derivatives of $S$, the determinant could be rewritten alternatively as
\begin{equation} \label{van Vleck det prime}
    {\det}'\qty(\frac{\partial^2S(p; X)}{\partial p_\mu \partial X^\bmu}) = {\det}'\qty(\frac{\partial P_\bmu}{\partial p_\mu}) = {\det}'\qty(\frac{\partial x^\mu}{\partial X^\bmu}).
\end{equation}
These can be interpreted respectively as the evolution of momentum or position perturbations along the geodesic. These are both controlled by the geodesic deviation equation. Our immediate task is to make sense of the determinant of this $d\times (d+1)$ dimensional matrix, whence the notation ${\det}'$.

The dimensionality mismatch has a simple explanation: the mass shell constraint $P^2+m^2=0$ in the bulk. Thus, $P_\bmu$ really only has $d$ independent components.
This observation suggests a natural way to define $\cD$ as the Jacobian between the covariant measures in the bulk and boundary:
\begin{equation}\label{P p Jacobian}
    \frac{\diff^{d+1}P\,\delta\qty(P^2+m^2)}{\sqrt{-g(X)}} \equiv \cD(p;X) \frac{\diff^d p}{\sqrt{-g(x)}},
\end{equation}
where $P$ and $p$ are compared at fixed $X$. Concretely, this can be evaluated by introducing a covector field\footnote{Not to be confused with the ADM normal vector in \eqref{ADM}.} $n_\bmu$ normalized to:
\begin{equation} \label{norm n}
    2 n_\bmu P^\bmu = 1.
\end{equation}
We can then evaluate \eqref{P p Jacobian} by expanding the variations $\diff P_\bmu$ into $d$ components orthogonal to $P^\bmu$ and one component along $n_\bmu$.
With this precise meaning for $\det'$, the van Vleck-Morette determinant corresponding to $S(p;X)$ is thus:
\begin{equation} \label{van Vleck-Morette}
    \cD(p; X) = \frac{\sqrt{-g(x)}}{\sqrt{-g(X)}}
     \det\begin{pmatrix}\frac{\partial P_\bmu}{\partial p_\mu}& n_\bmu\end{pmatrix}
    = \frac{\sqrt{-g(x)}}{\sqrt{-g(X)}}
    \det\begin{pmatrix}\frac{\partial x^\mu}{\partial X^\bmu}& n_\bmu\end{pmatrix}.
\end{equation}
Because it is linear in $n_\bmu$ and all but the last column are orthogonal to $P^\bmu$, the determinant is guaranteed to be proportional to $n_\bmu P^\bmu$.
The definition thus does not depend upon the choice of the covector $n$ satisfying \eqref{norm n}.

As an example, in empty AdS in Poincar\'e coordinates,
$(P_\mu,P_z)=(p_\mu,\sqrt{-p^2-m^2/z^2})$ and
\begin{equation}\label{eq:AdS det}
\cD(p; X) = z^{d+1}
    \det\begin{pmatrix} \delta_\mu^\nu & \partial P_z / \partial p_\nu \\ n_\mu & n_z \end{pmatrix}
    = z^{d+1}\qty(n_z-n_\mu \frac{\partial P_z}{\partial p_\mu}) = \frac{z^{d-1}}{2P_z} \qquad \mbox{(empty AdS).}
\end{equation}
This result will be used below. Notice that it indeed does not depend on $n$.

Physically, this determinant ensures that wavefunctions remain properly normalized when nearby trajectories focus or disperse. The standard position space determinant $\cD(X_i;X_j)$, for example, diverges at ``caustic'' points that are connected by a continuous family of geodesics. The mixed van Vleck-Morette determinant \eqref{van Vleck-Morette} similarly blows up when many initially parallel paths converge onto the same bulk point.\footnote{A classic example of a singular van Vleck-Morette determinant in momentum space is the rainbow. When light reflects off the back of spherical water droplets, many impact parameters exit with nearby angles, causing a divergence in the semiclassical amplitude $\cM\propto \sqrt{\cD(p_i;p_f)}\sim \sqrt{\det \partial x_i / \partial p_f}$ when $\theta\approx (180-42)^\circ$ \cite{Glauber:2019roq}.}

To summarize, our proposal for the bulk wavefunction is
\begin{equation} \label{Phi WKB ansatz}
    \Phi_p(X) \propto  \sqrt{\cD(p; X)} \exp(\ic S(p; X)/\hbar) \times (1+\cO(\hbar)),
\end{equation}
up to a proportionality factor independent of $X$ (determined below in \eqref{eq:Phi X with normalization}).
The action is computed by finding a classical solution with prescribed initial momentum along the boundary and final position in the bulk. The proportionality factor cannot be calculated semiclassically because the WKB approximation breaks down near the AdS boundary, but it will now be fixed using the exact AdS solutions. However, we can still be confident about the $X$ dependence of \eqref{Phi WKB ansatz} (within the limits of the physical optics) since one could imagine starting the path integral from a finite depth $z\neq0$. The validity of the ansatz \eqref{Phi hbar expansion} is confirmed in \S\ref{app:EOM} by verifying that it satisfies the wave equation in the bulk.

We have focused here on a particle action that depends only on the background metric. However, the fact that the determinant \eqref{van Vleck det prime} is expressed entirely in terms of the particle action makes it clear that the form \eqref{Phi WKB ansatz} remains valid for more general linearized equations of motion, incorporating for example background gauge fields or other forces.

This discussion highlights only the basics of the geometric optics approximation, which offers insights into bulk physics in terms of (null) geodesics. In particular, the lightcone of a bulk point extends to the boundary of asymptotically AdS spacetime, providing an observational tool for exploring bulk geometry \cite{Caron-Huot:2025she}.

\subsection{Connecting the AdS boundary to the WKB region}\label{s:dynamics:bdrytobulk}
In Poincar\'e coordinates near the AdS boundary, the WKB approximation is only valid for $z \gtrsim 1 / \abs{p}$. Here, we bridge the WKB region to the boundary using the exact solution to the wave equation in the asymptotically AdS region.

The main simplifying feature of the near-AdS region is the translation symmetry, which allows us to focus on plane-wave solutions:
\begin{equation}
    \Phi_p(X) = \ep^{\ic p_\mu X^\mu} \phi_p(z). \label{factorized Phi}
\end{equation}
We explain the method here for a scalar field with action \eqref{eq:scalaraction}, which is dual to a scalar operator of scaling dimension $\Delta$ where $m^2=\Delta(\Delta-d)$; spinning fields are discussed in \S\ref{s:spin}. The scalar equation of motion reads
\begin{equation}\label{eq:eom-scalar-pbh}
    z^{d+1} \partial_z \qty(z^{-d+1} \partial_z \phi_p(z)) - \qty(z^2p^2 + m^2) \phi_p(z) = 0, \qquad p^2=\eta^{\mu\nu}p_\mu p_\nu.
\end{equation}
Here, we use the same notation (lowercase $p$) for bulk and boundary momenta, since the Killing symmetries of the geometry identify them.

The radial equation \eqref{eq:eom-scalar-pbh} admits well-known solutions in terms of Bessel/Hankel functions: $\phi_p \propto z^{\frac{d}{2}} \rH_{\Delta-\frac{d}{2}}(\sqrt{-p^2}z)$. For a timelike momentum, the Hankel function is oscillatory and has the limits
\begin{equation} \label{H asympt}
    \rH^{(1)}_{\Delta - \frac{d}{2}} (\abs{p} z) \to 
    \begin{cases}
        \frac{1}{\ic \pi} \Gamma\qty(\Delta - \frac{d}{2})\qty(\frac{\abs{p}z}{2})^{\frac{d}{2}-\Delta}, & \abs{p}z\ll 1, \\[4pt]
        \sqrt{\frac{2}{\pi \abs{p} z}}\exp\qty(\ic \abs{p}z +
        \frac{\ic\pi}{4}\qty(d-1-2\Delta)
        +\cO\qty(\frac{\Delta}{\abs{p}z})
        ), & \abs{p}z\gg \Delta.
    \end{cases}
\end{equation}
The idea of our method is simple: we fix the correct normalization of the bulk-to-boundary propagator using the standard AdS/CFT dictionary as $z\to 0$, then match its large-$z$ asymptotics to the WKB solution \eqref{Phi WKB ansatz}. Here we will treat $\Delta$ as a fixed (not large) quantity, which means that bulk excitations are ultrarelativistic and effectively massless in the matching region. For large $\Delta$, the WKB approximation also covers a nonrelativistic region, which is discussed in \S\ref{s:ms:mass}.

Let us thus briefly review the standard AdS/CFT recipe and its Lorentzian version. Inserting a CFT operator is equivalent to turning on a boundary condition for the bulk field so that in Euclidean signature the bulk-to-boundary correlator on the left of \eqref{path integral} satisfies
\begin{equation} \label{limit phi}
    \lim_{z\to 0} z^{\Delta-d}\phi_p(z) = \qty[(2\Delta - d)C_\Delta]^{-\frac{1}{2}}, \qquad C_\Delta = \frac{\Gamma(\Delta)}{\pi^{\frac{d}{2}} \Gamma(\Delta - \frac{d}{2})}.
\end{equation}
The square bracket is a normalization factor which is often taken to be 1, here we are making a choice such that a canonically normalized scalar in the bulk will be dual to a canonically normalized CFT operator:\footnote{When checking the normalization of the bulk-to-bulk two-point there is a well-known subtlety that one should impose the boundary condition \eqref{limit phi} at finite $z=\epsilon$, i.e.,
$$\phi(z)\mapsto\phi_\epsilon(z)= \frac{\qty[(2\Delta - d)C_\Delta]^{-\frac{1}{2}} \phi(z)}{\epsilon^{\Delta-d} \phi(\epsilon)}.$$
The correlator is then obtained from the limit of the on-shell action integrated over $z\geq \epsilon$, see \cite{Freedman:1998tz}, and matches with the correlator of a canonically normalized boundary operator:
$$ \lim_{\epsilon\to 0} \epsilon^{1-d} \phi_\epsilon(\epsilon) \phi_\epsilon'(\epsilon) = \frac{\pi^{\frac{d}{2}}\Gamma\qty(\frac{d}{2}-\Delta)}{\Gamma(\Delta)}\qty(\frac{p^2}{4})^{\Delta-\frac{d}{2}} = \int \diff^d x_\rE \, \frac{\ep^{-\ic p \cdot x}}{(x^2)^\Delta}.$$}
$$\expval{O(x)O(y)}{0}=\abs{x - y}^{-2\Delta}.$$
In Euclidean signature, the bulk-to-boundary propagator is then given by \eqref{factorized Phi} with the modified Bessel function of the second type (which decays exponentially as $z\to\infty$):
\begin{equation}\label{eq:bare-bessel}
    \phi_p(z) = \frac{2 \qty[(2\Delta - d)C_\Delta]^{-\frac{1}{2}}}{\Gamma\qty(\Delta - \frac{d}{2})} \qty(\frac{\sqrt{p^2}}{2})^{\Delta - \frac{d}{2}} z^{\frac{d}{2}} \rK_{\Delta - \frac{d}{2}} \big( \sqrt{p^2} z \big)\qquad \mbox{(Euclidean signature)}.
\end{equation}
From the Euclidean correlator, one can readily obtain the time-ordered one using the standard Wick rotation or Feynman prescription of field theory, which takes $\sqrt{p^2}\mapsto \sqrt{p^2-\ic 0}=-\ic \sqrt{-p^2}$ for timelike momenta. In addition, momentum space correlators pick an extra $-\ic$ from the Wick rotation of the time integral in the Fourier transform. Anti-time-ordered correlators are similarly obtained via the opposite Wick rotation, altogether giving two Lorentzian solutions:
\begin{equation} \label{phi T}
    \phi_p^\pm(z) = \frac{\pi \qty[(2\Delta - d)C_\Delta]^{-\frac{1}{2}}}{\Gamma\qty(\Delta - \frac{d}{2})} \qty(\frac{\sqrt{-p^2}}{2})^{\Delta - \frac{d}{2}} z^{\frac{d}{2}} \rH^{(1\text{/}2)}_{\Delta - \frac{d}{2}} \big(\sqrt{-p^2} \, z\big).
\end{equation}
Here $\rH_\upnu^{(i)}$ denotes the Hankel function of the $i$\textsuperscript{th} kind. The time-ordered solution $\phi_p^+\propto \rH^{(1)}$ represents a wave moving from the boundary into the bulk as can be seen from the phase in \eqref{H asympt}. The anti-time-ordered solution $\phi_p^-\propto  \rH^{(2)}$ instead moves toward the boundary.

We discuss both solutions together, since, as explained in the introduction, we are interested in correlators that are not always time-ordered, such as inclusive observables that can probe bulk regions otherwise hard to reach. In fact, instead of time-ordered and anti-time-ordered cases, it is most useful to consider the Wightman function, which is simply their sum:
\begin{equation} \label{Wightman prop}
    \expval{\Phi(X) \, \Osmeared_{p}^\dagger }{0} = \ep^{\ic p_\mu X^\mu}\qty(\phi_p^+(z)+\phi_p^-(z))\qquad\mbox{(empty AdS)}.
\end{equation}
In this formula, we assume that $p$ is future timelike. The formula follows from \eqref{phi T} using a simple identity of time-ordered products:
\begin{equation}
    T(AB)+ \conj{T}(AB)= AB+BA,
\end{equation}
together with the fact that $\langle0|\Osmeared^\dagger_p=0$ for positive frequency. When discussing nontrivial excited states $|\Psi\rangle$, we will similarly assume that we are considering momenta $p$ sufficiently high that no pre-existing bulk excitations with that momentum exist so that $\langle \Psi|\Osmeared^\dagger_p=0$.

The physical interpretation of the Wightman propagator \eqref{Wightman prop} is that the state $\Osmeared_p^\dagger|0\rangle$ describes a wave that comes toward the boundary and reflects off it. It must contain both since it must solve the bulk equations of motion at all times without a source. It is proportional to the regular (normalizable) solution $$\phi^+_p(z)+\phi^-_p(z)\propto z^{\frac{d}{2}}J_{\Delta-\frac{d}{2}}(\sqrt{-p^2}z)\sim z^\Delta \quad \mbox{as } z\to 0.$$
When we superpose plane waves to make wavepackets that have a definite time, in the next subsection, only one of the $\phi^\pm_p$ solutions will survive, depending on whether $X$ is in the past or future of the boundary source.

Having found the exact solution, we can now expand it at large $z\abs{p}$ and compare it with the WKB approximation. Substituting \eqref{H asympt} into \eqref{phi T} we get:
\begin{equation}\label{eq:scalar-eads-wkb}
    \phi^\pm_p(z) = \cC^\pm_{\Delta,p} \, \qty(\frac{z^{d - 1}}{2 \sqrt{-p^2}})^{\frac{1}{2}} \exp(\pm \ic \sqrt{-p^2} z + \cO\qty(\frac{1}{\abs{p}})).
\end{equation}
We see that the $z$-dependence matches precisely the WKB expression \eqref{Phi WKB ansatz}, including the empty AdS van Vleck-Morette determinant \eqref{eq:AdS det}! The outcome of this exercise is the constant $\cC^\pm_{\Delta,p}$:
\begin{equation}\label{eq:unipre}
    \cC^\pm_{\Delta,p} = \ep^{\pm \ic \theta_\Delta}\sqrt{\frac{2 \pi^{\frac{d}{2}+1}}{\Gamma(\Delta) \Gamma\qty(\Delta - \frac{d}{2} + 1)}}\qty(\frac{\sqrt{-p^2}}{2})^{\Delta - \frac{d}{2}}, \qquad \text{with} \qquad \theta_\Delta = \frac{\pi}{4} \qty(d - 1-2\Delta).
\end{equation}
Thus, the full solution can be expressed as
\begin{equation}\label{eq:scalarwkb-ads}
    \phi^\pm_p(X) = \cC^\pm_{\Delta,p} \sqrt{\cD(p; X)} \exp(\ic p_\mu X^\mu \pm \ic \sqrt{-p^2} z + \cO\qty(\frac{1}{\abs{p}})).
\end{equation}
This is valid in empty AdS. However, it can be immediately generalized to an arbitrary asymptotically AdS spacetime provided that the WKB approximation deep in the bulk connects smoothly with the AdS region. The complete version of the proposal \eqref{Phi WKB ansatz} is thus:
\begin{equation}\label{eq:Phi X with normalization}
    \expval{\Phi(X) \, \Osmeared_{p}^\dagger }{\Psi} = \cC^\pm_{\Delta,p} \sqrt{\cD(p; X)} \exp(\ic S(p; X)),
\end{equation}
where $S(p; X)$ defined in \eqref{SXp} is the classical action of a null geodesic with momentum $p_\mu$ at the boundary and which reaches the bulk point $X$.

Here, $\cC^\pm_{\Delta,p}$ encapsulates both the normalization and the WKB phase shift, and depends solely on the dimension $\Delta$ of the boundary operator and its momentum $p$. This universality governs the near-boundary region, where the WKB approximation breaks down. Beyond this region and deeper into the bulk, the waveform can be evolved using standard WKB techniques and is fully determined by the $S$ and $\cD$ factors.

In a neighborhood of $X$ we can approximate the exponent as a plane wave:
\begin{equation}
    \exp(\ic S(p; X + \fdiff X)) = \exp(\ic S(p; X) + \ic P_\bmu \fdiff X^\bmu + \cO(\fdiff X^2)).
\end{equation}
Comparing with the local mode expansion \eqref{eq:a comm}, which gives
$$\expval{\Phi(X + \fdiff X) \  a^\dagger_{X,P}}{\Psi}=\ep^{\ic P \cdot \fdiff X},$$
we conclude that inserting $\Osmeared_p^\dagger$ in the boundary is equivalent (as far as the bulk dynamics near $X$ is concerned) to inserting a bulk creation operator near $X$:
\begin{equation}\label{eq:dictionary}
\boxed{
    \Osmeared^\dagger_{p} \;\simeq\;  \sqrt{\cD(p; X)} \, \ep^{\ic S(p; X)} 
    \times
    \begin{cases}
        \cC^+_{\Delta,p}a^\dagger_{X,P}, & \mbox{if $\Osmeared^\dagger$ is in the past of $X$}, \\
        \cC^-_{\Delta,p}b^\dagger_{X,P}, & \mbox{if $\Osmeared^\dagger$ is in the future of $X$}.
    \end{cases}}
\end{equation}

This is the crucial relation that we will use to relate boundary correlation functions to bulk scattering amplitudes. The Hermitian conjugate of this equation is also valid and relates $\Osmeared_p$ to absorption operators $a_{X,P}$ or $b_{X,P}$ that act in the near past/future of the bulk point $X$. The bulk momentum $P_\bmu$ is obtained from the boundary momentum $p_\mu$ by integrating the geodesic equation (see \eqref{eq:momentum-def}).

To be fully precise, the past and future cases in \eqref{eq:dictionary} should be viewed as distinct saddle point solutions to the same variational problem that defines $S(p; X)$ in \eqref{SXp}. This problem involves a fixed boundary momentum and a fixed bulk point; its solutions include both future-directed and past-directed geodesics. In general, correlators are the sum of all saddle points, as exemplified by \eqref{Wightman prop}. In \eqref{eq:dictionary} we treated each saddle separately since they will be easily distinguished once we integrate against wavepackets that are localized in time.

Table \ref{t:2pt-bdry-conditions} summarizes the connections between the different boundary and bulk operators and their associated phases.

\begin{table}[t]
    \centering
    \renewcommand{\arraystretch}{1.5}
    \setlength{\tabcolsep}{12pt}
    \arrayrulecolor{gray}
    \begingroup
    \small
    \begin{tabular}{|c|c|}
    \hline
    Boundary  & Bulk \\ \hline
    $\Osmeared_{x_-,p,\sigma}^\dagger$&
    $\cC_{\Delta,p}^+\sqrt{\cD}\,\exp(-\ic E T + \ic \mathbf{p} \cdot \mathbf{X} + \ic \sqrt{-p^2} z)\,a^\dagger_{X,P}$
    \\[1ex] \hline
    $\Osmeared_{x_+,p,\sigma}^\dagger$&
    $\cC_{\Delta,p}^-\sqrt{\cD}\,\exp(-\ic E T + \ic \mathbf{p} \cdot \mathbf{X} - \ic \sqrt{-p^2} z)\,b^\dagger_{X,P}$
    \\[1ex] \hline
    $\Osmeared_{x_+,p,\sigma}$&
    $\cC_{\Delta,p}^+\sqrt{\cD}\,\exp(\ic E T - \ic \mathbf{p} \cdot \mathbf{X} + \ic \sqrt{-p^2} z)\,b_{X,P}$
    \\[1ex] \hline
    $\Osmeared_{x_-,p,\sigma}$&
    $\cC_{\Delta,p}^-\sqrt{\cD}\,\exp(\ic E T - \ic \mathbf{p} \cdot \mathbf{X} - \ic \sqrt{-p^2} z)\,a_{X,P}$
    \\[1ex] \hline
    \end{tabular}
    \endgroup
    \caption{Map between boundary operators with $p_\mu = (-E,\mathbf{p})$ (or equivalently, $p^\mu = (E,\mathbf{p})$) and bulk operators corresponding to the various cases of \eqref{eq:dictionary} in empty AdS, including complete phases. Here $x_-$ and $x_+$ stand for insertion points in the past and future of the bulk point $X$.}\label{t:2pt-bdry-conditions}
\end{table}

Since our derivation treated $\Delta$ as finite but not large, the classical action $S(p; X)$ in \eqref{eq:dictionary} must be computed in the \emph{massless} theory. However, the formula \eqref{eq:dictionary} actually holds also for parametrically large $\Delta$ and nonrelativistic momenta. The essential point is to properly define the on-shell action $S(p;P)$ in the massive case (see \eqref{SXp massive}) by recognizing that such a particle must \emph{quantum tunnel} from the boundary.

\paragraph{Example: translation invariant geometry} Consider the radial equation of motion obtained using the metric \eqref{eq:pbhmetric} with general metric components $A(z)$ and $B(z)$. 
\begin{equation} \label{radialAB}
    \phi_p ''(z)-\frac{(d-1) \phi_p '(z)}{z} +\frac{\phi_p '(z) \left(A(z)B(z)\right)'+2 \phi_p (z) \left(E^2-\qty({\bf p}^2 +\frac{m^2}{z^2}) A(z)\right)}{2 A(z) B(z)} = 0.
\end{equation}
For the general case, we need to solve the radial equation with nontrivial functions $A(z)$ and $B(z)$, subject to the boundary condition \eqref{limit phi}. However, exact solutions to the radial equation are not straightforward except in special cases with symmetries (for instance, the planar BTZ black hole with $d=2$). We can nonetheless utilize the WKB method to solve the radial equation. Specializing to $m=0$, the one-dimensional WKB solution takes a similar form as in \eqref{eq:Phi X with normalization}:
\begin{equation} \label{black brane WKB}
    \phi^\pm_p(z) \propto  \sqrt{\frac{z^{d-1}}{\sqrt{A(z)B(z)}P_z}}
    \exp(\pm \ic \int_0^z \diff z \, P_z + \cO\qty(\frac{1}{P_z})),
\end{equation}
where the radial momenta $P_z$ is defined by
\begin{equation}\label{eq:pbh-rad-mom}
    P_z = \sqrt{- \frac{g^{\mu\nu}}{g^{zz}} p_\mu p_\nu} =  \sqrt{\frac{E^2 - \mathbf{p}^2 A(z)}{A(z)B(z)}},
\end{equation}
where $E$ is the energy, and $\mathbf{p}$ is the spatial momentum, and the $z$-dependent prefactor in \eqref{black brane WKB} was determined directly from \eqref{eq:eom-scalar-pbh} by expanding the solution as in \eqref{Phi hbar expansion}.

We now compare the normalization with the van Vleck-Morette determinant in \eqref{van Vleck-Morette}. Choosing the covector as $(n_\mu,n_z)=(0,1/(2P^z))$ (the result does not depend on $n$) we find
\begin{equation}
    \cD(p; X) = \frac{z^{d-1}}{2\sqrt{A(z) B(z)}P_z} \qquad \mbox{(black brane geometry)}.
\end{equation}
For the subcase of the black brane geometry where $A(z) = B(z) =1 - \frac{z^d}{\zh^d}$, this matches precisely the prefactor in \eqref{black brane WKB}, confirming \eqref{eq:dictionary} in a nontrivial excited geometry.

The massless WKB approximation is valid down to $z_{\rm min} \sim \Delta / \abs{p}$. Provided that this is sufficiently close to the boundary that $A(z) \approx 1$ for $0<z<z_{\rm min}$, then the exact solution in this region will agree with the one in empty AdS and the normalization in \eqref{eq:dictionary} automatically fulfilled.

\subsection{Wavepackets and their evolution in the bulk}\label{s:dynamics:wavepackets}
Having now clarified how to compute the boundary-to-bulk propagator for momentum eigenstates in terms of geodesics, it is now clear how to extend \eqref{eq:dictionary} to a wavepacket. We only need to superpose plane waves.

As discussed previously, localized wavepackets provide a useful way to probe the bulk when working in a generic background that may not admit any global isometry. Provided that the superposition of plane waves becomes small before leaving a neighborhood of the considered geodesic, it suffices to know the solutions inside this neighborhood. We stress that there is never any need to define plane waves or momentum eigenstates globally over the full geometry.

On the boundary, let us recall the definition of the Gaussian smeared operator \eqref{eq:bdryops}:
\begin{equation}
    \Osmeared_{x,p,\sigma}^\dagger = \int \diff^d \qty(\fdiff x) \, \exp(\ic p_\mu \fdiff x^\mu - \frac{1}{2} \qty(\sigma^{-1})_{\mu\nu} \fdiff x^\mu \fdiff x^\nu) O(x + \fdiff x),
\end{equation}
where $\sigma^{\mu\nu}$ is a positive matrix which defines the wavepacket's time and spatial widths. In Fourier space, this becomes equivalently
\begin{equation}\label{eq:transformed-operators-momentum-packets}
    \Osmeared_{x,p,\sigma}^\dagger = \int \frac{\diff^d k}{(2\pi)^d} \, \tilde\psi_{x,\sigma}(k{-}p) O_{k}^\dagger,
\end{equation}
where
\begin{equation}
    \tilde\psi_{x,\sigma}(k{-}p)\equiv\sqrt{\det 2\pi \sigma} \exp(-\ic (k{-}p)_\mu x^\mu - \frac{1}{2} \sigma^{\mu\nu} (k{-}p)_\mu (k{-}p)_\nu).
\end{equation}
These definitions conventionally set the phase to zero at the center of the wavepacket. Inserting the boundary-to-bulk map \eqref{eq:dictionary} inside the integral then gives
\begin{equation} \label{wavepacket first step}
    \Osmeared^\dagger_{x^-,p,\sigma} \simeq
    \int \frac{\diff^d k}{(2\pi)^d} \, \tilde{\psi}_{x,\sigma}(k{-}p) \, \cC^+_{\Delta,k} \sqrt{\cD(k,X)} \ep^{\ic S(k,X)} \,a^\dagger_{X,P(k,X)}.
\end{equation}
We see that the boundary wavepacket turns to a superposition of bulk creation operators in which the bulk momentum $P$ is itself expressed as a function of the boundary momentum $k$.

In order to do bulk computations, we would like to rewrite \eqref{wavepacket first step} as an integral over bulk momentum $P$. This is possible because, even though $P_\bmu$ has $(d+1)$ components, it is constrained by the mass-shell condition $P^2+m^2=0$. Thus, a one-to-one map between $k$ and $P$ generically exists, at least locally near a nonsingular saddle. The Jacobian was discussed in \eqref{P p Jacobian} and allows us to 
rewrite \eqref{wavepacket first step} as:
\begin{equation}\label{eq:dictionary wavepacket}
\Osmeared^\dagger_{x^-,p,\sigma}
    \simeq
    \int \frac{\diff^{d+1}P}{(2\pi)^d\sqrt{-g}}\delta\qty(P^2+m^2) \,
    \tpsib_{x, p, \sigma; X}(P)
    \, a^\dagger_{X,P},
\end{equation}
where the bulk wavepacket collects all other factors:
\begin{equation} \label{bulk wavepacket 1}
    \tpsib_{x,p,\sigma; X}(P) = \frac{\cC^+_{\Delta,k}\sqrt{\det2\pi\sigma}}{\sqrt{\cD(k;X)}} \exp(\ic S(k;X)-\ic (k{-}p)_\mu x^\mu - \frac{1}{2} \sigma^{\mu\nu} (k{-}p)_\mu (k{-}p)_\nu) \Bigg|_{k=k(X,P)}.
\end{equation}
Equation~\eqref{eq:dictionary wavepacket} is the main result of this section; we will shortly give a simplified form of this formula (see \eqref{psi bulk good} below). Let us first briefly highlight the data on which it depends. We have a fixed bulk point $X$, near which we expect a scattering process to happen. Given a covector $P_\bmu$ at $X$, the bulk geodesic with initial condition $(X,P)$ will reach the boundary with a momentum $k_\mu(X,P)$ which should be used to evaluate the other factors. The exponential in \eqref{bulk wavepacket 1} contains the classical action of the said geodesic. This geodesic reaches the boundary at a point $\partial S / \partial p_\mu$ which is separated from the center of the wavepacket by
\begin{equation}
    \Delta x^\mu \defeq x^\mu - \frac{\partial S(p;X)}{\partial p_\mu},
\end{equation}
which we are imagining is small compared with $\sqrt{\sigma}$ (otherwise the exponential will be rapidly oscillating and cancel out). Below we will treat $\Delta x^\mu$ and $(k{-}p)_\mu$ as being of the same order of smallness ($\sim \sqrt{\hbar}$ if one restores $1/\hbar$ in the exponent).

To simplify \eqref{bulk wavepacket 1} to a more usable form, we assume that the wavepacket is sufficiently narrow that we can Taylor-expand the exponent around $k\to p$, and drop all non-exponential dependence on $k$:
\begin{equation} \label{bulk wavepacket 2}
    \tpsib_{x,p,\sigma; X}(P) \approx \frac{\cC^+_{\Delta,p}\sqrt{\det2\pi\sigma}}{\sqrt{\cD(p;X)}} \, \exp(\ic S(p;X) - \ic (k{-}p)_\mu \Delta x^\mu - \frac{1}{2} \Sigma^{\mu\nu} (k{-}p)_\mu (k{-}p)_\nu) \Bigg|_{k=k(X,P)},
\end{equation}
where
\begin{equation}\label{complex Sigma}
    \Sigma^{\mu\nu}= \sigma^{\mu\nu}-\ic \frac{\partial^2S}{\partial p_\mu \partial p_\nu}.
\end{equation}
It remains to rewrite this in terms of $P$. Since $(k{-}p)$ only multiplies itself or the small quantity $\Delta x$, we see that we only need the map between $(k{-}p)$ and $\delta P$ to linear order. This map involves the matrix $\frac{\partial P_\bmu}{\partial k_\mu}=\frac{\partial x_\mu}{\partial X^\bmu}$. The only subtlety is that this matrix is degenerate, due to the mass shell constraint. The solution, by now familiar from \eqref{van Vleck-Morette}, is to introduce a reference $n_\bmu$ and separate the components of $\fdiff P$ into those perpendicular to $P^\mu$, and the one along $n_\bmu$. This leads to a formula involving a pseudo-inverse:
\begin{equation} \label{psi bulk good}
    \tpsib_{x,p,\sigma; X}(P+\delta P) \approx \frac{\cC^+_{\Delta,p}\sqrt{\det2\pi\sigma}}{\sqrt{\cD(p;X)}} \exp(\ic S(p;X) - \ic \Delta X^\bmu\fdiff P_\bmu-\frac12 \Sigma^{\bmu\bnu} \fdiff P_\bmu\fdiff P_\bmu),
\end{equation}
where
\begin{equation} \label{psi bulk X Sigma}
    \Delta X^\bmu = \Delta x^\mu\frac{\partial k_\mu}{\partial P_\bmu},\qquad
    \Sigma^{\bmu\bnu} = \Sigma^{\mu\nu}\frac{\partial k_\mu}{\partial P_\bmu}\frac{\partial k_\nu}{\partial P_\bnu},
\end{equation}
and
\begin{equation}
    \frac{\partial k_\mu}{\partial P_\bmu} = \qty[\begin{pmatrix}\frac{\partial P_\bmu}{\partial p_\mu}& n_\bmu\end{pmatrix}^{-1}]_{{\rm top\,} d {\rm\,rows}}.
\end{equation}
These quantities satisfy $n_\bmu \Delta X^\bmu = 0=n_\bmu \Sigma^{\bmu\bnu}$, and are defined modulo shifts proportional to $P^\bmu$ (since they are dotted into $\fdiff P$ which satisfies $\fdiff P \cdot P=0$). All functions and derivatives should be evaluated around the geodesic that has boundary momentum $p$ and passes through the bulk point $X$. The essential point is that \eqref{psi bulk good} gives us a Gaussian approximation to the bulk wavefunction.

The vector $\Delta X^\bmu$ has a simple physical interpretation: to the accuracy of our approximations, it measures by how much the geodesic with boundary data $(x,p)$ misses the bulk point $X$. The imaginary part of $\Sigma$ (see \eqref{complex Sigma}) will be interpreted shortly in terms of dispersive spreading.

The last general formula we present, anticipating its application in the next section, is the Fourier-transform of equation \eqref{psi bulk good} to position space:
\begin{equation}\begin{split} \label{psi X}
    \psib_{x,p,\sigma}(X + \fdiff X) &\defeq \int  \frac{\diff^{d+1}\fdiff P}{(2\pi)^d\sqrt{-g}} \delta\qty(2P \cdot \fdiff P) \, \tpsib_{x, p, \sigma; X}(P + \fdiff P) \, \ep^{\ic(P+\fdiff P) \cdot \fdiff X} \\
    &\,\approx \cF_\Delta(p;X) \exp(\ic P \cdot \fdiff X - \frac{1}{2} \Sigma^{-1}_{\bmu\bnu} (\fdiff X -\Delta X)^\bmu (\fdiff X -\Delta X)^\bnu ),
\end{split}\end{equation}
where we have defined
\begin{equation} \label{psi X definition F}
    \cF_\Delta(p;X) \defeq \frac{\cC^+_{\Delta,p}\sqrt{\cD(p;X)}}{\sqrt{\det \Sigma^{\mu\nu}/\det \sigma^{\mu\nu}}} \, \ep^{\ic S(p;X)}.
\end{equation}
Here again, we have defined a pseudo-inverse $\Sigma^{-1}_{\bmu\bnu} \defeq (\Sigma^{\bmu\bnu} + \infty \times P^\bmu P^\bnu)^{-1}$, which is naturally motivated by writing the constraint $\delta(2P \cdot \fdiff P)$ as a narrow Gaussian. It satisfies $P^\bmu \Sigma^{-1}_{\bmu\bnu}=0$: as physically expected, the bulk wavefunction is invariant under geodesic flow along $P^\bmu$.

\paragraph{Examples} In Poincar\'e coordinates we can identify boundary and bulk momenta due to the Killing isometries: $p_\mu=P_\mu$. Choosing the reference $n$ to be in the $z$ direction, the displacement and effective width \eqref{psi bulk X Sigma} for a massless particle are
\begin{align} \label{eq:delta-x section 3}
\Delta X^\mu &= x^\mu-X^\mu-\int \diff z \, \frac{\partial P_z}{\partial p_\mu},\qquad \Delta X^z=0, \\
\Sigma^{\mu\nu}&= \sigma^{\mu\nu}-\ic \int \diff z \, \frac{\partial^2P_z}{\partial p_\mu \partial p_\nu},
\qquad \Sigma^{z\mu}=0=\Sigma^{zz},
\end{align}
with $P_z=\sqrt{-g_{zz}(z)g^{\mu\nu}(z)p_\mu p_\nu}$. In empty AdS, these reduce to $P_z=\abs{p}=\sqrt{- \eta^{\mu\nu} p_\mu p_\nu}$ and
\begin{equation} 
\label{X Sigma AdS}
    \Delta X^\mu = x^\mu - X^\mu + z \, \frac{p^\mu}{\abs{p}}, \qquad \Sigma^{\mu\nu}=\sigma^{\mu\nu}+\frac{\ic z}{\abs{p}} \qty(\eta^{\mu\nu}-\frac{p^\mu p^\nu}{p^2}).
\end{equation}
To illustrate the physics of the pseudo-inverse $\Sigma^{-1}$, consider a wavepacket that is rotationally invariant around the time direction $P_\mu=(-\omega,0)$, with independent widths in time and space:
\begin{equation}
    \sigma^{\mu\nu} = \sigma_{t}^2 \, \frac{p^\mu p^\nu}{\abs{p}^2} + \sigma_x^2 \, \qty(\eta^{\mu\nu}-\frac{p^\mu p^\nu}{p^2}).
\end{equation}
Then
\begin{equation}
    \Sigma^{-1}_{\bmu\bnu} \diff X^\bmu \diff X^\bnu = \frac{(\diff T-\diff Z)^2}{\sigma_t^2} + \frac{\diff \mathbf{X} \cdot \diff \mathbf{X}}{\sigma_x^2+\ic z/\omega}.
\end{equation}
The first term is not surprising: a Gaussian uncertainty in the shooting time at the boundary directly translates into an uncertainty in the (retarded) arrival time in the bulk. The second term is more interesting. Its real part gives (see \eqref{psi X}):
\begin{equation}
\abs{\psib}^2 \propto \exp(- \Re \frac{\diff \mathbf{X} \cdot \diff \mathbf{X}}{\sigma_x^2+\ic z/\omega}) =
\exp(-\frac{\diff \mathbf{X} \cdot \diff \mathbf{X}}{\sigma_x^2+\frac{z^2}{\omega^2 \sigma_x^2}}).
\end{equation}
The uncertainty in the bulk spatial position is thus the combination of two factors: uncertainty in the shooting position $\sigma_x$, and uncertainty in the shooting angle $z/(\omega \sigma_x)$, also known as dispersive spreading (similar to a free particle in quantum mechanics whose position $x=x_0+v_0t$ is affected by uncertainty in both $x_0$ and $v_0$). The net uncertainty is minimized when these are of the same size, so that in all cases
\begin{equation}
    \langle (\diff {\bf X})^2\rangle
    \gtrsim \frac{z}{\omega}.
\end{equation}
In other words, there is a fundamental limit $\sqrt{z / \omega}$ on the spatial localization that can be achieved for a single particle of energy $\omega$ after it has moved a depth $z$ into the bulk. An identical formula controls the spreading of a Gaussian laser beam in Minkowski space. This happens because this example considers null geodesics in empty AdS, and empty AdS is conformally flat.

There is no analogous limit on time resolution: the retarded time is not affected by propagation, and for a single particle it is only limited by the spread $\sigma_{t}$ of the wavepacket. (For heavy particles, spectral distortions will weaken this to $\abs{\diff T} \gtrsim \Delta / \omega$, see \S\ref{s:ms:mass}.)

\section{Boundary correlators from bulk amplitudes}\label{s:dynamics:sing}
Above we considered the insertion of boundary operators integrated against energetic wavepacket aimed toward a bulk point $X$ inside some asymptotically AdS spacetime. As far as physics near the point $X$ is concerned, the result is equivalent to inserting creation or absorption operators near $X$. We repeat the main formula \eqref{eq:dictionary wavepacket} for convenience:
\begin{equation} \label{eq:dictionary wavepacket 2}
\Osmeared^\dagger_{x_j^-,p_j,\sigma_j}
    \simeq
    \int \frac{\diff^{d+1}P_j}{(2\pi)^d\sqrt{-g(X)}} \delta\qty(P_j^2+m_j^2)\,
    \tpsib_{x_j, p_j, \sigma_j; X}(P_j)
    \, a^\dagger_{X,P_j},
\end{equation}
where the bulk wavepacket (see \eqref{psi bulk good}) is peaked around the momentum $P(x_j,p_j)$ of a classical geodesic connecting $x_j$ to (near) $X$.

An idealized bulk scattering process then occurs if the following two conditions are met. First, the bulk wavepackets should not overlap except in a neighborhood of $X$. Second, long-range bulk interactions should be negligible. These assumptions ensure that we can propagate the excitations freely and independently up to $X$; generalizations will be discussed below.

Under these ideal conditions, it is straightforward to generalize \eqref{eq:dictionary wavepacket 2} to multipoint correlators. For example, to realize a $m \to n$ process near $X$, it suffices to choose $m$ shooting positions and momenta $(x_j,p_j)$ in the past of $X$, and $n$ similar pairs in its future, and to consider the boundary correlator:
\begin{equation}\label{m to n factorization step}
    \expval{\prod_{j=m+1}^{m+n} O_{x_j^+,p_j,\sigma_j} \prod_{j=1}^m O_{x_j^-,p_j,\sigma_j}^\dagger}{\Psi}
    \approx \int \textstyle{\dnPpsi}
    \expval{b_{X,P_{m+n}} \dots b_{X,P_{m+1}}
    a_{X,P_n}^\dagger \dots a_{X,P_1}^\dagger}{0}^{\text{bulk}},
\end{equation}
where the right-hand-side follows simply from the iterated application of \eqref{eq:dictionary wavepacket 2}. To be fully explicit, we have abbreviated here the measure which collects all factors in \eqref{eq:dictionary wavepacket 2}:
\begin{equation}
\dnPpsi=
    \prod_{j=1}^{m+n} \frac{\diff^{d+1}P_j}{(2\pi)^d\sqrt{-g}} \delta\qty(P_j^2)\theta(P^0)\,
 \tpsib_{x_j, p_j, \sigma_j; X}(P_j)^{(*)}.
\end{equation}
Here and below we use the superscript $(*)$ to denote quantities that must be complex-conjugated, corresponding to the absorption (non-dagger) operators $O$.

The expectation value in \eqref{m to n factorization step} gives the scattering amplitude in \eqref{eq: ordinary amp}, so altogether:
\begin{equation}\label{m to n factorization}
\boxed{\expval{\prod_{j=m+1}^{m+n} \!O_{x_j^+,p_j,\sigma_j} \prod_{j=1}^m O_{x_j^-,p_j,\sigma_j}^\dagger}{\Psi} \approx \int \textstyle{\dnPpsi} \, \sqrt{-g} \, (2\pi)^{d+1}\delta^{d+1}\qty(\textstyle{\sum \alpha_j P_j})\, \ic \cM(\{P_j\})}
\end{equation}
where the bulk amplitude $\cM$ is a function of local Mandelstam invariants $s_{ij} =(\alpha_iP_i+\alpha_jP_j)^2$; again we set $\alpha_i = \pm 1$ to distinguish outgoing/incoming momenta.

The \emph{factorization formula} \eqref{m to n factorization} is a key result of this paper. It predicts sharp features in boundary correlators in terms of a bulk scattering amplitude when the shooting data $(x_j,p_j,\sigma_j)$ are tuned to reach a common bulk point. We now highlight its features in various examples.

\subsection{\texorpdfstring{Situations with slowly varying $\cM$}{{Situations with slowly varying M}}}
Suppose the amplitude $\cM$ is a slowly varying function of momenta, which can be treated as constant over the range probed by the wavepackets. Then the only nontrivial factor in \eqref{m to n factorization} is the momentum-conserving delta function. The individual $P_j$ integrals can be done using the Fourier transform \eqref{psi X}, which yields a final Gaussian integral in $X$-space that can also be done exactly: 
\begin{align}
    \label{effective delta function}
    &\int \dnPpsi \, \sqrt{-g} (2\pi)^{d+1}\delta^{d+1}\qty(\textstyle{\sum_j \alpha_j P_j}) = \sqrt{-g}\int \diff^{d+1} \fdiff X \, \prod_j \psibb_{x_j,p_j,\sigma_j}(X + \fdiff X) \\
    &= \textstyle{\qty[\prod_j \cF_j^{(*)}]} \sqrt{\det 2\pi \Sigma_{\rm tot}^{\bmu\bnu}} \exp(-\frac{1}{2} \sum_{j} \Sigma^{-1}_{j\,\bmu\bnu}\Delta X_j^\bmu\Delta X_j^\bnu - \frac{1}{2} \Sigma_{\rm tot}^{\bmu\bnu} (P^{\rm tot}_\bmu{+}\ic\Pi^{\rm tot}_\bmu) (P^{\rm tot}_\bnu{+}\ic\Pi^{\rm tot}_\bnu)), \nonumber
\end{align}
where $\cF_j \defeq F_{\Delta_j}(p_j; X)$ is the single-particle factor in \eqref{psi X definition F} and:
\begin{equation}
    P^{\rm tot}_\bmu = \sum_{j=1}^{m+n} \alpha_j P_{j\,\bmu}, \qquad
   \qty(\Sigma_{\rm tot}^{\bmu\bnu})^{-1} = \sum_{j=1}^{m+n} \Sigma_{j\,\bmu\bnu}^{-1\,(*)}, \qquad
   \Pi^{\rm tot}_\bmu = \sum_{j=1}^{m+n} \Sigma_{j\,\bmu\bnu}^{-1\,(*)}\Delta X_j^\bnu.
\end{equation}

While \eqref{effective delta function} is still somewhat bulky, all terms in the exponent have a transparent physical interpretation. The correlator can be suppressed for either of two reasons: if bulk momentum is not conserved at the interaction point, or if the trajectories miss.

In fact, the only factors related to holography are the single-particle boundary-to-bulk normalizations and phase shifts $\cF_j$. The rest just account for the wavepackets and would appear in any discussion of Gaussian wavepacket scattering in Minkowski space.

The tolerance for momentum conservation $P^{\rm tot}$ is controlled by the effective width $\sqrt{\Sigma_{\rm tot}^{\bmu\bnu}}$ of the product of Gaussians. The tolerance for the $\Delta X_j$ is determined by whether \emph{all} the involved Gaussian factors overlap in position. Notice that the full expression \eqref{effective delta function} is invariant under an overall translation $\Delta X_j\mapsto \Delta X_j+K$, $X\mapsto X-K$. This had to be the case since boundary correlators cannot depend on $X$. The formula is also unaffected by excitations that have much wider position-space Gaussians than the others (and thus a small $\Sigma_{j\,\bmu\bnu}^{-1}$).

A nice additional simplification occurs when the centers of wavepackets are perfectly aimed at each other. Then the determinant and exponential in \eqref{effective delta function} becomes simply an approximation to the momentum-conserving delta function:
\begin{equation} \label{approx delta function}
    \sqrt{\det 2\pi \Sigma_{\rm tot}^{\bmu\bnu}} \exp(-\frac{1}{2} \Sigma_{\rm tot}^{\bmu\bnu} P^{\rm tot}_\bmu P^{\rm tot}_\bnu)\approx (2\pi)^{d+1}\delta^{d+1}\qty(P^{\rm tot}) \quad \mbox{(perfectly-aimed wavepacekts)}.
\end{equation}
Alternatively, setting $P^{\rm tot}=0$, the formula reduces to the effective spacetime volume of the interaction region: $\sqrt{\det2\pi\Sigma_{\rm tot}}$.

\begin{figure}[t!]
    \centering
    \begin{tikzpicture}[scale = 0.9]
    \draw (-2.0,2.25) -- (-2.0,-3.3)  (2.0,-3.3) -- (2.0,2.25);
    \draw[dashed] (2.0,-3.3) arc (0:180:2.0 and 0.25);
    \draw (2.0,-3.3) arc (0:-180:2.0 and 0.25);
    \draw (2.0,2.25) arc (0:360:2.0 and 0.25);
    \filldraw[mblue!50] (0,-0.5) -- (0.85,-2.08) -- (0.79,-2.07) -- (-0.06,-0.5);
    \draw[thick, dashed] (2.0,-2.3) arc (0:180:2.0 and 0.25);
    \filldraw[mblue!50] (0,-0.5) -- (0.98,0.98) -- (0.92,.97) -- (-0.06,-0.5);
    \filldraw[mblue!50] (0,-0.5) -- (-1.02,1.415) -- (-1.08,1.4) -- (-0.06,-0.5);
    \filldraw[mblue!50] (0,-0.5) -- (-0.82,-2.52) -- (-0.88,-2.51) -- (-0.06,-0.5);
    \draw[thick, dashed] (2.0,1.2) arc (0:180:2.0 and 0.25);
    \draw[thick] (2.0,-2.3) node[anchor=west]{$H^-(X)$} arc (0:-180:2.0 and 0.25);
    \draw[thick] (2.0,1.2) node[anchor=west]{$H^+(X)$} arc (0:-180:2.0 and 0.25);
    \filldraw[mblue] (-.03,-.5) circle (1.5pt);
    \node[anchor=west] at (0,-.5) {$X$};
    \end{tikzpicture}
    \caption{An exclusive $2 \to 2$ scattering process in global AdS using wavepackets, where the scattering happens at the center of global AdS.}\label{fig:2to2-global}
\end{figure}
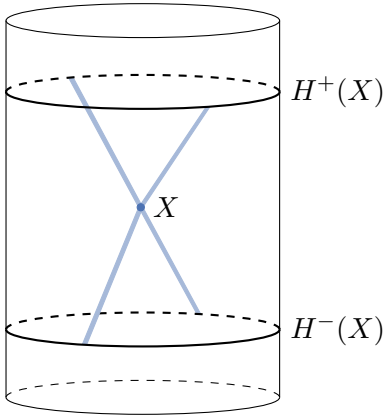

\paragraph{Empty AdS}
This setup includes as a special case the definition of (massless) bulk scattering amplitude in \cite{Polchinski:1999ry, Li:2021snj, Alday:2024yyj} in empty global AdS, illustrated in figure \ref{fig:2to2-global}. Aiming perfectly at the center of AdS is easy, by symmetry. If the desired momenta at the center of AdS are $(P_j^0, \vec{P}_j)= E_j(1,\vec{n}_j)$ where  $\vec{n}_j \in \bbR^d$ is a unit vector near the bulk point, then it suffices to shoot radially inward from corresponding points
$\pm \mathbf{n}_j \in \rS^{d-1}$ in the boundary at times $\pm \pi/2$:
\begin{equation}
    x_j^\mu=\alpha_j(\pi/2,\mathbf{n}_j), \qquad p_j^\mu = (E_j,\mathbf{0}).
\end{equation}
The classical action for this path vanishes and the van Vleck-Morette determinant is just $1/(2E_j)$. \footnote{As noted below \eqref{SXp}, the definition of $S(p;X)$ for a curved boundary depends on the choice of a local basis of covector fields. For AdS in global coordinates, a useful choice to describe geodesics ending near a boundary point $(t,\mathbf{n}_j)$ is to project the bulk spatial momentum to its $(d-1)$ components orthogonal to $\mathbf{n}_j$: $p_\mu=(-E,\mathbf{p}_{{\rm bulk} \perp \mathbf{n}_j})$. Then up to a quadratic error in each variable:
$$S(p;X+\fdiff X) = -E(\fdiff T + \mathbf{n}_j {\cdot} \fdiff \mathbf{X}) + \vec{p}_{{\rm bulk} \perp \mathbf{n}_j}{\cdot}\fdiff\mathbf{X} +O((\fdiff X)^2,(\mathbf{p}_{{\rm bulk} \perp \mathbf{n}_j})^2)$$
which gives the claimed determinant around the normal trajectory with $\mathbf{p}_{{\rm bulk} \perp \mathbf{n}_j}=0$.} Here we are assuming that $E_j\gg 1$ and that the wavepackets have a time spread $\sqrt{\sigma^{tt}}\ll 1$ in global time so that the sources are localized near times $\pm \pi/2$ at the boundary, see figure \ref{fig:2to2-global}. Choosing wavepackets that are sufficiently close to momentum eigenstates so that we can ignore dispersive spreading (this means angular size $\sqrt{\sigma_j^{\theta\theta}} \gtrsim 1/\sqrt{E_j}$), the single-particle factors \eqref{psi X definition F} then reduces to
\begin{equation}
    \cF_j= \frac{\cC_{\Delta_j,E_j}^{+}}{\sqrt{2E_j}},
\end{equation}
with $\cC$ defined in \eqref{eq:unipre}. For the described wavepackets, our formula \eqref{m to n factorization} with \eqref{approx delta function} thus amounts to:
\begin{equation} \label{amplitude AdS}
    \expval{\prod_{j=m+1}^{m+n} \!O_{x_j^+,p_j,\sigma_j} \prod_{j=1}^m O_{x_j^-,p_j,\sigma_j}^\dagger}{0} \approx \qty(\prod_{j=1}^n {\cal F}_j^{(*)})
    \sqrt{\det 2\pi\Sigma_{\rm tot}}
    \,\ic \cM(\{P_j\}).
\end{equation}
This seems equivalent to equation~(12) of \cite{Polchinski:1999ry}, although we haven't checked the precise normalization conventions.

\subsubsection{Connection with position-space and Mellin formulations} \label{ssec:literature}
The presence of ``sharp signals'' corresponding to bulk point singularities has been extensively studied in the literature (see \cite{Polchinski:1999ry, Gary:2009ae, Heemskerk:2009pn, Penedones:2010ue, Okuda:2010ym, Maldacena:2015iua, Chandorkar:2021viw}). In position space, a common starting point is the Schwinger representation of the bulk-to-boundary propagator (see \eqref{K schwinger app}),
\begin{equation}\label{K schwinger}
    K_\Delta(x;X) = \frac{\tilde{C}_\Delta^{\frac{1}{2}} \ep^{-\frac{\ic \pi \Delta}{2}}}{\Gamma(\Delta)} \int_0^\infty \diff s \, s^{\Delta - 1} \ep^{\ic s\,\hat{x} \cdot X},
\end{equation}
where $\hat{x}_M$ and $X$ are embedding coordinates for the boundary and bulk point, respectively (satisfying $\hat{x}^2=0$ and $X^2=-1$). Near a point $X$ that is lightlike from $x$ (i.e., $\hat{x} \cdot X=0$), this can be interpreted as a superposition of plane waves whose $(d+1)$-dimensional momentum $s \hat{x}$ is proportional to the direction of the geodesic connecting $x$ and $X$.

An $n$-point correlator for a bulk contact vertex $-\ic \lambda_n$ can then be written as \cite{Maldacena:2015iua,Chandorkar:2021viw}
\begin{equation}\label{n point position}
\expval{ T O(x_1)\cdots O(x_n)}{0}
=\left[\prod_{j=1}^n
\frac{\tilde{C}_\Delta^{\frac{1}{2}} \ep^{-\frac{\ic \pi \Delta}{2}}}{\Gamma(\Delta)} \int_0^\infty \diff s_j\right]
\int \diff^{d+1}X\,\sqrt{-g}\,
\ep^{\ic \sum_j  s_j\,\hat{x}_j{\cdot} X}(-\ic \lambda_n).
\end{equation}
In order to observe an apparent singularity, one needs the $n$ local momenta to become large in a way that conserves total momentum, which requires a linear relation $\det(\hat{x}_i \cdot \hat{x}_j)=0$. This determines the $s_j$'s up to an overall scale, which can be interpreted as the center-of-mass energy of the bulk process and which has to be integrated over. The contribution from arbitrary high energies then produces a singularity as $\det(\hat{x}_i \cdot \hat{x}_j)\to 0$.

As argued in \cite{Maldacena:2015iua}, the contact model \eqref{n point position} is only reasonable at low energies and one expects (in string theory realizations and conjecturally for any unitary boundary CFT) the bulk amplitude to soften in this high energies, fixed-angle regime. Thus, an actual CFT correlator displays no mathematical singularity but rather a narrow peak as $\det(\hat{x}_i \cdot \hat{x}_j) \to 0$.\footnote{For $n=(d+2)$ points satisfying the determinant constraint, the signal comes from a unique bulk point, otherwise it comes from an extended submanifold whose volume diverges in as $\det(\hat{x}_i \cdot \hat{x}_j)\to 0$. For $n<d+2$, one should divide by this volume before assessing the presence or absence of a bulk-point singularity \cite{Maldacena:2015iua}.}

The main difference in our approach is that we start by integrating \eqref{K schwinger} against boundary wavepackets, one for each $\hat{x}$. One effect of the wavepackets is that they determine each $s_j$: the center-of-mass energy of the bulk scattering process is determined directly by the wavepackets rather than through the size of $\det(\hat{x}_i \cdot \hat{x}_j)$. The resulting peak will have a comparable spread if the energies are chosen to be close to the softening scale in the bulk (ie. the string scale) and the wavepackets suitably narrow, otherwise, our signals will be more blurred than those in \cite{Maldacena:2015iua} in a way controlled by the width of the wavepackets.

The other effect of wavepackets is to control directionality and to remove the degeneracy in the bulk point for $n<(d+2)$. By considering inclusive (as opposed to exclusive) scattering amplitudes, we also make the bulk momentum conservation constraint easier to satisfy, as discussed below.

Finally, we briefly comment on the relation between the flat space amplitude ${\cal M}$ and Mellin amplitude $M$, which for light operators also includes an integral over overall energies:\cite{Penedones:2010ue, Fitzpatrick:2011hu, Li:2021snj}:
\begin{equation}
    \cM(s_{ij})=R_\AdS^{\frac{n(d-1)}{2}-d-1} \Gamma\qty(\frac{\Delta_\Sigma-d}{2}) \int_{-\ic \infty}^{\ic \infty} \frac{\diff \alpha}{2\pi \ic} \, \ep^{\alpha} \alpha^{\frac{d-\Delta_\Sigma}{2}}M\qty(\delta_{ij}=-\frac{R_\AdS^2}{4\alpha}s_{ij}),
\end{equation}
where $\Delta_\Sigma=\sum_{i=1}^n \Delta_i$. This was conjectured in \cite{Penedones:2010ue} and proved later \cite{Fitzpatrick:2011hu} using wavepacket arguments and a saddle point approximation similar to those discussed above \eqref{amplitude AdS}. In this sense, in empty AdS, the physical content of the Mellin formula is the same as that of wavepacket amplitudes. The main advantage of the wavepacket approach is its flexibility to describe processes in other geometries.

\subsection{Stretching diagrams: bulk Landau singularities} \label{ssec:landau}
A natural situation in which $\cM$ is \emph{not} slowly varying is when momenta are tuned to approach a Landau singularity. Consider for example a pole:
\begin{equation}
    \cM \propto \frac{\ic}{S_I-m^2+\ic 0} =
    \int_0^\infty \diff a\, \ep^{\ic a(S_I-m^2)}, \label{pole factorization}
\end{equation}
where $S_I=-P_I^2$ and $P_I=\sum_{j\in I} \alpha_j P_j$ is some Mandelstam invariant. We again have a Gaussian exponent and to the linear order in $\fdiff P$, it can be easily absorbed into the wavepacket \eqref{psi bulk good}, leading to a minimal modification of \eqref{effective delta function}:
\begin{equation}
\label{displaced vertex}
        \Delta X_j^\bmu \mapsto \Delta X_j^\bmu + a P_I^\bmu,\qquad (j\in I),
\end{equation}
together with multiplying the amplitude by an overall phase $\ep^{-\ic a m^2}$.

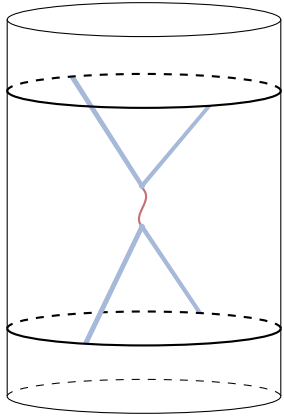
\begin{figure}[t!]
    \centering
    \begin{tikzpicture}[scale = 0.9]
    \draw (-2.0,2.25) -- (-2.0,-3.3)  (2.0,-3.3) -- (2.0,2.25);
    \draw[dashed] (2.0,-3.3) arc (0:180:2.0 and 0.25);
    \draw (2.0,-3.3) arc (0:-180:2.0 and 0.25);
    \draw (2.0,2.25) arc (0:360:2.0 and 0.25);
    \filldraw[mblue!50] (0,-0.8) -- (0.85,-2.08) -- (0.79,-2.07) -- (-0.06,-0.8);
    \draw[thick, dashed] (2.0,-2.3) arc (0:180:2.0 and 0.25);
    \filldraw[mblue!50] (0,-0.2) -- (0.98,0.98) -- (0.92,.97) -- (-0.06,-0.2);
    \filldraw[mblue!50] (0,-0.2) -- (-1.02,1.415) -- (-1.08,1.4) -- (-0.06,-0.2);
    \filldraw[mblue!50] (0,-0.8) -- (-0.82,-2.52) -- (-0.88,-2.51) -- (-0.06,-0.8);
    \draw[thick, dashed] (2.0,1.2) arc (0:180:2.0 and 0.25);
    \draw[thick] (2.0,-2.3) arc (0:-180:2.0 and 0.25);
    \draw[thick] (2.0,1.2) arc (0:-180:2.0 and 0.25);
    \draw[mred!70, thick] (-.06,-.2) .. controls (.2,-.4) and (-.2,-.6) .. (-.03,-.8);
    \filldraw[mblue!50] (-.03,-.2) circle (1pt);
    \filldraw[mblue!50] (-.03,-.8) circle (1pt);
    \end{tikzpicture}
    \caption{A $2 \to 2$ scattering in global AdS with bulk center-of-mass energy tuned to be close to a resonance.
    The time separation between the circles is then larger than $\pi$.}\label{fig:2to2-stretch-global}
\end{figure}

The physical interpretation of \eqref{displaced vertex} is simple: the correlator \eqref{effective delta function} is now peaked when all trajectories for particles $j\in I$ meet at some vertex $X_I$, \emph{and}  all trajectories with $j\notin I$ meet at another vertex $X_{\bar{I}}$, with displacement $X_I-X_{\bar{I}}= a P_I$. In other words, we have two displaced vertices in the bulk, see figure \ref{fig:2to2-stretch-global}.

Strictly speaking, an amplitude cannot exhibit a pole like \eqref{pole factorization} for real physical momenta. At best, one can produce a narrow resonance with some decay width $\Gamma$ (``if it can be produced, it can decay''). The corresponding complex shift $m\mapsto m - \ic \Gamma/2$ will make the above signal decay like $\ep^{-\tau\Gamma/2}$ where $\tau$ is the proper time separating the two vertices. If the displacement between the vertices is sufficiently small compared with the AdS radius, then we can still describe the whole process using the flat space factorization \eqref{m to n factorization}, otherwise \eqref{m to n factorization} breaks down and one should instead use the full AdS amplitude (which could still factorize into a product of flat-space amplitudes times a bulk-to-bulk propagator).

An identical analysis applies near a multi-particle normal threshold, corresponding to a branch cut in $\cM$, e.g., from a one-loop bubble diagram. In fact, the only modification is to insert a power of $a$ in \eqref{pole factorization}.

We expect a similar analysis to apply near more complicated Landau singularities, e.g., anomalous thresholds. In flat space, these singularities can be expressed as integrals like \eqref{pole factorization} but involving multiple Schwinger parameters, whose effect is to replace \eqref{displaced vertex} by the general Coleman-Norton formula for displaced vertices \cite{Coleman:1965xm}. When the vertices get displaced by too great a distance, our flat space factorization formula will break down and the diagram will presumably morph into an AdS Landau diagram \cite{Komatsu:2020sag}. Note that here we are considering real external momenta, and we do not have anything new to add about complex singularities.

\subsection{Inclusive correlators from generalized amplitudes}\label{ss:inclusive}
The preceding ``exclusive amplitudes'' all require the ability to reach $X$ from opposite directions, otherwise one cannot conserve bulk momentum. In particular, radial momentum can never be conserved starting from a compact region of a single Poincar\'e patch. However, as discussed in the introduction, inclusive expectation values can be nontrivial even in such situations.

Inclusive expectation values are described by non-time-ordered correlators, which we can treat using \eqref{eq:dictionary wavepacket 2} in exactly the same fashion as exclusive amplitudes. We discuss the result for the simplest nontrivial case, where three operators are inserted in the past of a bulk point $X$ and one in its future (say, $x_3$), as depicted in figure \ref{fig:incl-exp}.
\begin{equation}\label{inclusive factorization}
\begin{split}
    \expval{
    \Osmeared_{x_4^-,p_4,\sigma_4}
    \Osmeared_{x_3^+,p_3,\sigma_3}
    \Osmeared^\dagger_{x_2^-,p_2,\sigma_2}
    \Osmeared^\dagger_{x_1^-,p_1,\sigma_1}}{\Psi}
    \approx \int \dnPpsi
    \expval{a_{X,P_4} b_{X,P_3} a^\dagger_{X,P_2}a^\dagger_{X,P_1}}{0}^{\text{bulk}}
    \\ = \int \dnPpsi \, \sqrt{-g} \, (2\pi)^{d+1}\delta^{d+1}\qty(\textstyle{\sum_j} \alpha_j P_j) \, \ic \cM(s,t),
\end{split}
\end{equation}
where $s=-(P_1+P_2)^2$ and $t=-(P_2-P_3)^2$. We recall that in our notations all momenta in the subscripts of operators have a positive energy, and we followed the same steps as in \eqref{m to n factorization}. In the last step, we used the relation explained below \eqref{otoc example} to express the result in terms of a perfectly conventional (exclusive) $2\to 2$ on-shell bulk amplitude.

The physics of \eqref{inclusive factorization} remains qualitatively distinct from that in \eqref{m to n factorization}, and in particular, bulk momentum conservation is now easier to satisfy. To be fully explicit, consider empty AdS in Poincar\'{e} coordinates, where the bulk momenta $(p_\mu,P_z)$ corresponding to \eqref{inclusive factorization} are (see table \ref{t:2pt-bdry-conditions}): 
\begin{equation}
    P_1=(p_{1\mu},+\abs{p_1}),\quad
    P_2=(p_{2\mu},+\abs{p_2}),\quad
    P_3=(p_{3\mu},-\abs{p_3}),\quad
    P_4=(p_{4\mu},+\abs{p_4}),
\end{equation}
where $\abs{p_i} = \sqrt{-p_i^2}$. The crucial feature is the positive sign in $P_{4z}$, which makes it possible to now conserve all $(d+1)$-dimensional momenta:
\begin{equation}\label{inclusive mom conservation}
P^{\rm tot}=0 \quad\mbox{with}\quad
    P^{\rm tot} = P_3+P_4-P_1-P_2.
\end{equation}

\begin{figure}[t]
    \centering
    \raisebox{3em}{
    \begin{subfigure}{0.4\textwidth}
        \centering
        \begin{tikzpicture}[scale=0.7]
        \draw[mblue!65, line width=2pt]  (0,-.2) -- (.2,1.3) arc (172:82:.06) coordinate (turn);
        \draw[mblue!35, line width=2pt] (turn) arc (82:-8:.06) -- (.12,-.17) (.09,-.44) -- (-0.13,-2.1);
        \draw[dashed] (turn)+(-0.5,0.5)--++(0.5,-0.5);
        \fill[mblue!35] (-0.08,-2.1) -- (-.084,-2.14) -- (-.186,-2.14) --  (-0.1775,-2.08) -- cycle;
        \fill[mblue!65] (-2.6,-2.77) -- (-0.08,-.2) -- (0.08,-.2) -- (-2.4,-2.69) -- cycle;
        \fill[mblue!65] (2.4,-2.69) -- (-0.08,-.2) -- (0.08,-.2) -- (2.6,-2.77) -- cycle;
        \fill[mblue!65] (-1.5,2.33) -- (-0.08,-.2) -- (0.08,-.2) -- (-1.3,2.29) -- cycle;
        \draw[thick] (-3,3) node[anchor=north east]{\footnotesize $H^+(X)$} to[out=-30,in=-150] (3,3);
        \draw[thick] (-3,-3) node[anchor=south east]{\footnotesize $H^-(X)$} to[out=30,in=150] (3,-3);
        \draw (-.2,-.2) node[anchor=east]{$X$};
        \filldraw[mblue] (0,-0.2) circle (2pt);
        \draw (-.4,2.14) node[anchor=south]{\footnotesize $O_3$}
                (-2.5,-2.73) node[anchor=north]{\footnotesize $O^\dagger_1$}
                (-.05,-2.14) node[anchor=north]{\footnotesize $O_4$}
                (2.4,-2.69) node[anchor=north]{\footnotesize $O^\dagger_2$};
    \end{tikzpicture}
    \end{subfigure}}
    \begin{subfigure}{0.5\textwidth}
    \centering
    \input{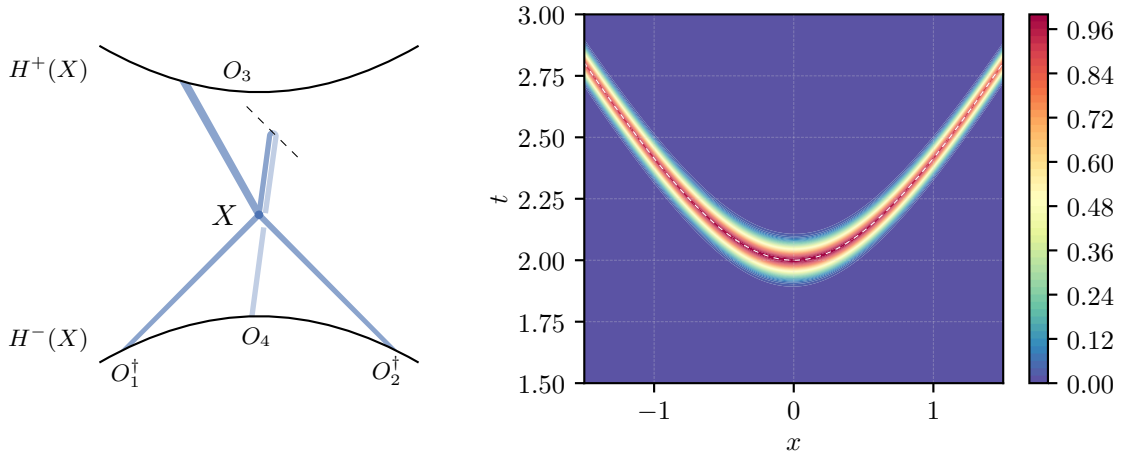}
    \end{subfigure}
    \caption{Absolute value of the inclusive correlator \eqref{inclusive factorization} recorded by a detector at $x_3$, assuming fixed Gaussian wavepackets for all operators. The meaning of the dashed line is as in figure \ref{fig:incl-exp}. Up to the width of the wavepackets, the signal peaks on the future lightcone of a single bulk point. The horizontal stretch, along which a signal simultaneously appears at spacelike separated boundary points, is a smoking gun of local bulk dynamics. Here we used $\theta = \pi/3$ (see appendix \ref{app:incl}).} \label{fig:ads-gaussian}
\end{figure}

Let us describe these kinematics in more detail for the thought experiment shown in figure \ref{fig:ads-gaussian}. There we consider a coherent superposition of one and two-particle states,
\begin{equation}
    |\Psi\rangle = O_{x_1,p_1,\sigma_1}^\dagger O_{x_2,p_2,\sigma_2}^\dagger|0\rangle + O_{x_4,p_4,\sigma_4}^\dagger|0\rangle,
\end{equation}
where all three particles are aimed toward a common bulk point $X$. The expectation value \eqref{inclusive factorization} then captures an interference contribution to the one-point function $\expval{O_{x_3,p_3,\sigma_3}}{\Psi}$. We choose the momenta symmetrically as follows, in flat boundary coordinates $(p^0,p^x,p^y)$:
\begin{equation} \label{kinematics incl}
    p_1^\mu= E_1 (1,\sin \theta,0), \quad
    p_2^\mu= \omega_2 (1,-\sin\theta,0),\quad
    p_3^\mu= \omega_3 (1,v_3^x,v_3^y),\quad
    p_4^\mu= E_4(1, 0, 0)
\end{equation}
where $\mathbf{v}_3^2<1$ is some arbitrary boundary velocity and $0 < \theta < \frac{\pi}{2}$. The trajectories can meet at a finite point provided that $\theta \neq 0$.

In figure \ref{fig:ads-gaussian} we display the magnitude of the Gaussian envelope \eqref{effective delta function} as a function of the observation point $x_3$, for a specific fixed value of the $E_i$ and $\omega_i$ and fixed coordinates for $x_1$, $x_2$ and $x_4$. A full description of the chosen parameters can be found in appendix \ref{app:incl}. A clear peak is visible along the intersection of the future lightcone of $X$ with the boundary, which is a hyperbola of curvature radius equal to the depth $z$ of the bulk point.

It would be very difficult to explain the sudden appearance of a signal along this hyperbola (whose points are spacelike-separated from each other) without the bulk picture: we view this signal as a smoking gun of local bulk dynamics.

The full correlation function \eqref{inclusive factorization} is the Gaussian envelope \eqref{effective delta function} times a bulk amplitude, which in general depends on the bulk dynamics. At a minimum, even in the absence of direct couplings between the (say, scalar) probes, we expect gravitational bulk interactions, which for example in the perturbative domain to give the one-graviton-exchange amplitude
\begin{equation} \label{grav}
    \cM\approx 8\pi G_N \qty( \frac{su}{t}+ \frac{st}{u}+ \frac{tu}{s})\qquad\mbox{(one graviton exchange)}.
\end{equation}
The Mandelstam invariants corresponding to figure \ref{fig:ads-gaussian} depend weakly on $x_3$ so as long as the bulk amplitude is a slowly varying function of Mandelstam invariants, it will not qualitatively affect the shape of the peak.

In a holographic theory, perturbative expressions like \eqref{grav} are only valid for bulk center-of-mass energies below the higher-spin scale, $\sqrt{s}R_\AdS\lesssim \Delta_{\rm h-s}$. As noted above, the softening of amplitudes at higher energies prevents bulk-point singularities from becoming infinitely sharp. In a non-holographic theory, such as the 3D Ising model (see \cite{Caron-Huot:2022lff}), $\Delta_{\rm gap}$ is not large and correlators effectively behave as if the bulk amplitudes decay as soon as $\sqrt{s} R_\AdS\gtrsim 1$. This leads to a time smearing in figure \ref{fig:ads-gaussian} that is as large as the curvature radius of the hyperboloid, which becomes indistinguishable from merely the future lightcone of a boundary point. Thus, in a non-holographic theory, no bulk dual needs to be invoked to explain the signal.

The electron-photon scattering experiment sketched at the beginning of the introduction is a slight modification of this setup where the wavepacket data for particle 1 (the ``electron'') is the complex conjugate of that of particle 4. In these kinematics, we find that the energy of the reflected photon (particle 3) is typically somewhat smaller than in the kinematics \eqref{kinematics incl}, and the geometry is more affected by dispersive spreading. This leads to a blurrier signal for the same center-of-mass energy, but otherwise qualitatively similar. How these CFT correlators can naturally model interactions with an ambient space is further discussed in appendix \ref{app:portal}.

The bulk center-of-mass energy for the correlator displayed in figure \ref{fig:ads-gaussian} is $\sqrt{s} R_\AdS=100\sqrt{3}$. Requiring that this is below the higher-spin scale, for example in ${\cal N}=4$ super-Yang-Mills where this coincides with the mass of first stringy excitation ($\Delta_{\rm h-s}\approx 2\lambda^{\frac14}$), requires a sufficiently large 't Hooft coupling $\lambda$ and number of colors:
\begin{equation}
    \lambda\gtrsim (\sqrt{s} R_\AdS/2)^4\approx 10^8,\qquad  N_c\gtrsim \frac{\lambda}{4\pi}\approx 10^7\qquad\mbox{(for figure \ref{fig:ads-gaussian})}.
\end{equation}
For smaller 't Hooft coupling, the particular correlation function plotted here will become blurrier due to the softening of the bulk amplitude. It would be interesting to explore if a modified experiment could produce sharp images of $H^+$ using bulk center-of-mass energies that are not so large (and therefore achievable with a smaller 't Hooft coupling and smaller $N_c$).

\section{Extensions}\label{s:mass-spin}
So far, we have explored the map between the bulk and boundary observables, through the semiclassical approximation for massless scalar fields in general background geometries. This captures most of the key features of our method, with the exception of mass and spin. In this section, we aim to address some subtleties when the particle is massive or spinning. In \S\ref{ssec:discussion} we discuss other subtleties with the factorization formula, notably long-range interactions.

\subsection{Heavy fields} \label{s:ms:mass}

\begin{figure}
    \centering
    \begin{tikzpicture}[scale = 1.5]
    \draw[fill=mblue!10,opacity=0.3] (0.5,-0.5) -- (3,-1) -- (3,4) -- (0.5,4.5) -- cycle;
    \draw[fill=mgray, draw=none, opacity=.5] (1.93,-0.18) -- (2.07,-0.28) -- (2.07,0.62) -- (1.93,0.72) -- cycle;
    \draw[color=mblue,decorate,decoration={snake,amplitude=0.5mm,segment length=1mm}] 
    (-0.75,3) .. controls (-.25,1.75) and (.75,1.5) .. (1.25,1) .. controls (1.9,0.35) and (1.9,0.15) .. (1.25,-.5);
    \draw[dashed,thick,color=mblue] (-0.75,3) -- (-1,4);
    \draw[dashed] (2,1) -- (2,0.3) (1.74,1) -- (1.74,0.45);
    \draw[<->] (2,1.03) -- (1.74,1.03);
    \draw (1.87,1.03)node[anchor=south,font=\scriptsize,color=black]{$z = \frac{\Delta}{\abs{p}}$};
    \filldraw (2,0.25) circle (1pt) node[anchor=south west,font=\footnotesize,color=black]{$x,p$};
    \filldraw (-0.75,3) circle (1pt) node[anchor=south west,font=\footnotesize,color=black]{$X,P$};
    \end{tikzpicture}
    \caption{Creating a massive bulk excitation from the boundary: the wavy line shows a real timelike geodesic which remains separated from the boundary by a radial distance $z \sim \Delta / \abs{p}$. In order to interact with the boundary, the excitation must tunnel across this distance. The boundary signal can only be time-localized with an uncertainty  $\abs{\diff t} \geq \Delta/\abs{p}$, as illustrated by the shaded area. }
    \label{fig:wavepacketmassive}
\end{figure}
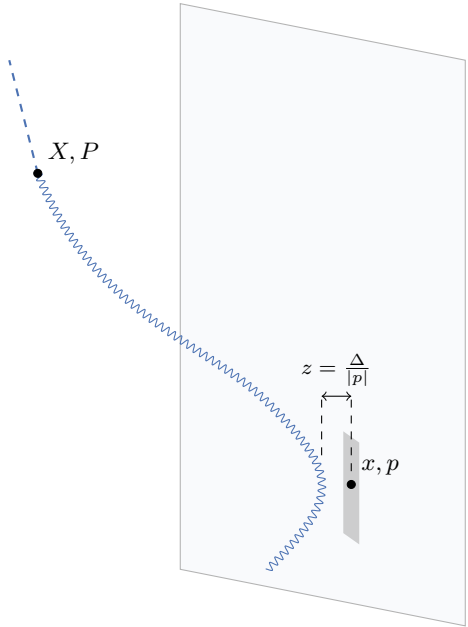

For massive bulk fields, the treatment of \S\ref{s:dynamics} remains valid in the regime where $P_z^2 \gg m^2$, where, the bulk excitation is ultrarelativistic.
However, there is now also a nonrelativistic regime in which the WKB approximation is valid, so it is interesting to extend the formulas to this additional regime. This requires treating the mass and momenta as large and parametrically of comparable size.

An important subtlety is that no real timelike geodesic connects the boundary to the bulk's interior. This can be seen from the mass shell condition for a massive particle in the Poincar\'{e} AdS metric:
\begin{equation}
     P_z = \sqrt{\abs{p}^2 - \frac{m^2}{z^2}}, \qquad \abs{p}^2 = - \eta^{\mu\nu} p_\mu p_\nu.
\end{equation}
This reveals the presence of a potential barrier for massive particles near the boundary of the spacetime. The region $z<m/\abs{p}$ is not classically accessible, and any manipulation of a massive bulk excitation from the boundary will require tunneling across this region. Although the resulting boundary observables are necessarily exponentially small, heavy fields can still be interesting and are the subject of a vast literature.

\begin{figure}
    \centering
    \input{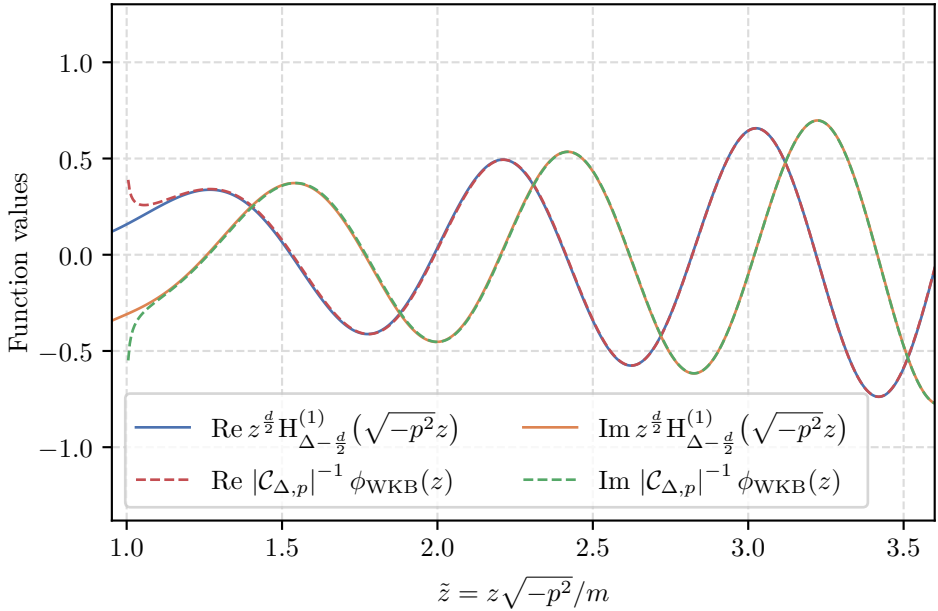}
    \caption{Comparison of the real and imaginary parts of the WKB solution \eqref{phi massive WKB} with the exact solution (Bessel function) for a massive particle, using the parameter $\Delta = 10$ in $d = 3$, i.e., $m = \sqrt{70}$. The plot demonstrates that the WKB approximation becomes valid in the non-relativistic region almost immediately after the particle tunnels through the potential barrier at $\tilde{z}<1$ near the boundary of spacetime.}\label{fig:massive-behavior}
\end{figure}

In the tunneling region, we can write the WKB ansatz (from the boundary to the bulk) as
\begin{equation}
    \phi(z) \propto
    \frac{z^{\frac{d}{2} - \frac{1}{2}}}{\sqrt{2 \abs{P_z}}} \exp(- \int^{z} \diff \zeta \, \sqrt{\frac{m^2}{\zeta^2} - \abs{p}^2}).
\end{equation}
At $z\to 0$, this behaves like $z^{\frac{d}{2}-m}$ which agrees with the expected form $z^{d-\Delta}$ up to $1/m$ corrections using the relation $\Delta = \frac{d}{2} + \sqrt{\frac{d^2}{4} + m^2}$. We can thus apply the normalization \eqref{limit phi} to this order (including the extra $-i$ from Lorentzian signature), and performing the integral we find:
\begin{equation}
    \phi(z) = \qty[(2\Delta - d) C_\Delta]^{-\frac{1}{2}} \qty(\frac{\abs{p}\ep}{2m})^{m} \frac{-\ic z^{\frac{d}{2}}}{(1-\abs{p}^2z^2/m^2)^{1/4}} \, \ep^{-\sqrt{m^2 - \abs{p}^2 z^2} + m \arctanh \frac{\sqrt{m^2 - \abs{p}^2 z^2}}{m}},
\end{equation}
which is expected to be valid up to corrections of order $1/m$ or $1/p$. We can now analytically continue into the oscillatory regime (with the usual $p^2\mapsto p^2-\ic 0$) to get:
\begin{equation}
    \phi(z) = \qty[(2\Delta - d) C_\Delta]^{-\frac{1}{2}} \qty(\frac{\abs{p}\ep}{2m})^{m} \frac{z^{\frac{d}{2}}\ep^{-\ic \frac{\pi}{4}}}{(\abs{p}^2z^2/m^2-1)^{1/4}} \, \ep^{\ic \sqrt{m^2 - \abs{p}^2 z^2} -\ic m \arctan \frac{\sqrt{\abs{p}^2 z^2-m^2}}{m}}.
\end{equation}
For large $m$, we have $\Delta\approx \frac{d}{2}+m \gg 1$. In this regime, we can use Stirling's approximation to write this WKB answer in a more evocative form which matches the constant in \eqref{eq:unipre}:
\begin{equation} \label{phi massive WKB}
\phi(z) = \cC^+_{\Delta,p} \qty(\frac{z^{d-1}}{2\abs{P_z}})^{\frac12}
\exp(\ic \int_{\frac{m}{\abs{p}}}^z \diff \zeta \, \sqrt{\abs{p}^2- \frac{m^2}{\zeta^2}}+\ic\frac{m\pi}{2})\times (1+O(1/m)).
\end{equation}
If we expand at large $z\gg m/\abs{p}$ we reproduce precisely \eqref{eq:scalarwkb-ads}. This calculation confirms that by applying the WKB approximation all the way to the boundary, following a complex trajectory that accounts for the tunneling region, we reproduce precisely the boundary-to-bulk propagator that can be calculated exactly. The agreement includes the precise normalization and phase, up to $1/m$ corrections at large $m$. A numerical comparison with the exact solution in empty AdS in the non-relativistic regime is shown in figure \ref{fig:massive-behavior}.

This agreement makes it straightforward to extend the boundary to bulk dictionary \eqref{eq:dictionary} to a general asymptotically AdS metric. The parenthesis in \eqref{phi massive WKB} generally becomes the van Vleck-Morette determinant, while the exponent is simply the action of the massive particle along a suitable complex trajectory. The only subtlety is that this action should be defined using a holographic renormalization scheme which subtracts the large-$m$ limit of the factors already included in $\cC$:
\begin{equation}\label{SXp massive}
    S(p;X) \equiv \lim_{\epsilon\to 0}\qty[S_\epsilon(p;X) \mp \ic m \log\frac{2m}{\epsilon\abs{p}\ep}+\frac{m\pi}{2}]\qquad\mbox{(massive case)},
\end{equation}
where $S_\epsilon$ is the on-shell action integrated from a boundary at $z=\epsilon$. Note that on-shell geodesics that contribute to $S(p;X)$ have a complex endpoint $\Im x^\mu \sim \pm p^\mu \Delta / p^2$ where the sign depends on whether the trajectory tunnels in or out of the boundary. The definition is such that $S(p;X)$ reduces to the action of a massless particle in empty AdS when $\abs{p}z\gg m$.

\paragraph{Time resolution for massive particles}
Due to the complex endpoint just mentioned, one can anticipate that it is difficult to achieve good time resolution when heavy bulk fields are involved. Mathematically, this effect can be seen from the energy-dependent factor $\abs{p}^\Delta$ in \eqref{eq:unipre}, which can lead to heavy spectral distortions and can significantly influence the saddle point location when $\Delta$ is large. Consider for example a wavepacket with small spatial momentum. If the boundary wavepacket has a Gaussian energy dependence, then the transmitted amplitude
$$\propto E^{\Delta} \ep^{- \frac{1}{2} \sigma_t^2 \qty(E - E_0)^2}$$
peaks at an energy that is far into the tail of the original Gaussian, unless  $\sigma_t^2 \gtrsim \Delta^2/ E_0^2$. The temporal width of the wavepacket is then $\Delta/E_0$. This means that it is effectively impossible, from the boundary, to control the time of a single heavy bulk excitation to better than $\Delta/E_0$. This is comparable to the size of the imaginary part of the saddle point time noted below \eqref{SXp massive} and is larger than the naive uncertainty relation by a factor of $\Delta$.

This temporal spreading is also comparable to the curvature radius of the hyperbolic trajectory followed by the particle in the bulk, see \ref{fig:wavepacketmassive}. If better time resolution could be achieved in practice, one could witness a particle moving instantaneously across a bulk depth $\Delta/\abs{p}$: a causality violation that could be easily diagnosed from spacelike commutators at the boundary.

We strongly advise against using Gaussian wavepackets that have a smaller temporal spread. In effect, one would simply be fooling oneself with  ``signals'' that come from the deep tail of the boundary wavepackets rather than from near their peaks.

\subsection{Spinning fields}\label{s:spin}
Here, we extend the WKB method to spinning fields. We begin with general kinematics and then discuss explicit wave equations for spin-1 and spin-2 bulk fields. One motivation is for eventual bootstrap applications, where the spin of excitations typically leads to stronger constraints on their dynamics.

We consider a (real) boundary operator $O^{\mu_1\cdots \mu_J}(x)$ that is a symmetric traceless tensor of spin $J$ and transforms like a conformal primary of dimension $\Delta$. The form of its vacuum two-point function is fully fixed by conformal symmetry, and we take it to be canonically normalized:
\begin{equation} \label{two-point spin}
    \expval{O^{\mu_1\cdots \mu_J}(x)O^{\nu_1\cdots \nu_J}(0)}{0} = \frac{\cI^{\mu_1(\nu_1}(x) \cdots \cI^{\mu_J\nu_J)}(x) - {\rm traces}}{x^{2\Delta}},
\end{equation}
where the $\nu$ indices are symmetrized and $$\cI^{\mu\nu}(x) \equiv \eta^{\mu\nu}-2\frac{x^\mu x^\nu}{x^2}.$$

In the bulk, the field dual to the above boundary operator is a symmetric traceless tensor field $H_{\bmu_1\cdots \bmu_J}(X)$. We would like to understand how $O^{\mu_1\cdots \mu_J}(x)$ integrated against a wavepacket turns into local bulk creation/annihilation operators of the field $H$. Following the logic of \S\ref{s:dynamics}, we start by understanding plane waves in translation-invariant states.

From the boundary perspective, the (timelike) momentum $p_\mu$ of the plane wave determines a frame, using which we can naturally distinguish transverse and longitudinal polarizations. We package all choices into an index-free notation:
\begin{equation} \label{def bdy pol}
(\bdypol^T+\bdypol^L \hat{p})^J \cdot O^\dagger_p
\equiv  (\bdypol^T_{\mu_1}+\bdypol^L \hat{p}_{\mu_1})
\cdots (\bdypol^T_{\mu_J}+\bdypol^L \hat{p}_{\mu_J})
\int \diff^dx\, \ep^{\ic p \cdot x} O^{\mu_1\cdots\mu_J}(x),
\end{equation}
where
\begin{equation}
    \bdypol^T_\mu p^\mu =0=\bdypol^T_\mu\bdypol^{T\mu},\qquad \hat{p}=\frac{p^\mu}{\abs{p}}, \qquad \abs{p}=\sqrt{-p^2}.
\end{equation}
We take the transverse vector $\bdypol^T_\mu$ to be null to avoid writing explicit trace subtractions, with no loss of information. Note that $\bdypol^L$ is not a vector, it is just a bookkeeping device to track how many longitudinal indices we take in \eqref{def bdy pol}.

From symmetry considerations, in empty AdS, bulk solutions that correspond to different numbers of transverse and longitudinal indices will not mix with each other. 
This motivates the introduction of bulk versions of transverse and longitudinal polarizations.

In fact, it will be useful to think ahead and to define polarization vectors that make sense along an arbitrary geodesic in an asymptotically AdS spacetime.
Physically, we expect the polarization $\bulkpol_\bmu(\lambda)$ of a bulk particle to be always orthogonal to its momentum $P_\bmu(\lambda)$ (defined in \eqref{eq:momentum-def}) and to satisfy the parallel transport equation:
\begin{equation}\label{transport equation}
    P_\bmu g^{\bmu\bnu}\bulkpol_\bnu(\lambda)=0,
    \qquad \frac{\diff \bulkpol_\bmu(\lambda)}{\diff \lambda} = \Gamma_{\brho \bmu}^{\bnu}\bulkpol_\bnu(\lambda)
    \frac{\diff X^\brho}{\diff \lambda}.
\end{equation} 
Indeed, this follows readily from the WKB expansion of equations of motion (see appendix \ref{app:EOM}). It is thus natural to separate the transverse and longitudinal polarizations according to their boundary condition in the asymptotically AdS region by defining the following bulk vectors:
\begin{equation} \label{AdS pols}
    \lim_{z\to 0} \bulkpol^T_\bmu=\frac{1}{z}(\bdypol^T_\mu,0)\equiv \bulkpol^{T,\rm AdS}_\bmu,
    \qquad \lim_{z\to 0}\tilde{P}_\bmu=
    \frac{1}{m}\qty(P_z \hat{p}_\mu,\abs{p})\equiv \tilde{P}^{\rm AdS}_\bmu \qquad\mbox{(AdS limit)}.
\end{equation}
Both vectors $\bulkpol^{T,\rm AdS}_\bmu$ and $\tilde{P}^{\rm AdS}_\bmu$ satisfy the parallel transport condition \eqref{transport equation} in empty AdS, where we recall that in Poincar\'e coordinates $P_\bmu=(p_\mu,P_z)$ with $P_z=\sqrt{\abs{p}^2-m^2/z^2}$. These boundary conditions are essentially the only vectors we can write down given the symmetries of AdS, where we have chosen to normalize the longitudinal vector so that $\tilde{P}^2=1$. The boundary conditions can then be evolved along any bulk geodesic by integrating the parallel transport equation \eqref{transport equation}.

Having defined bulk polarizations, we can now state a natural proposal for the spinning version of the WKB boundary-to-bulk dictionary \eqref{eq:dictionary}, now relating the boundary operator \eqref{def bdy pol} to a bulk creation operator involving the polarizations $\bulkpol^T_\bmu$ and $\tilde{P}_\bmu$. In the WKB region, bulk polarizations are expected to follow the parallel transport equation \eqref{transport equation}, however, in the non-WKB region near the AdS boundary, they can evolve differently. Our proposal must thus allow for a different normalization factor depending on how many boundary indices are transverse or longitudinal. Taking a generic term in \eqref{def bdy pol}, our proposal is thus:
\begin{equation} \label{eq:dictionary spin}
    \bdypol^T_{\mu_1}\cdots \bdypol^T_{\mu_n} \hat{p}_{\mu_{n+1}}\hat{p}_{\mu_J} O_p^{\dagger\, \mu_1 \cdots\mu_J}\simeq
    \sqrt{\cD(p;X)}\ep^{\ic S(p;X)}
    \,\bulkpol^T_{\bmu_{1}}\cdots \bulkpol^T_{\bmu_n}\,
    \tilde{P}_{\bmu_{n+1}}\cdots \tilde{P}_{\bmu_J}\,
    \times \cC_{\Delta,J,p}^{(n)+}\,a^{\dagger\,\bmu_1\cdots \bmu_J}_{X,P}
\end{equation}
where the van Vleck-Morette determinant and geodesic action are the same as for scalar fields discussed in \eqref{eq:dictionary} and the only change is the normalization $\cC_{\Delta,J,p}^{(n)+}$, which now depends on the number $n$ of transverse indices (similar to \eqref{eq:dictionary}, the equation has a second case with $\cC_{\Delta,J,p}^{(n)-}b^\dagger_{X,P}$ when $O^\dagger_p$ is inserted in the future of $X$).

Note that we treat the bulk creation operator $a^{\dagger\,\bmu_1\cdots \bmu_J}_{X,P}$ as traceless, which is why we do not write trace subtractions in \eqref{eq:dictionary spin}. Its commutation relation with $a^{\bmu_1\cdots \bmu_J}_{X,P}$ is the residue of the pole in \eqref{propagator spin}.

Below we demonstrate this dictionary explicitly in empty AdS using the field equations for vector and spin-2 tensors, for which we give explicit formulas for the coefficients $\cC_{\Delta,J,p}^{(n)}$.

For massless bulk fields, i.e., dual to conserved boundary operators with $\Delta-J=d-2$, we will find that the coefficients with $n<J$ vanish and the correct interpretation of \eqref{eq:dictionary spin} is to simply discard to the right-hand-side when $n<J$, i.e., all longitudinal modes. Note however that this is not equivalent to taking the $m\to 0$ limit of \eqref{eq:dictionary spin} since the longitudinal polarizations diverge as $\tilde{P}\propto 1/m$. This reflects the well-known fact that the massless limit is not smooth for spinning particles.

\subsubsection{Free vector field}\label{s:ms:maxwell}
The action for a (free) canonically normalized massive vector field $A(X)$ in the bulk is
\begin{equation}
    S = \int \diff^{d+1}X \, \sqrt{-g} \, \qty(-\frac14 g^{\bmu\brho} g^{\bnu\bsigma} F_{\bmu\bnu} F_{\brho\bsigma} -\frac{m^2}{2} g^{\bmu\bnu} A_\bmu A_\bnu),
\end{equation}
where $F_{\bmu\bnu}=\partial_\bmu A_\bnu-\partial_\bnu A_\bmu$. It leads to the equation of motion
\begin{equation} \label{maxwell EOM}
    \frac{1}{\sqrt{-g}} \partial_\bsigma \qty( \sqrt{-g} g^{\brho\bmu} g^{\bsigma\bnu} \qty(\partial_\bmu A_\bnu - \partial_\bnu A_\bmu) ) + m^2 g^{\brho\bsigma} A_\bsigma = 0.
\end{equation}
The components $A_M$ are not all independent. Upon applying a derivative to the equations of motion, we obtain the constraint equation:
\begin{equation}\label{eq:massivemaxwellconstraint}
   m^2 \partial_K \qty( \sqrt{-g} g^{\brho\bmu} A_\bmu) = 0.
\end{equation}
This constraint is valid only for the massive case and is trivial for a massless photon. We start by discussing the massive case and then comment on the massless case.

Consider empty AdS in Poincar\'e coordinates. On symmetry grounds, the bulk field induced by the operator \eqref{def bdy pol} must take the form
\begin{equation}\begin{aligned} \label{spin1 ansatz} \expval{A_\bmu(X)\,\qty(\bdypol^T_\mu +\bdypol^L \hat{p}_\mu)O^{\dagger\,\mu}_p}{0} &=
    \ep^{\ic p_\mu X^\mu}\qty( \bulkpol_\bmu^{T,\rm AdS}\alpha_T(z) + \bdypol^L
    (p_\mu\alpha_L(z),\alpha_z(z))) \\ &\equiv  \ep^{\ic p_\mu X^\mu} A_\bmu(z).
\end{aligned}\end{equation}
Substituting into the equation of motion immediately gives a decoupled equation for the transverse mode:
\begin{equation}\label{spin1 transverse eq}
     z^{d} \partial_z \qty(z^{-d+3} \partial_z (z^{-1}\alpha_T)) + \qty(z^2\abs{p}^2- m^2) \alpha_T=0.
\end{equation}
For the longitudinal mode, we have a coupled system involving $\alpha_L$ and $\alpha_z$. The most useful combinations are the constraint equation \eqref{eq:massivemaxwellconstraint}, and the $z$ equation after eliminating $\alpha_L$ using a derivative of the constraint:
\begin{align} \label{spin1 cons}
    m^2\qty(z^{d-1}\partial_z\qty(z^{-d+1}\alpha_z)-\ic \abs{p}^2\alpha_L)&=0,\\ \label{spin1 z eq}
    z^2 \partial_z \qty(z^{d-1} \partial_z \qty(z^{-d+1} \alpha_z)) + \qty(z^2\abs{p}^2- m^2) \alpha_z &= 0.
\end{align}
The equations \eqref{spin1 transverse eq} and \eqref{spin1 z eq} are in fact identical, and solved by Hankel functions:
\begin{equation} \label{radial sol spin1}
    \alpha_T, \alpha_z \propto z^{\frac{d}{2}}
    H^{(1/2)}_{\Delta-\frac{d}{2}}(\abs{p}z),
    \qquad
    \Delta = \frac{d}{2} +\sqrt{\frac{(d-2)^2}{4} + m^2}.
\end{equation}
Here we used the mass-dimension relation for a massive spin-1 field \cite{lYi:1998trg, Aharony:1999ti}. The longitudinal solution $\alpha_L$ is then obtained by differentiating $\alpha_z$ according to the constraint equation \eqref{spin1 cons}. Finally, we should normalize the solution by imposing at the AdS boundary
\begin{equation} \label{boundary condition vector}
    \lim_{z\to 0} \ic z^{d-1-\Delta} A_\mu(z) = \qty[(2\Delta - d)C_{\Delta,1}]^{-\frac{1}{2}} \qty( \bdypol^T_\mu+\bdypol^L \hat{p}_\mu),
\end{equation}
where $C_{\Delta,1}=C_{\Delta}\Delta/(\Delta-1)$. This normalization ensures that canonically normalized fields in the bulk match to canonically normalized operators at the boundary and was obtained by multiplying the scalar normalization in \eqref{limit phi} by a factor determined from the boundary two-point function \eqref{Cspin} with $n=1$. The $\ic$ is because we are in Lorentzian signature frequency space as noted below \eqref{limit phi}. In this way, all the scalar functions entering the bulk ansatz \eqref{spin1 ansatz} are determined:
\begin{equation}\begin{split}
\label{alpha sols}
    \qty(\alpha_T,\alpha_z,\alpha_L) &=
    \qty(1,\frac{-\partial_z+(d{-}1)/z}{\abs{p}(\Delta-1)}, \frac{-\ic\abs{p}}{\Delta-1})
    \frac{\pi \qty[(2\Delta - d)C_{\Delta,1}]^{-\frac{1}{2}}}{\Gamma\qty(\Delta - \frac{d}{2})} \qty(\frac{\abs{p}}{2})^{\Delta - \frac{d}{2}} z^{\frac{d}{2}} \rH^{(1)}_{\Delta - \frac{d}{2}} \big(\abs{p} z\big).
\end{split}\end{equation}
Finally, taking the large-$z$ limit, we find a scalar-like plane wave factor times bulk polarizations:
\begin{equation} \label{spin1 result}
    A_\bmu(z) \to
    \cC_{\Delta,1,p}^{(1)+}
    \qty(\frac{z^{\frac{d-1}{2}}}{\sqrt{2\abs{p}}}) \ep^{\ic \abs{p} z}
    \times \qty(
    \tfrac{1}{z}(\bdypol^T_\mu,0)
    -\frac{\ic}{\Delta-1}
    \bdypol^L (p_\mu,\abs{p})+\ldots).
\end{equation}
The transverse term in the parenthesis is precisely $\bulkpol^{T,\rm AdS}_{\bmu}$, and the longitudinal polarization is $m$ times the large-$z$ limit of $\tilde{P}^{\rm AdS}$ in \eqref{AdS pols}, confirming the general form \eqref{eq:dictionary spin} and determining the coefficients in it. For example, the form of the parenthesis directly fixes the ratio:
\begin{equation}
    -\ic (p_\mu,\abs{p})
    =-\ic m \tilde{P}^{\rm AdS}_\bmu+O(1/z)\quad\Rightarrow\quad
    \frac{\cC_{\Delta,1,p}^{(0)+}}{\cC_{\Delta,1,p}^{(1)+}}
    = -\ic \frac{m}{\Delta-1}.
\end{equation}
The two coefficients are finally related to the scalar one \eqref{eq:unipre} as
\begin{equation}
    \cC_{\Delta,1,p}^{(1)+} = 
    \sqrt{\frac{\Delta-1}{\Delta}}\,\cC_{\Delta,p}^{+},
    \qquad
    \cC_{\Delta,1,p}^{(0)+}
    = -\ic \sqrt{\frac{\Delta-d+1}{\Delta}}\,\cC_{\Delta,p}^{+},
\end{equation}
where we used the (exact) relation $m=\sqrt{(\Delta-1)(\Delta-d+1)}$. As a check, the absolute value of these quantities agrees precisely with the boundary two-point functions in \eqref{Cspin}.

It is significant that the longitudinal term in \eqref{spin1 result} is enhanced by a relative factor $z\abs{p}\gg 1$ (we only displayed the leading term for each polarization). This effect will be further discussed below for spin-two fields, where it is even more pronounced.

Finally, we briefly comment on the massless case. The equation and solution for the transverse mode are exactly the same, however, the equation \eqref{spin1 z eq} can no longer be used to determine $\alpha_z$ since it was obtained from the constraint \eqref{eq:massivemaxwellconstraint}. This is not surprising since the massless Maxwell equation exhibits a gauge redundancy, and in fact, one could simply impose the radial gauge: $A_z(z) = 0$. The other Maxwell equation then imposes that $\alpha_L(z)$ is constant: the bulk solution corresponding to the longitudinal mode is pure gauge, non-oscillatory, and will not contribute to any wavepacket amplitude. It can thus safely be dropped from the right-hand side of \eqref{spin1 result}.

\subsubsection{Linearized gravity and longitudinal subtleties}\label{s:ms:gravity}
We now move on to studying boundary insertions of the stress tensor $T^{\mu \nu}$ or of a non-conserved spin-two operator $O^{\mu\nu}$. The former corresponds to a metric fluctuation $h_{\bmu\bnu}$, whose study should be essential to understanding the emergence of bulk geometry. The latter is also interesting as it could represent a Kaluza-Klein excitation. In both cases, we consider only the linearized minimally coupled action, which takes the form \cite{Hinterbichler:2011tt}:
\begin{equation}
\begin{split}
    S[h] = \frac{1}{2} \int \diff^{d+1} X \, \sqrt{-g} &\Bigg(-\frac{1}{2}\nabla_\brho h_{\bmu\bnu}  \nabla^\brho h^{\bmu\bnu} +  \nabla_\brho h_{\bmu\bnu} \nabla^\bmu h^{\brho\bnu} - \nabla_\bmu h  \nabla_\bnu h^{\bmu\bnu} \\
    &\quad + \frac{1}{2} \nabla_\bmu h \nabla^\bmu h - \qty(d + \frac{m^2}{2}) h_{\bmu\bnu} h^{\bmu\bnu} + \qty(\frac{d}{2} + \frac{m^2}{2}) h^2 \Bigg),
\end{split}
\end{equation}
where we have used the value for the cosmological constant $\Lambda = - \frac{d(d-1)}{2}$ and the covariant derivatives are with respect to the background metric. The equations of motion are
\begin{equation} \label{spin2 EOM}
    \qty(\nabla^2
    -m^2-2d)h_{\bmu\bnu} -2\nabla_\brho \nabla_{(\bmu} h^\brho_{\bnu)} + \nabla_\bmu \nabla_\bnu h +g_{\bmu\bnu}\qty(\nabla_\brho \nabla_\bsigma h^{\brho \bsigma}-\nabla^2 h + (m^2+d)h) = 0.
\end{equation}
Assuming that the background satisfies the vacuum Einstein equations (with cosmological constant), the divergence and trace of the equation give respectively \cite{Porrati:2000cp}:
\begin{equation}\label{divstress}
    m^2 (\nabla_\bmu h^{\bmu\bnu}-\nabla_\bnu h) = 0,
    \qquad (d-1)(\nabla_\bmu\nabla_\bnu h^{\bmu\bnu}-\nabla^2 h+dh)-(d+1)m^2h = 0.
\end{equation}
In the massive case, these combine to the simpler equations: $h=0$ and $\nabla_\bmu h^{\bmu\bnu}=0$. These constraints leave us with the expected number of transverse and longitudinal polarizations for a massless spin-two particle in $(d+1)$-dimensions:
\begin{equation} \label{nd}
    n =  \frac{(d+2)(d-1)}{2}.
\end{equation}
In the massive gravity case, the constraints \eqref{divstress} close the second-class constraint algebra \cite{Chakravarty:2023cll}, which explains the $d$ additional polarizations as compared to the massless case, where $d$ solutions are diffeomorphisms that decouple.

As done above, we would like to calculate the boundary-to-bulk propagator in Poincar\'e AdS and compare it with the general WKB ansatz \eqref{eq:dictionary spin}:
\begin{equation}\label{spin2 ansatz} \expval{h_{\bmu\bnu}(X)\,\qty(\bdypol^T_\mu +\bdypol^L \hat{p}_\mu)
    \qty(\bdypol^T_\nu +\bdypol^L \hat{p}_\nu)
    O^{\dagger\,\mu\nu}_p}{0} =
    \ep^{\ic p_\mu X^\mu} h_{\bmu\bnu}(z).
\end{equation}
By exploiting the symmetries we can make an ansatz similar to \eqref{spin2 ansatz} involving seven scalar functions: one for each of the transverse, longitudinal, and radial cases for the two indices, and additionally a part proportional to $\eta_{\mu\nu}$. Substituting into \eqref{spin2 EOM} immediately gives an equation for the transverse-traceless mode:
\begin{equation}\label{spin2 transverse eq}
    \qty(z^2\partial_z^2-(d-1)z\partial_z+z^2\abs{p}^2-m^2)h_{TT}(z)=0.
\end{equation}
This is identical to the equation for a scalar field and is solved by Hankel functions $z^{\frac{d}{2}}H_{\Delta-\frac{d}{2}}(\abs{p}z)$ where $\Delta(\Delta-d)=m^2$. As in the vector case, by combining the equations of motion in the $z$ directions with the constraints \eqref{divstress}, we obtain an identical equation for  $h_{zT}(z)$ and for $h_{zz}(z)$, from which all other components can be recovered by taking derivatives. Finally, fixing the normalizations using the analog of \eqref{boundary condition vector} gives an exact solution in terms of Hankel functions, whose large-$z$ limit we find to be:
\begin{equation}\label{h limit}\begin{split}
    h_{\bmu\bnu}(z) \to
    \cC_{\Delta,2,p}^{(2)+}\qty(\frac{z^{\frac{d-1}{2}}}{\sqrt{2\abs{p}}}) \ep^{\ic \abs{p} z}
    \times &\Bigg(
    \bulkpol^{T,\rm AdS}_\bmu\bulkpol^{T,\rm AdS}_\bnu-\frac{\ic m}{\Delta}\bdypol^L(\tilde{P}_\bmu \bulkpol^{T,\rm AdS}_\bnu
    + \bulkpol^{T,\rm AdS}_\bmu\tilde{P}_\bnu) 
    \\ &\qquad
    -\frac{m^2}{\Delta(\Delta-1)} (\bdypol^L)^2\qty(\tilde{P}_\bmu\tilde{P}_\bnu
    -\Pi_{\bmu\bnu}/d)\Bigg),
\end{split}\end{equation}
with $\Pi_{\bmu\bnu}=g_{\mu\nu}+P_\mu P_\nu/m^2$. Again, we display only the leading term at large $z$ for each polarization, even though the transverse and longitudinal contributions do not scale with $z$ in the same way (see \eqref{AdS pols}). The form of this equation precisely confirms the proposed formula \eqref{eq:dictionary spin} and gives the coefficients:
\begin{equation} \label{C spin 2} 
    \cC_{\Delta,2,p}^{(2)+} = 
    \sqrt{\frac{\Delta-1}{\Delta+1}}\,\cC_{\Delta,p}^{+},
    \quad
    \cC_{\Delta,2,p}^{(1)+}
    = -\ic \sqrt{\frac{(\Delta-1)(\Delta-d)}{\Delta(\Delta+1)}}\,\cC_{\Delta,p}^{+},
    \quad
    \cC_{\Delta,2,p}^{(0)+}
    =-\frac{\Delta-d}{\sqrt{\Delta^2-1}}
    \,\cC_{\Delta,p}^{+}.
\end{equation}
In the massless case ($\Delta=d$), one should again simply discard the terms with $\bdypol^L$: one can always choose a gauge where $h_{z\bmu}=0$. The longitudinal modes are then pure gauge and do not oscillate with $z$.

\paragraph{Longitudinal subtleties} The magnitude of the first two coefficients in \eqref{C spin 2} agrees precisely with the boundary two-point function in \eqref{Cspin}, confirming the calculation. However, for massive spin-two fields, the outcome of the calculation is that the third coefficient does not quite match:
\begin{equation} \label{C 2 comparison}
    \abs{\cC_{\Delta,2,p}^{(0)\rm bdy}}^2
    =\abs{\cC_{\Delta,2,p}^{(0)+}}^2\times \qty(1+\frac{d-1}{\Delta(\Delta-d)}).
\end{equation}
This is an issue only when the massive particle is light and therefore ultra-relativistic in the bulk WKB regime. When a nonrelativistic regime exists, $\Delta\sim m \gg 1$ is necessarily large, and the mismatch can be neglected.

We attribute this somewhat puzzling result to the high-energy growth of the longitudinal modes. Namely, the longitudinal components in \eqref{h limit} are enhanced by $(\abs{p}z)^2$ compared with the transverse modes, which makes the subleading corrections not shown in \eqref{h limit} also contribute to the two-point function calculation. To test this hypothesis, we have calculated the ($z$-independent) Wronskian between $h$ and $h^*$ accounting for the subleading terms as $z\to\infty$, and we find the same factor:
\begin{equation}
\Im \qty[ \sqrt{-g} h_{\bmu\bnu}^*
\qty(\nabla^z h^{\bmu\bnu}-\nabla^\bmu h^{z\bnu}-\nabla^\bnu h^{\bmu z})_{(\bdypol^L)^2}]
\propto \qty(1+\frac{d-1}{\Delta(\Delta-d)}),
\end{equation}
which confirms that our exact solution in fact produces the correct boundary two-point function for any value of $\Delta$. Thus, the problem lies not with \eqref{h limit} but is caused by the necessity of keeping subleading corrections to it. This is an instance of a phenomenon previously observed in \cite{Chandorkar:2021viw}: it can be necessary to keep subleading curvature corrections in the bulk when dealing with longitudinal modes.

The high-energy growth of longitudinal polarizations also happens in flat space, however, the important fact is that it does not necessarily imply stronger scattering.
This depends on the explicit form of scattering amplitudes. In fact, cancellation of the energy growth is expected if the massive states reorganize at high energies into those of a sensible massless theory. (For example, longitudinal $WW\to WW$ scattering famously grows like a power of $s/M_W^2$ in the Standard Model without the Higgs particle, but in the full Standard Model this growth disappears.)

In general, we expect the leading WKB term described by \eqref{eq:dictionary spin} and \eqref{C spin 2} to give the correct correlator if there are no amplitude-level cancellations for the longitudinal-longitudinal modes, otherwise curvature corrections will be important. The two-point function calculation is an example where significant cancellations occur because $\tilde{P}^2=1$ is much smaller than the individual component of $\tilde{P}_\bmu$.

\subsection{Discussion: amplitudes versus hard functions and further issues}\label{ssec:discussion}
We derived the factorization formula \eqref{m to n factorization} by making an unrealistic assumption: that all bulk interactions are short-ranged. This cannot be exact. At a minimum, there has to be gravity in the bulk. Nonetheless, we are hopeful that the formula can be corrected. For gauge theories in flat space, it is indeed well understood how to separate (``factorize'') the degrees of freedom into hard, soft, and collinear modes, allowing to write scattering amplitudes as products of corresponding matrix elements:
\begin{equation}
	\cM = \cH \times \cS \times \prod_j \cJ_j.
\end{equation}
We refer to equation~(186) of \cite{Feige:2014wja} for a precise formula. Schematically, the hard function (also often called $C$ or infrared-renormalized amplitude) is an ordinary function of the momenta and polarizations of partons that meet at a point; the $\cS$ operator is a product of straight Wilson lines connecting this point to infinity; and $\cJ_j$ are jet creation/absorption operators (in practice, field insertions connected to infinity by reference Wilson lines). The only point we would like to make here is that all these ingredients can be naturally embedded into AdS space or any other geometry. The hard function should be unchanged, while the straight Wilson lines can be naturally upgraded to follow boundary-anchored geodesics, and similarly for the jet operators.

In other words, we are proposing that \eqref{m to n factorization} is actually incomplete, but that it should be possible to correct it by replacing $\cM$ by $\cH^{\text{bulk}}$ and multiplying by appropriate matrix elements of Wilson lines and jet functions. These matrix elements can be nontrivial even when there are no infrared divergences, due to finite contributions from bulk-to-bulk propagators of light fields stretching over AdS-scale distances. The relevant dimensionless parameters are $\ep^2/R_\AdS^{d-3}$ and $Gs/R_\AdS^{d-3}$ for gauge and gravity interactions, respectively.\footnote{These effects can be suppressed by considering a sequence of theories where e.g. $N_c\to\infty$ \cite{Polchinski:1999ry}. However, we would like a factorization formula that holds within a given theory.} We do not expect these multiplicative effects to affect the qualitative features of the correlators we studied, but they could affect precise numerical values at the loop level.

It could be interesting to pursue the WKB expansion to further subleading orders, building on AdS-corrected Virasoro-Shapiro amplitudes of \cite{Alday:2023mvu}.
Being able to account for curvature corrections in a general bulk geometry could have practical applications in other situations. These corrections are likely related to flat space amplitudes involving additional soft gravitons, but we do not have a precise statement to offer.

CFT correlators define bulk scattering amplitudes (or hard functions) with some accuracy. We would like to stress that this is not the same as defining an S-matrix. By ``scattering amplitude'' we mean broadly any matrix element of operators that create or absorb well-separated excitations. However, arguably, such amplitudes only define an S-matrix if the corresponding asymptotic states form a complete basis of the Hilbert space. We do not expect that the states we have considered form a useful basis of a CFT's  Hilbert space. Nonetheless, incompleteness does not prevent one from computing various amplitudes which inclusively sum over all final states, since these can be defined from non-time-ordered correlation functions.

How about compact extra dimensions? The wavepackets we considered only localize in the $(d+1)$-dimensional sense but involve delocalized normal modes in the extra dimensions. In $AdS_5\times S^5$, it is known how to extract local ten-dimensional features using correlators with large R-charge, which represent high Kaluza-Klein harmonics on the $S^5$ \cite{Aprile:2020luw}. It would be very interesting to learn how to do so in less symmetrical situations. A potentially relevant observation is that in some cases when multiple correlators are assembled into a matrix, the eigenvectors appear to automatically single out combinations that localize in the extra dimensions \cite{Aprile:2018efk, Caron-Huot:2018kta}.

\section{Summary and conclusion}\label{s:conclusion}
This work presented a refined dictionary to study bulk singularities of holographic correlators in arbitrary asymptotically AdS spacetimes. The framework is based on correlators integrated against Gaussian wavepackets, eschewing the use of complicated integral transforms. For sufficiently focused wavepackets, these correlators display singular features that reveal (the boundary imprints of) the past and future lightcones of bulk points, which we call boundary hyperboloids. Details of these features measure the scattering amplitudes of local bulk processes. We discussed some smoking gun signals of local bulk dynamics (see figure~\ref{fig:ads-gaussian}).

From the bulk perspective, this paper is essentially an exercise in geometrical optics. We used this approximation to study the propagation of high-frequency waves in terms of the geodesic motion of point particles, extending the WKB analysis of \cite{Polchinski:1999ry} to arbitrary geometries. Classical trajectories can be characterized by either their canonical boundary data $(x,p)$ or bulk data $(X,P)$. In the Hamilton-Jacobi formalism, one should fix either the position or the momentum at each endpoint; we found that the most useful choice is to fix a bulk point $X$ and the boundary momentum $p$ of a geodesic. The resulting Hamilton-Jacobi function $S(p;X)$, which was studied in greater detail in our companion paper \cite{Caron-Huot:2025she}, encodes all relevant optical properties of the bulk as seen from the boundary. This includes the shape of boundary hyperboloids and the normalization of signals through the van Vleck-Morette determinant \eqref{van Vleck-Morette}.

Our main result, the factorization formula \eqref{m to n factorization}, captures local physics near $X$ by combining the exact solutions of the wave equation near the AdS boundary, the WKB approximation on geodesics toward $X$, and a flat-space scattering amplitude near $X$. We demonstrated that the overall normalizations and phases can be universally predicted across asymptotically AdS geometries, for massless and massive fields, with or without spin.

The sharpness of features can be limited by two factors: optics and dynamics. On the optics side, we discussed the effects of the position and momentum resolution of the wavepackets, which are captured by an overall effective Gaussian factor \eqref{effective delta function} built from the product of the individual Gaussians (see \eqref{psi bulk good} and \eqref{psi X}). The optical properties of light bulk fields in AdS share many similarities with those in Minkowski space. For example, there is an absolute lower limit $\sim \sqrt{z/\omega}$ on the spatial localization that can be achieved with a certain energy at a certain depth into the bulk, which is similar to what can be achieved with ordinary laser beams. Heavy bulk fields are harder to localize since interactions between the bulk and boundary involve quantum tunneling.

Of course, none of this will happen if the dynamics of the CFT are not favorable. Necessary conditions for  CFT correlators to display singular features have been discussed in \cite{Caron-Huot:2022lff}. These include the requirements of \cite{Heemskerk:2009pn}: the CFT must have a large central charge and a large higher-spin gap. Even for theories that do satisfy these properties, perfect point-like resolution is not expected to be achievable due to the softening of bulk amplitudes above stringy or Planckian energies \cite{Maldacena:2015iua}.

Wavepacket correlators naturally appear in a simple physical setup: when a lower-dimensional CFT is coupled to a higher-dimensional theory, for example, the putative holographic CFT sheet interacting with ambient electromagnetic fields discussed at the beginning of the introduction. While the body of this paper analyzes wavepacket correlators themselves, for completeness, in appendix \ref{app:portal} we discuss such couplings. The main message is that a holographic CFT behaves in many ways like any optical interface. An incoming photon would partly reflect and partly enter the AdS space dual to the CFT, within which it would follow geodesics and scatter as studied in this paper.

If a holographic CFT could be realized in a laboratory, the considered experiments would enable direct measurements of its bulk metric and scattering amplitudes. In contrast with some other reconstruction procedures (see for example \cite{Fefferman:1985con, Hamilton:2006az}), these measurements would require no knowledge of bulk equations of motion. How to read off the bulk conformal metric (in a boundary-anchored coordinate system) from the shape of boundary hyperboloids was detailed in \cite{Caron-Huot:2025she}, building on the work of \cite{Engelhardt:2016wgb, Engelhardt:2016vdk, Engelhardt:2016crc}. The present paper demonstrates how the past and future hyperboloids of a bulk point $X$ can be measured using an open subset of each hyperboloid. 

Awaiting actual experiments, we hope that these thought experiments can shed light on the mechanism behind the emergence of extra dimensions of space, and of gravity, from strongly coupled dynamics. It is now fairly well understood that, in any CFT with a large central charge and large higher-spin gap, four-point stress-tensor correlators must agree with the predictions of Einstein's theory in a dual AdS space up to small corrections \cite{Heemskerk:2009pn, Afkhami-Jeddi:2016ntf, Caron-Huot:2021enk, Caron-Huot:2022jli, Chang:2023szz}. The approach outlined here lays the essential groundwork for extending these results to more complex backgrounds.

A number of questions for numerical and analytical bootstrap naturally suggest themselves. Can the bulk Raychaudhuri equations and focussing theorems be derived from properties of boundary correlators?  Can energy-momentum conservation or other principles in the boundary theory explain the equivalence principle--why must the bulk scattering amplitudes be the same around every bulk point?

\section*{Acknowledgements}
We thank Nima Afkhami-Jeddi for valuable discussions. We also thank Diksha Jain, Alex Maloney, Viraj Meruliya, and Aron Wall for related discussions. The work of S.C.H. and J.C. is supported by the National Science and Engineering Council of Canada (NSERC) and the Canada Research Chair program, reference number CRC-2022-00421. K.N. is supported in part by the Natural Sciences and Engineering Research Council of Canada (NSERC), funding reference number SAPIN/00047.

\appendix\section{Wave equation and van Vleck-Morette determinant} \label{app:EOM}
Here we describe an alternative derivation of the WKB wavefunction \eqref{Phi WKB ansatz} starting from the wave equations in the bulk. Consider a massive scalar field:
\begin{equation}
    \frac{\hbar^2}{\sqrt{-g}} \partial_\bmu \qty(\sqrt{-g} g^{\bmu\bnu} \partial_\bnu \Phi(X)) -m^2\Phi(X)= 0.
\end{equation}
The WKB method provides an order-by-order solution:
\begin{equation} \label{Phi hbar expansion}
    \Phi(X) = \exp\qty(\frac{\ic}{\hbar} \sum_{n=0}^\infty \hbar^n S_n(X)).
\end{equation}
At leading order, also known as the geometric optics approximation, we obtain:
\begin{equation} \label{HJ 0}
    g^{\bmu\bnu} \partial_\bmu S_0 \partial_\bnu S_0 +m^2= 0.
\end{equation}
This is readily solved by choosing $S_0(X) = S(p; X)$ in \eqref{SXp}. In fact, this is the only option, since \eqref{HJ 0} implies the bulk geodesic equations (see \cite{Caron-Huot:2025she}). At the next order in the WKB expansion—the physical optics approximation—both the normalization and phase shift are determined. At this order, the equation is
\begin{equation}
    \partial_\bmu \qty(\sqrt{-g} g^{\bmu\bnu} \partial_\bnu S_0) + 2\ic \sqrt{-g} g^{\bmu\bnu}\partial_\bmu S_0 \partial_\bnu S_1  = 0.
\end{equation}
By setting $S_1 = - \ic \log \cA$, we can rewrite our WKB ansatz and second-order equation as
\begin{equation}\label{eq:dynamics:pops}
    \Phi_p(X) = \cA(X) \, \exp(\frac{\ic}{\hbar} S(p; X)),\qquad
    \partial_\bmu \qty(\sqrt{-g} g^{\bmu\bnu} P_\bnu \cA(X)^2) = 0.
\end{equation}
The meaning of this equation is that the particle flux $P^\bmu \abs{\cA(X)}^2$ is conserved along the geodesic. We would like to check that this is satisfied by $\abs{\cA(X)}^2$ equal to the van Vleck-Morette determinant \eqref{van Vleck-Morette}. We can prove this by differentiating the mass-shell condition with respect to $p$, $P^\bmu \partial P_\bmu/\partial p=0$, and then with respect to $X$:
\begin{equation}
    0= \frac{\partial}{\partial X^\bnu}
    \qty[P^\bmu \begin{pmatrix}\frac{\partial P_\bmu}{\partial p_\mu}& n_\bmu\end{pmatrix}]
    =\frac{\partial P^\bmu}{\partial X^\bnu}
    \begin{pmatrix}\frac{\partial P_\bmu}{\partial p_\mu}& n_\bmu\end{pmatrix} +
    P^\bmu\frac{\partial}{\partial X^\bnu}
    \begin{pmatrix}\frac{\partial P_\bmu}{\partial p_\mu}& n_\bmu\end{pmatrix}.
\end{equation}
Multiplying on the right by the inverse of the matrix in the parenthesis and taking a trace this becomes
\begin{equation}
    \partial_\bmu P^\bmu+ P^\bmu \partial_\bmu \log \det\begin{pmatrix}\frac{\partial P_\bmu}{\partial p_\mu}& n_\bmu\end{pmatrix}  = 0.
\end{equation}
Note that this expression is not covariant, but it nicely becomes so upon including the $1/\sqrt{-g(X)}$ factor in the van Vleck-Morette determinant \eqref{van Vleck-Morette}:
\begin{equation}
    \partial_\bmu \qty(\sqrt{-g} g^{\bmu\bnu} P_\bnu \cD) = 0.
\end{equation}
This confirms that \eqref{Phi WKB ansatz}
satisfies the next-to-leading order WKB equation \eqref{eq:dynamics:pops} and is, therefore, the correct WKB bulk wavefunction.

\paragraph{Spinning fields} It is interesting to apply the same approach to spinning fields. The massive Maxwell equations of motion \eqref{maxwell EOM} for example take the covariant form:
\begin{equation} \label{maxwell covariant}
    (D^2-m^2)A_\bmu(X) - D^\bnu D_\bmu A_\bnu(X)=0,\qquad m^2 D^\bnu A_\bnu=0,
\end{equation}
where the second equation follows from taking the divergence of the first. It is natural to make a semi-classical ansatz using the \emph{same} scalar ingredients as in \eqref{eq:dynamics:pops} but adding a polarization:
\begin{equation}
 A_\bmu(X)=a_\bmu(X) \sqrt{\cD(p;X)} \, \ep^{\ic S(p;X)/\hbar}.
\end{equation}
Substituting into \eqref{maxwell covariant} we find at order $\hbar^{-2}$ the Hamilton-Jacobi equation \eqref{HJ 0}, which is satisfied by the ansatz, together with the constraint $P{\cdot}a=0$. At the order $\hbar^{-1}$ we obtain an interesting equation for $a_\bmu$. Using the transverse constraint and noting that the commutator $[D^\bnu,D_\bmu]$ is $O(\hbar^0)$, we can ignore the second term in \eqref{maxwell covariant} (in the massless case we could reach the same conclusion by working in covariant gauge $D^\bnu A_\bnu=0$). We then get
\begin{equation}
   \qty(\ic\hbar^{-1}  P^\bnu  D_\bnu a_\bmu)\sqrt{\cD} \, \ep^{\ic S/\hbar} +
   a_\mu D^2 \qty(\sqrt{\cD}\ep^{\ic S/\hbar})=O(\hbar^0).
\end{equation}
The second term is $O(\hbar^0)$ thanks to the scalar equation \eqref{eq:dynamics:pops}, so we conclude that $P{\cdot}  D a_\bmu=0$ at leading order: the polarizations must satisfy the parallel transport equation along each geodesic. This is a geometrically natural result used in the main text. We expect it to apply to other wave equations, although we have not verified this explicitly.

\section{Bulk-boundary correlators and propagators}\label{s:b2bpropagator}
\subsection{Fourier transform of bulk-to-boundary propagator in empty AdS}
Here we give an alternative derivation of the bulk plane wave solutions in empty AdS, by Fourier-transforming the exact bulk-to-boundary propagator:
\begin{equation}\label{eq:b2bpwprop}
    \Phi_p(X) = \int \diff^d x \, K_\Delta(x;X) \, \ep^{\ic p_\mu x^\mu},
\end{equation}
where $K_\Delta$ is the standard bulk-to-boundary propagator in empty AdS \cite{Freedman:1998tz}:
\begin{equation} 
    K_\Delta(x;X) = \tilde{C}_\Delta^{\frac{1}{2}} \qty(\frac{z}{z^2 + \eta_{\mu\nu}(X^\mu - x^\mu) (X^\nu - x^\nu)})^\Delta, \qquad \tilde{C}_\Delta = \frac{\Gamma(\Delta)}{2 \pi^{\frac{d}{2}} \Gamma(\Delta - \frac{d}{2} + 1)}.
\end{equation}
A convenient way to make contact with the bulk WKB approximation is to utilize the Schwinger parametrization:
\begin{equation}\label{K schwinger app}
    K_\Delta(x;X) = \frac{\tilde{C}_\Delta^{\frac{1}{2}} \ep^{-\frac{\ic \pi \Delta}{2}}}{\Gamma(\Delta)} \int_0^\infty \diff s \, s^{\Delta - 1} \exp(\ic s \frac{z^2 + \eta_{\mu\nu}(X^\mu - x^\mu) (X^\nu - x^\nu)}{z}).
\end{equation}
This representation is particularly helpful when comparing with the existing literature on bulk point singularities, see \S\ref{ssec:literature}. Here, starting with the plane wave \eqref{eq:b2bpwprop} at the boundary, the integral over $x$ is now Gaussian and can be evaluated explicitly, yielding:
\begin{equation} \label{Phi X app}
    \Phi_p(X) =\frac{\tilde{C}_\Delta^{\frac{1}{2}} \ep^{-\frac{\ic \pi}{4}(2\Delta-d+2)}}{\Gamma(\Delta)} (\pi z)^{\frac{d}{2}} \ep^{\ic p_\mu X^\mu} \int_0^\infty \diff s \, s^{\Delta - \frac{d}{2} - 1} \exp(\ic \qty(s - \frac{p^2}{4s}) z).
\end{equation}
The $s$ integral may then be computed explicitly. For $p$ timelike with the usual Feynman $p^2-i0$ (and a slight tilt $s\mapsto s(1+i0)$ at \emph{large} $s$), the integral yields the result:
\begin{equation}
    \Phi_p(X) = 
\pi \qty(\frac{\pi^{\frac{d}{2}}}{2\Gamma(\Delta)\Gamma(\Delta+1-\frac{d}{2})})^{\frac12}
 \qty(\frac{\sqrt{-p^2}}{2})^{\Delta-\frac{d}{2}} z^{\frac{d}{2}}
    H^{(1)}_{\Delta-\frac{d}{2}}(z\sqrt{-p^2}).
\end{equation}
This matches the formula in \eqref{phi T}, confirming its normalization. The WKB approximation discussed in \S\ref{s:dynamics} amounts to evaluating \eqref{Phi X app} via a saddle-point approximation, which replaces the Hankel function with its large-$z$ asymptotics. By integrating against a boundary Gaussian wavepacket, one also reproduces the bulk Gaussian in \eqref{psi X} and \eqref{X Sigma AdS}.

\subsection{Normalization check using scalar two-point function}
As a straightforward consistency check, we verify that the map \eqref{eq:dictionary wavepacket} correctly reproduces the past-past two-point functions on the boundary:
\begin{equation}\label{norm check}
    \begin{split}
    \expval{\Osmeared_{x_-,p,\sigma}\Osmeared^\dagger_{x_-,p,\sigma}}{\Psi}
    &\simeq \int \frac{\diff^{d+1}P}{(2\pi)^d\sqrt{-g}} \frac{\diff^{d+1}P'}{(2\pi)^d\sqrt{-g}} \, \delta(P^2) \delta(P'^2) \psib_{x,p,\sigma}(P) \psibs_{x,p,\sigma}(P') \, [a_{X,P'},a^\dagger_{X,P}]    
    \\ &=
    \int \frac{\diff^{d+1}P}{(2\pi)^d\sqrt{-g}} \delta(P^2)
    \abs{\psib_{x,p,\sigma}(P)}^2
    \\ &=
    \int \frac{\diff^{d+1}P}{(2\pi)^d\sqrt{-g}} \, \delta(P^2) \frac{\abs{\cC_{\Delta,k}}^2}{\cD(k,X)} \abs{\tilde{\psi}_{p,\sigma}(k)}^2 \Bigg|_{k=k(P,X)} \\ &= \int \frac{\diff^d  k}{(2\pi)^d} \, \abs{\cC_{\Delta,k}}^2 \abs{\tilde\psi_{p,\sigma}(k)}^2,
\end{split}
\end{equation}
where to pass to the second line we inserted the canonical commutation relation \eqref{eq:a comm}, which simply cancels the $P'$ integral, on the third line we inserted the explicit expression \eqref{bulk wavepacket 1}, and finally we used the Jacobian \eqref{P p Jacobian} backward to express the result as a boundary integral. Finally, we observe that $\abs{\cC_{\Delta,k}}^2$ is precisely the vacuum Wightman function of $O$ in momentum space (equal to twice the imaginary part of its time-ordered correlator for $k^\mu$ is future timelike):
\begin{equation}
\label{Csquared interpretation}
    \abs{\cC_{\Delta,k}}^2 = \frac{\pi^{\frac{d}{2}}}{\Gamma(\Delta)\Gamma\qty(\Delta+1-\frac{d}{2})} \qty(\frac{-k^2}{4})^{\Delta - \frac{d}{2}}  = \int \diff^d x \frac{\ep^{-\ic k \cdot x}}{x^2-\ic 0 x^0}.
\end{equation}
By Plancherel's formula, we conclude that \eqref{norm check} is just the Fourier transform of the two-point function
\begin{equation}
\expval{\Osmeared_{x_-,p,\sigma}\Osmeared^\dagger_{x_-,p,\sigma}}{\Psi}
\simeq \int \diff^d x\,\diff^d  y \,
\psi^*_{p,\sigma}(x)\psi_{p,\sigma}(y) \expval{O(x)O(y)}{0},
\end{equation}
as it should. This boundary correlator is simple since both operators are inserted at nearly equal times, so the correlator can be evaluated without doing any significant time evolution. This is also why we get a vacuum correlation function since we assume that the state does not contain any pre-existing excitations at that frequency; the short-distance product is then dominated by identity exchange. This calculation confirms the role of the van Vleck-Morette determinant in providing the Jacobian of the transformation between bulk and boundary momenta that ensures probability conservation.

The relation \eqref{Csquared interpretation} gives a simple way to determine the norm of $\cC_{\Delta,p}$ without having to work out the precise normalization of bulk-to-boundary propagators. We now describe the analogous calculation for symmetric traceless tensors.

\subsection{Normalization for spinning fields}
\label{s:normalization}
As in the main text, let $O^{\mu_1\cdots \mu_J}$ be a traceless symmetric primary operator in the CFT, with dimension $\Delta$ and spin $J$. We take it to be canonically normalized:
\begin{equation} \label{two-point spin app}
    \expval{O^{\mu_1\cdots \mu_J}(x)O^{\nu_1\cdots \nu_J}(0)}{0}=
    \frac{\cI^{\mu_1(\nu_1}(x) \cdots \cI^{\mu_J\nu_J)}(x) - {\rm traces}}{x^{2\Delta}},
\end{equation}
where the $\nu$ indices are symmetrized and
$$\cI^{\mu\nu}(x)=\eta^{\mu\nu}-2\frac{x^\mu x^\nu}{x^2}.$$
In the bulk, its dual is a symmetric traceless tensor field $H^{\bmu_1\cdots \bmu_J}(X)$. We take it to be canonically normalized, so its momentum space propagator in the flat space limit is
\begin{equation} \label{propagator spin}
G^{\bmu_1\cdots\bmu_J,\bnu_1\cdots \bnu_J}(P) =
    \frac{\Pi^{\bmu_1(\bnu_1}(P)\cdots \Pi^{\bmu_J\bnu_J)}(P)- {\rm traces}}{P^2+m^2},
\end{equation}
which is again symmetrized in $\bnu$ indices and where
$$\Pi^{\bmu\bnu}(P)=g^{\bmu\bnu}+\frac{P^\bmu P^\bnu}{m^2}.$$
In this case, the traces should be subtracted using $\Pi^{\bmu_i\bmu_j}$. This ensures that the residue on the $P^2=-m^2$ pole is transverse: $P_{\bmu_1}G^{\bmu_1\cdots\bmu_J,\cdots}(P)$ is pole-free.

We would like to understand the normalization of the boundary-to-bulk map which relates the preceding two expressions. This exercise amounts to computing the Fourier transform of \eqref{two-point spin app}.

The Fourier transform can be calculated straightforwardly using \eqref{Csquared interpretation} and its derivatives (see for example \cite{Gillioz:2018mto, Karateev:2018oml}). To describe the result, it is useful to separate the polarizations into transverse and longitudinal with respect to the (timelike) momentum $p$. We can package all choices into an index-free notation:
\begin{equation}
    (\chi^T+\chi^L \hat{p})^J \cdot O_p
    \equiv  (\chi^T_{\mu_1}+\chi^L \hat{p}_{\mu_1})
    \cdots (\chi^T_{\mu_J}+\chi^L \hat{p}_{\mu_J})
    \int \diff^dx\, \ep^{-\ic p \cdot x} O^{\mu_1\cdots\mu_J}(x),
\end{equation}
where
\begin{equation}
    \chi^T_\mu p^\mu =0=\chi^T_\mu\chi^{T\mu},\qquad \hat{p}=\frac{p^\mu}{\abs{p}}, \qquad \abs{p}=\sqrt{-p^2}.
\end{equation}
We take the transverse vector $\chi^T$ to be null to avoid writing explicit trace subtractions.

We will be interested in the Wightman function, which is nonzero for $p$ future-timelike. On symmetry grounds, the result must take the general form 
\begin{equation}\begin{split} \label{OO spin}
    \expval{\,
    (\chi^T+\chi^L \hat{p})^J \cdot O_p\,
    (\chi'^T+\chi'^L \hat{p})^J \cdot O_{p'}^\dagger\,
    }{0} &= \theta(p^0)\theta(-p^2)(2\pi)^d\delta^d(p-p') 
    \\ & \times
    \sum_{n=0}^J
    \abs{\cC_{\Delta,J,p}^{(n)\rm bdy}}^2 \cN_{J}^{(n)} 
    (\chi^T \cdot \chi'^T)^n (\chi^L\chi'^L)^{J-n}.
\end{split}\end{equation}
Here $n$ counts the number of transverse indices. The coefficients depend on $n$ and have actually been computed before. We read off these coefficients from equation~(2.28) of \cite{Gillioz:2020wgw}:
\begin{align}\label{Cspin}
    \abs{\cC_{\Delta,J,p}^{(n)\rm bdy}}^2&=
    \qty(\frac{-p^2}{4})^{\Delta - \frac{d}{2}}
    \frac{2 \pi^{\frac{d}{2}+1} (\Delta-1)_n(\Delta-J-d+2)_{J-n}}{\Gamma(\Delta+J) \Gamma\qty(\Delta - \frac{d}{2} + 1)}, \, \\
    \cN^{(n)}_J &= \frac{J!}{n!(J-n)!}
    \frac{\qty(d-2+2n)_{J-n}}{2^{J-n}\qty(\tfrac{d-2}{2}+n)_{J-n}}.
\end{align}
We have verified explicitly \eqref{OO spin}-\eqref{Cspin} up to spin $J=4$.

For $J=0$, the normalization in \eqref{Cspin} reduces to the scalar coefficient in \eqref{eq:unipre}. For $J>0$ the coefficients are positive above the \emph{unitarity bound} $\Delta-J\geq d-2$ and all longitudinal modes $(n<J)$ decouple at the unitarity bound, in agreement with the operator then being conserved.

The factor $\cN^{(n)}_J$, which is slightly different in our basis than in the one considered in \cite{Gillioz:2020wgw}, naturally normalizes our structure. It was obtained as follows. Suppose that $\chi^T_M$ and $\tilde{P}_M$ are transverse and longitudinal bulk vectors, satisfying:
\begin{equation}
    P \cdot \chi^T=\tilde{P} \cdot \chi^T=P \cdot \tilde{P}=0=\chi^T \cdot \chi^T,\quad \tilde{P} \cdot \tilde{P}=1.
\end{equation}
In empty AdS in Poincar\'e coordinates, these will be naturally related to the boundary polarizations as:
\begin{equation}
    \chi^T_\bmu(z)=\frac{1}{z}(\chi^T_\mu,0),
    \quad \tilde{P}_\bmu(z)=
    \frac{1}{m}\qty(P_z \hat{p}_\mu,\abs{p}),
\end{equation}
where we recall that $P_z=\sqrt{\abs{p}^2-m^2/z^2}$. A combinatorial calculation (guessed from the first few cases) then gives the pole in the bulk propagator \eqref{propagator spin} dotted into this basis:
\begin{equation} \label{HH bulk}
    (\chi^T+\chi^L \tilde{P})^J\cdot
    G(P) \cdot (\chi'^T+ \chi'^L\tilde{P})^J
    =\frac{1}{P^2+m^2}\sum_{n=0}^J
    \cN_{J}^{(n)} 
    (\chi^T \cdot \chi'^T)^n (\chi^L\chi'^L)^{J-n}.
\end{equation}
Comparison between the boundary and bulk two-point functions \eqref{OO spin} and \eqref{HH bulk} then suggests that the spinning boundary-to-bulk dictionary \eqref{eq:dictionary spin} must multiply each term by $\cC^{(n)}_{\Delta,J,p}$, e.g.,
\begin{equation}
    (\chi^T+\chi^L \hat{p})^J \cdot O_p^\dagger =
    \sqrt{\cD(p;X)}\ep^{iS(p;X)}
    \sum_{n=0}^J 
    \frac{J!\,\cC_{\Delta,J,p}^{(n)+}}{n!(J-n)!}
    \,\chi^T_{\bmu_{1}}\cdots \chi^T_{\bmu_n}\,
    (\chi^L)^{J-n}
    \tilde{P}_{\bmu_{n+1}}\cdots \tilde{P}_{\bmu_J}\,
    a^{\dagger\,\bmu_1\cdots \bmu_J}_{X,P}
\end{equation}
where $\abs{\cC_{\Delta,J,p}^{(n)+}} = \abs{\cC_{\Delta,J,p}^{(n)\rm bdy}}$. As discussed below \eqref{C 2 comparison}, this relation between bulk and boundary coefficients works for transverse modes but fails by $O(\Delta^{-2})$ curvature corrections when more than two longitudinal indices are involved $(n\leq J-2)$.

\section{Details of inclusive experiment kinematics}\label{app:incl}
Here we describe the inclusive experiment plotted in figure \ref{fig:ads-gaussian} in more detail. To recall the kinematics, the bulk momenta in this experiment are given by
\begin{equation}
\begin{aligned}
    P_{1 \bmu} &= E_1 \qty(-1,\sin \theta,0, \sqrt{1 - \sin^2 \theta}), &\quad
    P_{2 \bmu} &= \omega_2 \qty(-1,-\sin\theta,0, \sqrt{1 - \sin^2 \theta}), \\
    P_{3 \bmu} &= \omega_3 \qty(-1,v_3^x,v_3^y, - \sqrt{1 - \mathbf{v}_3^2}), &\quad
    P_{4 \bmu} &= E_4 \qty(-1, 0, 0, 1).
\end{aligned}
\end{equation}
For particles 1 and 2 we take a symmetric configuration of sparks such that the trajectories meet at a radial depth $z_0$:
\begin{equation}
\begin{split}
    x_1^\mu = \frac{-z_0}{\cos\theta}(1-\cos\theta,\sin\theta,0),
    \qquad x_2^\mu = \frac{-z_0}{\cos\theta}(1-\cos\theta,-\sin\theta,0).
\end{split}
\end{equation}
We also set $x_4^\mu = (0,0,0)$. This data singles out the bulk point $X=z_0(1,0,0,1)$ where the trajectories of $1,2,4$ intersect. The expectation value of $O(x_3)$ then peaks on the future lightcone of $X$, which is the hyperboloid,
\begin{equation}
    x_3^\mu(v_3) = z_0\qty(1+\frac{1}{\sqrt{1-\mathbf{v}_3^2}},
    \frac{v_3^x}{\sqrt{1-\mathbf{v}_3^2}},\frac{v_3^y}{\sqrt{1-\mathbf{v}_3^2}}).
\end{equation}
In order to compute the signal at $x_3$ on the boundary however, we parameterize the location of the third particle as $x_3^\mu = (t_3, x_3, y_3)$, and find the signal as a function of $x_3^\mu$.

We can now proceed to study each of the particles and their evolution in the bulk. First, we compute the $\Delta X$ for each particle specified in equation (\ref{X Sigma AdS}). For the first and the second particles, we find
\begin{equation}
    \Delta X_1^\mu = x_1^\mu - X_1^\mu + \frac{z_0 p_1^\mu}{\abs{p_1}} = (0, 0, 0), \qquad \Delta X_2^\mu = x_2^\mu - X_2^\mu + \frac{z_0 p_2^\mu}{\abs{p_2}} = (0, 0, 0),
\end{equation}
For the third particle, since the particle moves towards the boundary, we have
\begin{equation}
    \Delta X_3^\mu = x_3^\mu - X_3^\mu - \frac{z_0 p_3^\mu}{\abs{p_3}} = \qty(t_3 - \qty(1 + \frac{1}{\sqrt{1-\mathbf{v}_3^2}}) z_0, x_3 - \frac{v_3^x}{\sqrt{1-\mathbf{v}_3^2}}, y_3 - \frac{v_3^y}{\sqrt{1-\mathbf{v}_3^2}}).
\end{equation}
and for the fourth particle, since it both moves towards the boundary and is anti-time-ordered, we find
\begin{equation}
    \Delta X_4^\mu = x_4^\mu - X_4^\mu + \frac{z_0 p_4^\mu}{\abs{p_4}} = (0,0,0).
\end{equation}
Next, we need to compute the widths inside the bulk, using equation (\ref{X Sigma AdS}). Starting with the boundary widths $\sigma_i$, for the first and the second particles we find
\begin{equation}
    \Sigma_1^{\mu\nu} = \sigma_1^{\mu\nu} + \ic \frac{z_0}{\abs{p_1}}\qty(\eta^{\mu\nu} - \frac{p_1^\mu p_1^\nu}{p_1^2}) = \sigma_1^{\mu\nu} + \ic \frac{z_0}{E_1} \begin{pmatrix}
        \tan^2 \theta \sec \theta & \tan \theta \sec^2 \theta & 0 \\
        \tan \theta \sec^2 \theta & \sec^3 \theta & 0 \\
        0 & 0 & \sec \theta
    \end{pmatrix},
\end{equation}
and
\begin{equation}
    \Sigma_2^{\mu\nu} = \sigma_2^{\mu\nu} + \ic \frac{z_0}{\abs{p_2}}\qty(\eta^{\mu\nu} - \frac{p_2^\mu p_2^\nu}{p_2^2}) = \sigma_2^{\mu\nu} + \ic \frac{z_0}{\omega_2} \begin{pmatrix}
        \tan^2 \theta \sec \theta & -\tan \theta \sec^2 \theta & 0 \\
        -\tan \theta \sec^2 \theta & \sec^3 \theta & 0 \\
        0 & 0 & \sec \theta
    \end{pmatrix}.
\end{equation}
The third particle is an outgoing particle that moves towards the boundary, so the width is given by
\begin{equation}
    (\Sigma_3^{\mu\nu})^* = \sigma_3^{\mu\nu} + \ic \frac{z_0}{\abs{p_3}} \qty(\eta^{\mu\nu} - \frac{p_3^\mu p_3^\nu}{p_3^2}) = \sigma_3^{\mu\nu} + \ic \frac{z_0}{\omega_3} \frac{1}{\qty(1 - \mathbf{v}_3)^{3/2}} \begin{pmatrix}
        \mathbf{v}_3^2 & v_3^x & v_3^y \\
        v_3^x & 1 - (v_3^y)^2 & v_3^x v_3^y \\
        v_3^y & v_3^y v_3^x & 1 - (v_3^x)^2
    \end{pmatrix},
\end{equation}
and the fourth particle is outgoing, moving towards the boundary in an anti-time-ordered fashion, so we find:
\begin{equation}
    \qty(\Sigma_4^{\mu\nu})^* = \sigma_4^{\mu\nu} - \ic \frac{z_0}{\abs{p_4}} \qty(\eta^{\mu\nu} - \frac{p_4^\mu p_4^\nu}{p_4^2})
    = \sigma_{4}^{\mu\nu} - \ic \frac{z_0}{E} \begin{pmatrix}
        0 & 0 & 0 \\
        0 & 1 & 0 \\
        0 & 0 & 1
    \end{pmatrix}.
\end{equation}

For the widths inside the bulk, we need to uplift the widths we obtained from the evolution of the boundary widths. 
\begin{equation}
    \Sigma^{-1(*)}_{i \, \bmu\bnu} = \begin{pmatrix}
        \Sigma^{-1(*)}_{i \, \mu\nu} & - \frac{\Sigma^{-1(*)}_{i \, \mu\nu} p_i^\nu}{\abs{p_i}} \\[5pt]
        - \frac{\Sigma^{-1(*)}_{i \, \mu\nu} p_i^\mu}{\abs{p_i}} & \frac{\Sigma^{-1(*)}_{i \, \mu\nu} p_i^\mu p_i^\nu}{\abs{p_i}^2}
    \end{pmatrix}, \qquad i \in \{1, 2, 4\}.
\end{equation}
For the first and the second particles, this is the case since they both move towards the bulk point ($\dot{z} > 0$), and for particle four, it is justified by the fact that the particle is moving towards the boundary in an anti-time-ordered manner. Finally, for the third particle, we find
\begin{equation}
    \Sigma^{-1*}_{3 \, \bmu\bnu} = \begin{pmatrix}
        \Sigma^{-1*}_{3 \, \mu\nu} & \frac{\Sigma^{-1*}_{3 \, \mu\nu} p_3^\nu}{\abs{p_3}} \\[5pt]
        \frac{\Sigma^{-1*}_{3 \, \mu\nu} p_3^\mu}{\abs{p_3}} & \frac{\Sigma^{-1*}_{3 \, \mu\nu} p_3^\mu p_3^\nu}{\abs{p_3}^2}
    \end{pmatrix}.
\end{equation}
These all have the property that
\begin{equation}
    \Sigma^{-1(*)}_{i \, \bmu \bnu} P_i^\bnu = 0.
\end{equation}

We can focus on the case where $v_3^y = 0$ to compute the Gaussian envelope of the signal (\ref{effective delta function}) on the boundary and plot it. Given an arbitrary point on the boundary with coordinates $(t, x, 0)$, for a particle that hits the bulk point at $X = z_0(1,0,0,1)$, we can solve $\Delta X_3 = 0$ to find
\begin{equation}
    v_3^x = \frac{2 t x}{x^2 + y^2}.
\end{equation}
This is the final ingredient in computing the signal (\ref{effective delta function}) which is plotted in figure \ref{fig:ads-gaussian}. The bulk point is located at $z_0 = 1$ and the momenta of the first and second particles are fixed by $\theta = \pi/3$. The energies used in this experiment are
\begin{equation}
    E_1 = 100, \quad \omega_2 = 100, \quad \omega_3 = 50, \quad \text{and} \quad E_4 = 150.
\end{equation}
The Mandlestam variables for this experiment read
\begin{equation}
    s = 30000, \qquad \text{and} \qquad t = 15000.
\end{equation}
The widths used for particles in the plot are given by $\sigma_i = \diag((\fdiff E)^{-2}, (\fdiff E)^{-2}, (\fdiff E)^{-2})$ for $\fdiff E = 150$.
All the numbers are given in AdS units.

\section{Entering and exiting a CFT} \label{app:portal}
Here we consider a holographic CFT in $d$ spacetime dimensions weakly coupled to the fields of an ambient $(d+1)$-dimensional flat space theory (``us''). This discussion is probably not new, and the goal is to explain how the wavepacket correlators discussed in the main text arise naturally when an ambient observer interacts with the CFT. The $d=3$ case, for example, would be relevant for a CFT on a sheet that interacts electromagnetically with our photon.

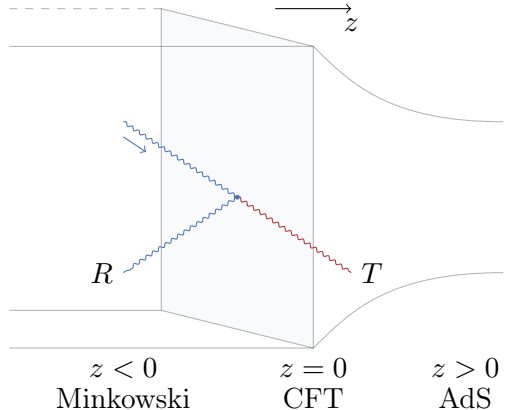
\begin{figure}
    \centering
    \begin{tikzpicture}
    \draw[fill=mblue!10,opacity=0.3] (0.5,-0.5) -- (2.5,-1) -- (2.5,3) -- (0.5,3.5) -- cycle;
    \draw[opacity=0.3] (2.5,3) .. controls (3, 2.5) and (3.5, 2) .. (5, 2);
    \draw[opacity=0.3] (2.5,-1) .. controls (3, -.5) and (3.5, 0) .. (5, 0);
    \draw[opacity=0.3] (2.5,3) -- (-1.5,3) (2.5,-1) -- (-1.5,-1) (0.5,-0.5) -- (-1.5,-.5);
    \draw[opacity=0.3,dashed] (0.5,3.5) -- (-1.5, 3.5);
    \draw[color=mblue,->] (0,1.8)--++(0.3,-0.2);
    \draw[color=mblue,decorate,decoration={snake,amplitude=0.2mm,segment length=1mm}] (0,2) -- (1.5,1) -- (0,0);
    \draw (0,0) node[anchor=east] {$R$} (3,0) node[anchor=west] {$T$};
    \draw (0,-1) node[anchor=north] {$z<0$} (0,-1.4) node[anchor=north] {Minkowski}
    (4.5,-1) node[anchor=north] {$z>0$} (4.5,-1.4) node[anchor=north] {AdS} (2.5,-1) node[anchor=north] {$z=0$} (2.5,-1.4) node[anchor=north] {CFT};
    \draw[color=mred,decorate,decoration={snake,amplitude=0.2mm,segment length=1mm}] (1.5,1) -- (3,0);
    \filldraw[color=mblue] (1.5,1) circle (.7pt);
    \draw[->] (2,3.5)--++(1,0) node[below] {$z$};
    \end{tikzpicture}
    \caption{A photon walks into a CFT and enters an AdS space.}
    \label{fig:entering-cft}
\end{figure}

Let us first consider the case where the CFT, at $z=0$, only interacts with the ambient modes coming from $z<0$, ie. the CFT rests on an insulating material. The action is
\begin{equation} \label{S0 CFT+us}
    S_{\rm CFT+us}= S_{\rm CFT}  +\int_{z<0} \diff^{d+1}x  \frac{-F_{\mu\nu}F^{\mu\nu}}{4e^2_{\rm us}}+
    \int d^dx J^\mu A_\mu(x,z=0)
\end{equation}
where $J^\mu$ is a conserved U(1) current in the CFT. Note that even though we are writing $S_{\rm CFT}$ it won't matter here whether the CFT admits a Lagrangian description. If the theory is holographic, we can identify the source $A_\mu\big|_{z=0}$ with the boundary limit of an AdS bulk field\: $\lim_{z\to 0^+} A_\mu^{\rm bulk}(z)$, and the relevant part of the action is equivalent to:
\begin{equation} \label{S CFT+us}
    S_{\rm CFT+us} \simeq 
    -\int_{z<0} \diff^{d+1}x \, \eta^{\bmu\bnu}
    \eta^{\brho\bsigma}\frac{
    F_{\bmu\brho}F_{\bnu\bsigma}}{4e_{\rm us}^2}
    -\int_{z>0} \diff^{d+1}x \, \sqrt{-g}
    g^{\bmu\bnu}g^{\brho\bsigma}\frac{ F_{\bmu\brho}F_{\bnu\bsigma}}{4e_{\rm AdS}^2}.
\end{equation}
In other words, for $z<0$ we have a gauge field that exists in our spacetime, for $z>0$ we have another gauge field that exists in the AdS spacetime representing the CFT, and at $z=0$ the two gauge fields are equated with each other. This geometry is shown in figure \ref{fig:entering-cft}.

We can now treat \eqref{S CFT+us} as a standard transmission-reflection problem. Suppose the right-moving photon plane wave approaches the interface from $z<0$. Assume the bulk geometry is empty AdS and the wave is transversely polarized. At short times (before the wave has time to bounce off any feature in the AdS space) we can make an ansatz
\begin{equation} \label{A transmission ansatz}
    A_\mu(x,z) = \bdypol_\mu\,\ep^{ip_\mu x^\mu} \times \begin{cases} 
    \ep^{\ic z\abs{p}} + R \ep^{-\ic z\abs{p}},\quad &z<0,\\
    T \ep^{\ic \pi \frac{d-1}{4}} \sqrt{\pi\abs{p}/2}\,z^{\frac{d-2}{2}} H^{(1)}_{\frac{d-2}{2}}(\abs{p}z), & z>0,
\end{cases}
\end{equation}
where $\abs{p}=\sqrt{-p^2}$, $p^\mu \bdypol_\mu=0$ and we used the radial solution \eqref{radial sol spin1} specialized to a conserved current (and normalized to become a simple plane wave $T \ep^{\ic \abs{p}z}z^{\frac{d-3}{2}}$ at large $z$). The boundary condition and equation of motion at $z=0$ then give the junction conditions 
\begin{equation}  \label{A transmission eqs}
    \lim_{z\to 0^-} A_\mu = \lim_{z\to 0^+} A_\mu,\qquad
    \lim_{z\to 0^-}
    \frac{1}{e_{\rm us}^2}\eta^{z\brho}\eta^{\mu\bsigma}F_{\brho\bsigma}=
    \lim_{z\to 0^+}
    \frac{1}{e_{\rm AdS}^2}\sqrt{-g} g^{z\brho}g^{\mu\bsigma}F_{\brho\bsigma},
\end{equation}
which can be solved for $R$ and $T$. For example, in $d=3$ the solution is
\begin{equation} \label{RT}
    R = \frac{e^2_{\rm AdS}-e^2_{\rm us}}{e^2_{\rm AdS}+e^2_{\rm us}},\qquad
    T = \frac{2e^2_{\rm AdS}}{e^2_{\rm AdS}+e^2_{\rm us}}.
\end{equation}
This is identical to the usual transmission-reflection coefficients for light entering a medium with a nontrivial dielectric constant. For $d+1=4$ this should not be too surprising since Maxwell's equation is conformally invariant: the action \eqref{S CFT+us} is insensitive to the distinction between the AdS and Minkowski geometry.

Similar formulas apply to the reverse problem of a plane wave moving within the AdS bulk toward its boundary. In the standard AdS/CFT setup one imposes a perfectly reflecting boundary condition, which makes (global) AdS akin to a box. Here instead we get a leaky box due to partial transmission of AdS excitations into the ambient space.

The $d$-dimensional version of \eqref{RT} is qualitatively similar, with the couplings replaced by the dimensionless combinations $e^2_{\rm us}\abs{p}^{d-3}$ and $e^2_{\rm AdS}R_\AdS^{3-d}$ up to constant factors. The standard AdS/CFT decoupling limit \cite{Maldacena:1997re} is recovered when $\abs{p}\to 0$ in $d>3$, in which case the AdS boundary becomes again perfectly reflecting for the waves coming from within AdS. However, at finite momentum $\abs{p}$, a CFT excitation can leak out with finite probability. For gravity, the analogous dimensionless couplings are $G_{\rm us}\abs{p}^{d-1}$ and $G_{\rm bulk} R_\AdS^{1-d}$.

It is amusing that there is no reflection in $d=3$ when the couplings match: $e^2_{\rm us}=e^2_{\rm AdS}$. The latter quantity determines the current-current correlator in the CFT (before it is coupled to our spacetime) \cite{Freedman:1998tz}:\footnote{The numerical factor also matches with $(2\Delta-d)C_{\Delta,1}$ from \eqref{boundary condition vector}.}
\begin{equation}
    \langle J^\mu(x)J^\nu(0)\rangle_{\rm CFT} =
\qty(\frac{1}{e^2_{\rm AdS}R_\AdS^{3-d}}
\frac{\Gamma(d)}{\pi^{\frac{d}{2}}\Gamma(\frac{d}{2}-1)})
    \frac{\eta^{\mu\nu}-2x^\mu x^\nu/x^2}{(x^2)^{d-1}}.
\end{equation}
Setting $\frac{e^2_{\rm us}}{4\pi}=\alpha \approx \frac{1}{137}$ and $d=3$, we see that perfect transmission from the vacuum into the CFT happens when the quantity within parenthesis
is equal to $1/(2\pi^3\alpha)$.
For comparison, a 2+1-dimensional free CFT with $N_f$ massless Dirac fermion would give a parenthesis equal to $N_f/(8\pi^2)$. In other words, perfect transmission occurs when the CFT has the same current-current two-point function as $\frac{4}{\pi\alpha}\approx 174$ free fermions. We do not know how to engineer such a system (the perfect-transmission condition can also change if an intermediate layer is inserted between the CFT and the ambient space).

The analogy with standard optical interfaces seems robust. For example, an experimentally realizable CFT might have an effective speed of light $c_{\rm CFT}\neq c_{\rm out}$ that differs from that in the ambient space, e.g., related to the Fermi velocity if the CFT is realized as a system of strongly interacting fermions.
In this case, it is easy to see that the transmitted photon will enter the bulk AdS space at a refraction angle that satisfies Snell's law: $\frac{\sin\theta_{\rm CFT}}{c_{\rm CFT}} = \frac{\sin\theta_{\rm out}}{c_{\rm out}}$. The formulas for the transmission and reflection coefficients then depend on the incidence angle and polarization and are slightly more complicated than \eqref{RT}, but still no more complicated than that for the interface between say air and glass.

One can also consider a codimension-one sheet of material coupled to a flat spacetime on both sides, e.g. a CFT sheet hanging in the vacuum. Then a wave impending upon the sheet from the left will partly reflect, partly transmit across the sheet, and partly ``enter'' the AdS space that describes the CFT sheet. In effect, we have a junction between three half-space universes. The ansatz \eqref{A transmission ansatz} then has three free parameters, which can again be determined since we have an additional boundary condition in \eqref{A transmission eqs} from equating the three fields at $z=0$. Similarly, a wave coming from within AdS toward its boundary can partly exit on either side of the sheet, as if going through a beam splitter.

After a CFT is coupled to an external system as in \eqref{S0 CFT+us}, it is of course generally no longer a CFT; at best it could flow to a so-called boundary conformal field theory (BCFT) if the external system is conformal. What we assumed above is that the interaction is sufficiently weak that we can use the undisturbed CFT as a starting point; for electromagnetism, the relevant small parameter is $\alpha$. One effect of interactions is that the current density on the sheet will cease to be conserved if the charge can be exchanged with the ambient space, and generally, the energy-momentum on the sheet is also no longer conserved. In any case, the bulk actions in the present setup are unmodified away from $z=0$.

\bibliographystyle{JHEP}
\bibliography{refs}

\end{document}